\documentclass[a4paper,onecolumn,showpacs,notitlepage,floatfix,nofootinbib,superscriptaddress,prd]{revtex4-2}

\usepackage{graphicx,amsmath,array,verbatim,setspace}
\usepackage{hyperref}       
\usepackage{url}            
\usepackage{amsfonts}       
\usepackage{color}
\usepackage{amssymb,amsthm}
\usepackage{tikz}
\usepackage[subrefformat=parens]{subcaption}
\usetikzlibrary{positioning}
\usetikzlibrary{shapes.geometric}
\usetikzlibrary{shapes.misc}
\usepackage{tikz-cd}

\allowdisplaybreaks

\newcommand{\be}{\begin{eqnarray}}
\newcommand{\ee}{\end{eqnarray}}

\newcommand{\nl}{\nonumber \\}
\newcommand{\pd}{\partial}

\newcommand{\bul}{\overset{\underset{\bullet}{}}}
\newcommand{\eqsym}{\overset{\underset{{\rm sym}.}{}}{=}}

\makeatletter
\newcommand{\xnrightarrow}[2][]{%
  \mathrel{%
    \vphantom{\xrightarrow[#1]{#2}}%
    \ooalign{\hidewidth\neg@arrow\hidewidth\cr$\m@th\xrightarrow[#1]{#2}$\cr}%
  }%
}
\newcommand{\neg@arrow}{%
  $\m@th\vcenter{\hbox{%
    \rotatebox[origin=c]{-45}{\scalebox{1.5}[1]{$\m@th\scriptscriptstyle|$}}%
  }}$
}
\makeatother

\definecolor{dgreen}{rgb}{0.0, 0.5, 0.0}
\definecolor{dred}{rgb}{0.82, 0.1, 0.26}
\definecolor{dblue}{rgb}{0.0, 0.0, 1.0}

\begin{document}
\title{
  Exact WKB analysis for ${\cal PT}$ symmetric quantum mechanics:\\
  Study of the Ai-Bender-Sarkar conjecture
}

\author{Syo Kamata}
\email{skamata11phys@gmail.com}
\affiliation{Department of Physics, The University of Tokyo,
  7-3-1 Hongo, Bunkyo-ku, Tokyo 113-0033, Japan}

\begin{abstract} 
  We consider exact WKB analysis to a ${\cal PT}$ symmetric quantum mechanics defined by the potential, $V(x) = \omega^2 x^2 + g x^2(i x)^{\varepsilon=2}$ with $\omega \in {\mathbb R}_{\ge 0}$, $g \in {\mathbb R} _{> 0}$.
  We in particular aim to verify a conjecture proposed by Ai-Bender-Sarkar (ABS), 
  that pertains to a relation between  $D$-dimensional ${\cal PT}$-symmetric theories and analytic continuation (AC) of Hermitian theories concerning the energy spectrum or Euclidean partition function.
  For the purpose, we construct energy quantization conditions by exact WKB analysis and write down their transseries solution by solving the conditions.
  By performing  alien calculus to the energy solutions, we verify  validity of the ABS conjecture and seek a possibility of its alternative form  by Borel resummation theory if it is violated.
  Our results claim that the validity of the ABS conjecture drastically changes depending on whether $\omega > 0$ or $\omega = 0$: If ${\omega}>0$, then the ABS conjecture is violated when exceeding the semi-classical level of the first non-perturbative order, but its alternative form is constructable by Borel resummation theory.
  The ${\cal PT}$ and the AC energies are related to each other by a one-parameter Stokes automorphism, and a median resummed form, which corresponds to a formal exact solution, of the AC energy (resp. ${\cal PT}$ energy) is directly obtained by acting Borel resummation to a transseries solution of the ${\cal PT}$ energy (resp. AC energy).
  If $\omega = 0$, then, with respect to the inverse energy level-expansion, not only perturbative/non-perturbative structures of the ${\cal PT}$ and the AC energies but also their perturbative parts do not match with each other.
  These energies are independent solutions, and no alternative form of the ABS conjecture can be reformulated by Borel resummation theory.
\end{abstract}

\maketitle

\tableofcontents


\section{Introduction} \label{sec:intro}

Non-Hermitian theory is an exciting topic in a broad area of theoretical and experimental physics. 
Although ${\cal CPT}$ invariance is a theoretically fundamental concept for field theories,  non-Hermitian theories show rich physics depending on relaxed symmetry.
Non-Hermitian effects were experimentally observed, such as in open systems and non-equilibrium systems, and such a non-Hermitian physics is expected to offer further insights into these interesting phenomena, providing us with a deeper understanding~\cite{el2018non,Ashida:2020dkc,Okuma:2022bnb}.

It is known that ${\cal PT}$ symmetric quantum mechanics have a special property in their energy spectrum~\cite{Bender:1998gh}.
When we consider the ${\cal PT}$ symmetric potential defined by
\be
&& V_{\cal PT} (x) = \omega^2 x^2 + g x^2 (i x)^\varepsilon, \qquad (\omega \in {\mathbb R}_{\ge 0}, \ g, \varepsilon \in {\mathbb R}_{>0}, \  x \in {\mathbb C}) \label{eq:VPT_def}
\ee
despite lack of ${\cal CPT}$ invariance, the ${\cal PT}$ symmetric Hamiltonian gives a real and bounded energy spectrum due to ${\cal PT}$ invariance under the ${\cal P}$ and ${\cal T}$ transforms,
\be
{\cal P}: \ x \rightarrow -x, \qquad {\cal T}: \ x \rightarrow \bar{x}, \ i \rightarrow - i, 
\ee
where $\bar{x}$ is complex conjugate of $x$~\cite{Bender:1998ke,Dorey:2001uw,Jones:2006qs,Bender:2019}.
Generalizations to $D$-dimensional ${\cal PT}$ symmetric field theories were also considered in Refs.\cite{Bender:2018pbv,Felski:2021evi}.
In such a generalization to field theories, it was suggested that Hermitian theories given by a quartic potential,
\be
V_{\cal H}(x) = \omega^2 x^2 + \lambda x^4,  \qquad (\omega \in {\mathbb R}_{\ge 0}, \lambda \in {\mathbb R})
\label{eq:VH_def}
\ee
with a negative coupling $\lambda$ relates to the ${\cal PT}$ symmetric theories with $\varepsilon = 2$ in Eq.(\ref{eq:VPT_def})~\cite{Bender:2021fxa,Mavromatos:2021hpe,Grunwald:2022kts,Ai:2022csx,Lawrence:2023woz}.
In particular, Ai-Bender-Sarkar (ABS) made a conjectured relation of Euclidean partition functions for $D \ge 1$ from a semi-classical analysis~\cite{Ai:2022csx}, that takes the form that
\be
\log Z_{\cal PT}(g) &=& {\rm Re} \log Z_{\cal H}(\lambda = -g + i 0_\pm), \qquad g \in {\mathbb R}_{>0}. \label{eq:conj_Ai} 
\ee
However, a contradiction was reported in Ref.~\cite{Lawrence:2023woz}, and the authors claimed violation of the ABS conjecture for the pure quartic potential by the study of quantum mechanics and the $D=0$ $N$-component scalar model.
By revealing validity and a limitation of the ABS conjecture, it is expected to provide us physical meaning of theories with an unbounded potential from the relation to ${\cal PT}$ symmetric theories.
Symmetry is quite crucial for any physical phenomena, and thus the relation would be possibly helpful to characterize their physical features.
Especially, applications to four-dimensional theories are important issues, such as their IR physics.
See Refs.~\cite{Romatschke:2022jqg,Romatschke:2022llf,Grable:2023paf,Weller:2023jhc,Romatschke:2023sce}, for example.
In order to theoretically investigate the ABS conjecture in a more rigorous way, however, a method beyond-semi-classical analysis must be necessary.

Since a Euclidean partition function can be written down by an energy spectrum, analysis of an energy spectrum is direct investigation of the ABS conjecture expressed by partition functions.
In quantum mechanics ($D=1$), exact WKB analysis (EWKB) is quite powerful for such a problem related to an energy spectrum~\cite{Berry,BPV,Voros1983,Silverstone, DDP2,DP1,Takei1,Takei2,Takei3,Kawai1,AKT1,Schafke1,Iwaki1,Alvarez1, Zinn-Justin:2004vcw, Zinn-Justin:2004qzw,Dunne:2013ada,Dunne:2014bca}. 
EWKB is not only a generalization of semi-classical analysis treating all order of $\hbar$ but also some sort of Borel resummation theory and resurgence theory, specialized in a one-dimensional Schr\"{o}dinger equation.
EWKB offers an energy quantization condition (QC) by performing analytic continuation on the complex $x$-plane, and then its energy spectrum can be found by solving it.
When a perturbative part of the energy solution is divergent series and Borel non-summable, one can access to non-perturbative informations from the perturbative expansion through Borel resummation. 
Such a relation among perturbative and non-perturbative parts is called as \textit{resurgence} or \textit{resurgent relation}~\cite{Ec1,delabaere1994introduction,Sauzin:1405,Marino:2012zq,Dorigoni2019,Aniceto:2018bis,delabaere2016divergent,Costin2008}.
Non-perturbative effects crucially affect many physical phenomena, and thus EWKB and related approaches were employed in many physical contexts of non-perturbative physics~\cite{Mironov:2009uv,Gaiotto:2009hg,ashok2016exact,taya2021exact,enomoto2022exact,kashani2015pure,Basar:2017hpr,Cavusoglu:2023bai, suzuki2023exact,Sueishi:2020rug,Sueishi:2021xti,Kamata:2021jrs,Bucciotti:2023trp,Grassi:2014uua}.
It is also an interesting topic in mathematical physics, such as a relation to integrable systems~\cite{takei2004toward,dunne2017wkb,hollands2020exact,imaizumi2020exact,Emery:2020qqu,Ito:2018eon,ito2023exact,vanSpaendonck:2022kit,Dorey:2007zx,Franco:2015rnr,Hatsuda:2015qzx}, and analyses of ${\cal PT}$ symmetric quantum mechanics were considered in Ref.~\cite{Emery:2019znd,Emery:2020qqu}.
In addition, ${\cal PT}$ symmetric eigenvalue problems with numerical approaches were studied in, e.g. Ref.~\cite{Noble2013,Noble2017,Khan:2022uyz}.

In this paper, we verify the ABS conjecture for $D=1$ in Eq.(\ref{eq:conj_Ai}) by EWKB which is a beyond-semi-classical analysis. 
We construct QCs of the theories by EWKB and then obtain formal transseries solutions of the energy spectra by solving them.
From those results, we consider validity of the ABS conjecture and seek a possibility of its alternative form by Borel resummation theory if it is violated.

\section{Summary of our results and structure of this paper}
As described in the end of Sec.~\ref{sec:intro}, our main purpose of this paper is verification of the ABS conjecture for $D=1$ by using EWKB and reformulation by Borel resummation theory if it is violated.
Since a Euclidean partition function can consist of an energy spectrum, the ABS conjecture expressed by partition functions can be known by investigation of the relation of the energies.
This study is a benchmark for generalizations to field theories.
\\ \, \par
\begin{figure}[tbp]
 \centering
 \includegraphics[width=140mm]{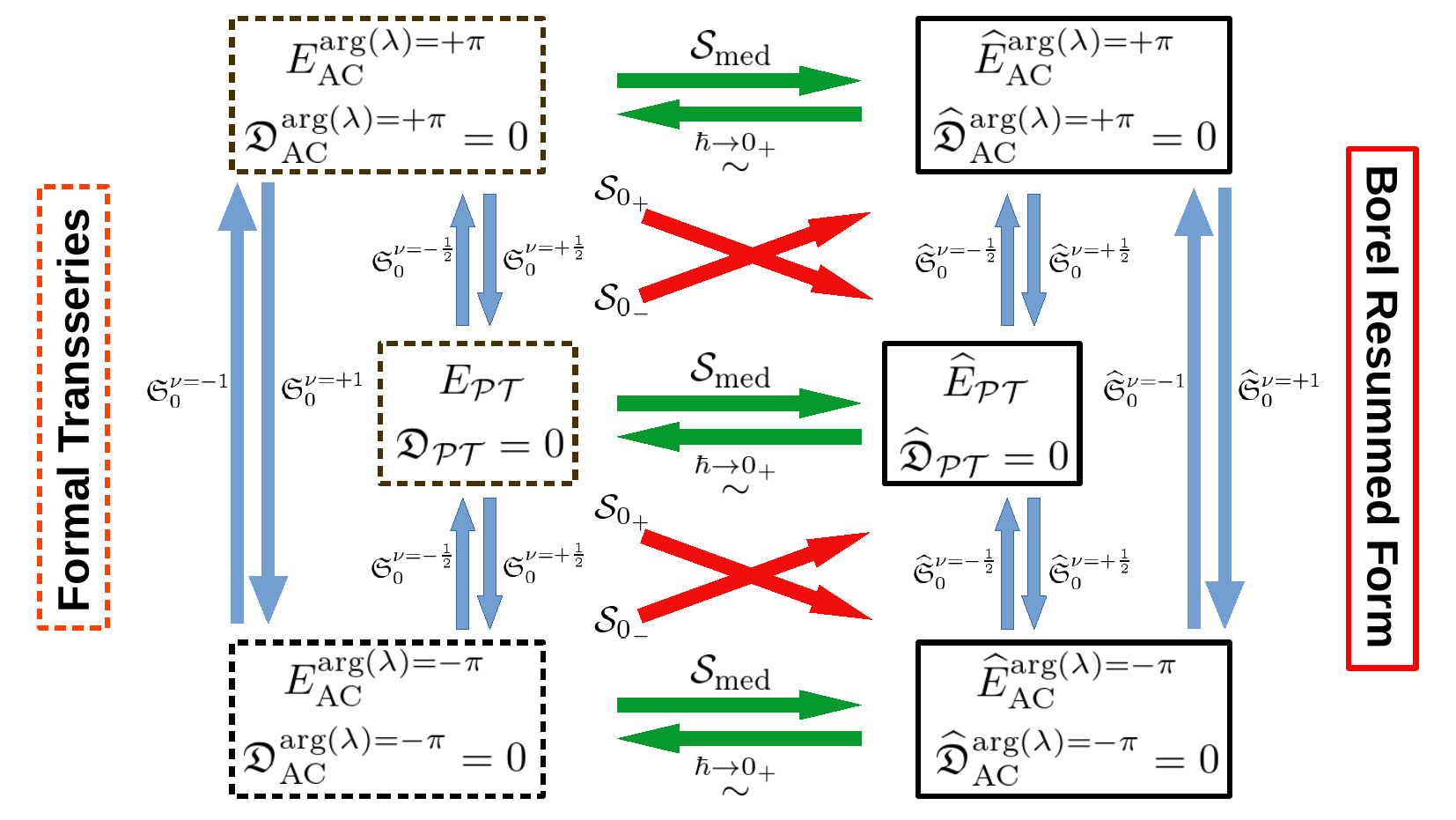}
\caption{
   Schematic figure of our modified ABS conjecture for $\omega>0$.
   Borel resummation ${\cal S}_{0_\pm}$ and a one-parameter Stokes automorphism ${\frak S}^{\nu \in {\mathbb R}}_0$ are defined around $\arg(g)=0$. 
   Formal transseries of the ${\cal PT}$ and the AC energies (resp. QCs), denoted by $E_{{\cal PT}/{\rm AC}}$ (resp. ${\frak D}_{{\cal PT}/{\rm AC}}$), are connected to each other by the one-parameter Stokes automorphism with $\nu = \pm 1/2$.
   The superscript, $\arg(\lambda) = \pm \pi$, corresponds to $ \lambda = -g + i 0_\pm$ in Eq.(\ref{eq:conj_Ai}). 
  Acting the median resummation ${\cal S}_{{\rm med}}$ to the formal transseries gives their median resummed forms, $\widehat{E}_{{\cal PT}/{\rm AC}}$ (resp. $\widehat{{\frak D}}_{{\cal PT}/{\rm AC}}$), that reproduce the original transseries by taking $\hbar \rightarrow 0_+$.
  By introducing another one-parameter Stokes automorphism acting to Borel resummed forms, $\widehat{\frak S}^{\nu \in {\mathbb R}}$, the similar relations among the Borel resummed forms of the ${\cal PT}$ and the AC energies (resp. QCs) hold formally.
  Borel resummation ${\cal S}_{0_\pm}$ makes a direct connection from $E_{{\cal PT}/{\rm AC}}$ (resp. ${\frak D}_{{\cal PT}/{\rm AC}}$) to $\widehat{E}_{{\rm AC}/{\cal PT}}$ (resp. $\widehat{\frak D}_{{\rm AC}/{\cal PT}}$).
}
\label{fig:summary_PT}
\end{figure}

We summarize our results below:
\begin{itemize}
\item If $\omega>0$, then the ABS conjecture is violated when exceeding the semi-classical level of the first non-perturbative order.
  While the ${\cal PT}$ energy is purely perturbative and Borel non-summable, the AC energy contains non-perturbative contributions in addition to the same perturbative part to the ${\cal PT}$ energy.
  The ABS conjecture is violated in the second non-perturbative sector and higher, which contains both real and imaginary parts in the AC energy, but an alternative form is constructable by a one-parameter Stokes automorphism and Borel resummation.
  The ${\cal PT}$ and the AC energies are related to each other by the one-parameter Stokes automorphism, and a median resummed form, which corresponds to a formal exact solution, of the AC energy (resp. the ${\cal PT}$ energy) is directly obtained by acting Borel resummation to a transseries solution of the ${\cal PT}$ energy (resp. the AC energy).  
  The relations of the energy solutions for the formal transseries and the Borel resummed forms are represented in Eqs.(\ref{eq:EPT_DDP}) and (\ref{eq:ABS_hatEPT}), respectively.
  The overall view is schematically shown in Fig.~\ref{fig:summary_PT}.
  The Euclidean partition functions also exhibit the same relations, and those are expressed by Eqs.(\ref{eq:ZPT_ZAC}) and (\ref{eq:ABS_hatZPT}).
\item If $\omega = 0$, then the ABS conjecture is not satisfied.
  With respect to the inverse energy level-expansion, the ${\cal PT}$ energy is purely perturbative and Borel non-summable, and the AC energy has non-perturbative parts.  
  In addition to the difference of the perturbative/non-perturbative structures between the ${\cal PT}$ and the AC energies, even their perturbative parts do not match with each other unlike the case of $\omega > 0$.
  Since their QCs can not take the same form by a one-parameter Stokes automorphism, those are independent solutions.
  No alternative form of the ABS conjecture can be reformulated by Stokes automorphism and Borel resummation.
\end{itemize}
In this paper, we mainly address the case of Eq.(\ref{eq:VPT_def}) with $\varepsilon = 2$, but this result is directly extendable to $V_{\cal PT} = \omega^2 x^{2} + g x^{2K} (i x)^{\varepsilon=2}$ with $K \in {\mathbb N}$.

This paper is organized as follows:
In Sec.~\ref{sec:Borel_EWKB}, we review Borel resummation theory and EWKB.
We firstly explain main concepts of Borel resummation and Stokes automorphism in a slightly generic point of view in Sec.~\ref{sec:Borel_Stokes}, and then introduce EWKB to provide QCs in Sec.~\ref{sec:EWKB}.
After the review part, we consider a quartic potential with a quadratic term in Sec.~\ref{sec:with_mass_term}.
As warm-up, we apply EWKB to the Hermitian potential with a positive coupling $\lambda$ in Sec.~\ref{sec:warm_up}. 
Then, in Sec.~\ref{sec:PT_sym_mass}, we derive the ${\cal PT}$ and the ${\rm AC}$ energy solutions from the negative coupling potential.
From the results, we modify the ABS conjecture for $\omega>0$ based on Borel resummation theory in Sec.~\ref{sec:reform_ABS}.
In Sec.~\ref{sec:without_mass_term}, we consider a quartic potential without a quadratic term.
In Sec.~\ref{sec:Herm_sym_wo_mass}, we begin with the Hermitian potential with a positive coupling $\lambda$ and obtain the energy solution by using the inverse energy level-expansion.
In Sec.~\ref{sec:PT_sym_wo_mass}, we consider the negative coupling potential.
In Sec.~\ref{sec:impossibility_PT_AC}, we show the fact that no alternative form of the ABS conjecture can be reformulated by Stokes automorphism and Borel resummation.
In Sec.~\ref{sec:add_remark}, we make some remarks about a generalization to $V_{\cal PT} = \omega^2 x^{2} + g x^{2K} (i x)^{\varepsilon=2}$ with $K \in {\mathbb N}$ and spectral reality from the viewpoint of EWKB.
Technical computations, such as derivation of non-perturbative parts and alien calculus for the energy spectrum, are summarized in Appendices~\ref{sec:der_G} and \ref{sec:DDP_energy}, respectively.

For QCs, denoted as ${\frak D} = 0$, we frequently use the symbol that ${\frak D}_1 \propto {\frak D}_2$ through this paper, which means that ${\frak D}_1$ and ${\frak D}_2$ are equivalent to each other except an overall factor not to affect their energy solutions.

\section{Preliminary: Borel resummation theory and exact WKB analysis} \label{sec:Borel_EWKB}
In this section, we introduce Borel resummation and EWKB.
In Sec.~\ref{sec:Borel_Stokes}, we review Borel resummation and Stokes automorphism.
Then, in Sec.~\ref{sec:EWKB}, we explain some basics of EWKB.

\subsection{Borel resummation and Stokes automorphism} \label{sec:Borel_Stokes}
We introduce concepts of Borel resummation (or Borel-\'{E}calle resummation) and Stokes automorphism.
Intuitively, Borel resummation is a method to reconstruct a function from a divergent series contained in transseries, and analyzing transseries is essentially identical to investigation of properties Borel resummed forms, i.e. their original functions associated to the transseries\footnote{
In the terminology of Ref.~\cite{Costin2008}, Borel resummation is a mapping from a transseries to an analyzable function;
\textit{Determining the transseries of a function $f$ is the ``analysis" of $f$ , and transseriable functions are ``analyzable," while the opposite process, reconstruction by BE summation of a function from its transseries is known as ``synthesis."}}.
There are many textbooks and reviews of Borel resummation theory and resurgence theory.
See Refs.~\cite{Ec1,delabaere1994introduction,Sauzin:1405,Marino:2012zq,Dorigoni2019,Aniceto:2018bis,delabaere2016divergent,Costin2008}, for example.

Let us  begin with the following formal power series expanded by $\hbar$:
\be
f(\hbar) \sim \sum_{n \in {\mathbb N}} c_{n}\hbar^n  \quad \mbox{as} \quad \hbar \rightarrow 0_+. \qquad (c_n \in {\mathbb C}) \label{eq:def_f_asym}
\ee
For simplicity, we assumed that $f$ does not have non-perturbative contributions and oscillations such as $e^{-S/\hbar}$ with ${\rm Re} (S) \ge 0$.
We also assume that $f$ is a divergent series and Gevrey-1 class, i.e.,
\be
c_{n} \sim  A S^{-n} n! \quad \mbox{as} \quad n \rightarrow \infty, \qquad  \lim_{n \rightarrow \infty} \left| \frac{c_{n}}{c_{n+1}} \right| = 0. \qquad (A,S \in {\mathbb C}) \label{eq:asym_f}
\ee
Borel resummation ${\cal S}_{\theta}$ is defined as an operator acting to a formal power series and being composition of Borel transform ${\cal B}$ and Laplace integral ${\cal L}_{\theta}$, i.e., ${\cal S}_{\theta} = {\cal L}_{\theta} \circ {\cal B}$.
Those are defined as
\be
&& {\cal B} [f](\xi) :=\sum_{n \in {\mathbb N}} \frac{c_n}{\Gamma(n)} \xi^{n-1} = f_B(\xi), \qquad
{\cal L}_\theta [f_B](\hbar) := \int_{0}^{\infty e^{i \theta}} d \xi \,e^{-\frac{\xi}{\hbar}} f_B(\xi). \label{eq:_def_Borel_Lap}
\ee
The Borel transformed expansion, $f_B$, is an analytic function in a neighborhood of $\xi = 0$, but it is not on the entire complex $\xi$-plane when $f$ is a divergent series.
If no singular point lays on the integration ray with $\arg(\xi) = \theta$, then the Laplace integral is formally performable.
In such a case, we say $f$ is Borel (re)summable along $\arg(\xi) = \theta$. Otherwise, $f$ is Borel non-(re)summable.
Borel resummation is a homomorphism:
\be
&&  {\cal S}_\theta [f_1+ f_2] = {\cal S}_\theta[f_1] + {\cal S}_\theta[f_2], \qquad  {\cal S}_\theta [f_1 f_2] = {\cal S}_\theta[f_1] \cdot {\cal S}_\theta[f_2]. \label{eq:S_homo}
\ee
Notice that the Borel resummed form, ${\cal S}_{\theta}[f]$, is a (formal) function, not a power series or transseries.

Suppose the case that $f$ is Borel non-summable along $\theta = 0$. 
When $f$ is Borel non-summable, one usually adds a infinitesimal phase to the integration ray to make $f$ Borel summable, as $\theta= 0_\pm$.
However, because of a discontinuity from the singular point, the Borel resummed form depends on the sign of the phase, namely,
\be
{\cal S}_{0_+}[f] \ne {\cal S}_{ 0_-}[f].
\ee
This implies that the resulting form of ${\cal S}_{ 0_\pm}[f]$ does not match with $f$ in the asymptotic limit, i.e.,\footnote{
  In this expression, we used a symbol ``$0_\mp$'' instead of ``$0_\pm$'' in $f^{0_\mp}$ to adjust to the notation in Eq.(\ref{S0pm_f}).
}
\be
   {\cal S}_{0_\pm}[f] \overset {\hbar \rightarrow 0_+} \sim f^{0_\mp} \ne f, \label{eq:simh}
\ee
where $\overset {\hbar \rightarrow 0_+} \sim$ denotes reduction of a Borel resummed form to a formal transseries by taking a small $\hbar$. 
Although our input in Borel resummation is a formal power series as Eq.(\ref{eq:asym_f}), $f^{0_\mp}$ in Eq.(\ref{eq:simh}) generally becomes a formal transseries taking the form, for example, that~\cite{Costin2008,Aniceto:2018bis}\footnote{
  This form is constrained by underlying mathematical structures such as a differential equation.
  Eq.(\ref{eq:f0mp_exp}) is just an example, and a resurgent structure of our problem slightly differs from Eq.(\ref{eq:f0mp_exp}).
}
\be
&& f^{0_\mp}(\hbar) \sim \sum_{n \in {\mathbb N}} c^{(0)}_{n}\hbar^n + \sum_{\ell \in {\mathbb N}} \sum_{n \in {\mathbb N}_0} c^{(\ell)}_{n} \left( \pm \frac{{\cal A}}{2} \right)^\ell e^{-\frac{\ell S}{\hbar}}
\hbar^{n+ \beta \ell}, \qquad (c^{(\ell)}_{n} \in {\mathbb C}, {\cal A} \in i {\mathbb R}) \label{eq:f0mp_exp}
\ee
with a non-zero ${\cal A}$.
The deviation from $f$ can be roughly estimated by a radius of convergence of $f_B$, and it is evaluated as $r_* = S$.
Hence, 
\be
{\cal S}_{0_+}[f] - {\cal S}_{0_-}[f] \overset {\hbar \rightarrow 0_+} \sim i e^{-\frac{S}{\hbar}} \left[ c  + O(\hbar) \right],
\ee
with a constant $c$.
Therefore, in general, Borel summability does not guarantee to give a Borel resummed form perfectly reproducing the original formal expansion.

The question is how to construct a Borel resummed form to reproduce the original formal power series, $f$, in the small $\hbar$ limit.
It is possible by introducing \textit{Stokes automorphism}, ${\frak S}_\theta$, which has a key role through this paper, defined as
\be
{\cal S}_{\theta + 0_+} = {\cal S}_{\theta + 0_-} \circ {\frak S}_{\theta}.  \label{eq:def_Stokes}
\ee
Roughly speaking, it is a mapping from a transseries to a transseries to compensate a difference due to the discontinuity along the integration ray with $\arg(\xi) = \theta$.
Stokes automorphism is a homomorphism:
\be
&&  {\frak S}_{\theta} [f_1+ f_2] = {\frak S}_{\theta}[f_1] + {\frak S}_{\theta}[f_2], \qquad {\frak S}_{\theta} [f_1 f_2] = {\frak S}_{\theta}[f_1] \cdot {\frak S}_{\theta}[f_2]. \label{eq:G_homo}
\ee
If $f$ is Borel summable along $\arg(\xi) = \theta$, then ${\frak S}_{\theta}$ is an identity mapping. 
Stokes automorphism defined in Eq.(\ref{eq:def_Stokes}) is just an automorphism determined by action of ${\cal S}_{\theta + 0_\pm}$ to $f$,  but it is extendable to a one-parameter group generated by a generator, \textit{alien derivative} $\bul{\Delta}_\theta$, as
\be
&& {\frak S}^\nu_{\theta} = \exp \left[ \nu \bul{\Delta}_{\theta}  \right], \qquad \bul{\Delta}_{\theta} = \sum_{w \in {\Gamma(\theta)}} \bul{\Delta}_w, \qquad (\nu \in {\mathbb R}) \label{eq:Stokes_alien}
\ee
where $\Gamma(\theta)$ denotes a set of singular points along the integration ray with $\theta$.
It satisfies
\be
&& {\frak S}^{\nu=0}_{\theta} = 1, \qquad {\frak S}^{\nu_1}_{\theta} \circ {\frak S}_{\theta}^{\nu_2} = {\frak S}_{\theta}^{\nu_2} \circ {\frak S}_{\theta}^{\nu_1} = {\frak S}_{\theta}^{\nu_1+\nu_2}, \qquad (\nu_1,\nu_2 \in {\mathbb R})
\label{eq:Stokes_prop_add}
\ee 
and taking $\nu = 1$ leads to ${\frak S}_{\theta}$ in Eq.(\ref{eq:def_Stokes}).
Intuitively, action of $\bul{\Delta}_{w \in \Gamma(\theta)}$ to a formal power series means pre-extracting a non-perturbative contribution originated by taking a Hankel contour clockwisely going around a singular point located at $\xi = w$ in the Laplace integral.
An alien derivative is additive and satisfies Leibniz rule:
\be
\bul{\Delta}_w[f_1+f_2] = \bul{\Delta}_w[f_1] + \bul{\Delta}_w[f_2], \qquad \bul{\Delta}_w[f_1f_2] = \bul{\Delta}_w[f_1] \cdot f_2 + f_1 \cdot \bul{\Delta}_w[f_2].
\ee
It is notable that, when action of the Stokes automorphism or alien derivative to a formal power series (or transseries) is non-trivial, it generates a non-perturbative contribution such as $e^{-S/\hbar} \sum_{n \in {\mathbb N}_0} c_n \hbar^n$, which means that non-perturbative informations are accessible from a perturbative expansion (or vice versa).
Such a relation among perturbative and non-perturbative sectors is called as a resurgent relation.

Let us come back to our question.
Again, we take $\theta = 0$.
We suppose that there exists formal transseries $f^{0_\pm}$ and a Borel resummed form $\widehat{f}$ such that ${\cal S}_{0_+}[f^{0_+}] = {\cal S}_{0_-}[f^{0_-}] = \widehat{f} \overset {\hbar \rightarrow 0_+} \sim f$.
Thus, knowing a relation between $f$ and $f^{0_\pm}$ is the solution.
This problem can be solved by introducing \textit{median resummation} defined as\footnote{
  These exponents of Stokes automorphism in this definition, $\mp 1/2$, are determined by the conditions that
\be
   {\cal C} \circ  {\cal S}_{\theta}   = {\cal S}_{-\theta}  \circ {\cal C}, \qquad {\cal C} \circ {\cal S}_{{\rm med},{\theta}} = {\cal S}_{{\rm med},{-\theta}} \circ {\cal C}, \label{eq:CS_SC}
\ee
where ${\cal C}$ is complex conjugate.
The first condition in Eq.(\ref{eq:CS_SC}) can be expressed by Stokes automorphism and alien derivative as
\be
{\cal C} \circ {\frak S}^{\nu}_{\theta} = {\frak S}^{-\nu}_{-\theta} \circ {\cal C}, \qquad {\cal C} \circ \bul{\Delta}_{\theta} = - \bul{\Delta}_{-\theta} \circ {\cal C}, \label{eq:CG_GC}
\ee
and requiring the second condition in Eq.(\ref{eq:CS_SC}) gives the exponent, $\mp 1/2$, using Eq.(\ref{eq:CG_GC}).
}
\be
&& {\cal S}_{{\rm med},0}[f]  := {\cal S}_{0_+} \circ {\frak S}^{-1/2}_0[f] = {\cal S}_{0_-} \circ {\frak S}^{+1/2}_0[f], 
\label{eq:def_Smed} 
\ee
where $f^{0_\pm}$ are related to $f$ as
\be
f^{0_\pm} = {\frak S}^{\mp 1/2}_0[f], \qquad f = {\frak S}^{\pm 1/2}_0[f^{0_\pm}], \label{eq:rel_fpm_f}
\ee
and thus, ${\cal S}_{{\rm med},0}[f]  = {\cal S}_{0_\pm}[f^{0_\pm}]$\footnote{
In the aspect of Lefschetz thimble decomposition, the Stokes automorphism in Eq.(\ref{eq:def_Stokes}) has a role of reconstruction of an integration-path to keep the same homology class against a discontinuity induced by a Stokes phenomenon.
In addition, ${\frak S}_0^{\mp 1/2}$ and ${\cal S}_{0_\pm}$ correspond to appropriately adding non-trivial saddles to reproduce its original integration-path and performing the integration along the thimbles, respectively.
}.
The median resummation directly provides a Borel resummed form which is continuous at $\arg(\hbar)=0$ from $f$.
Action of the alien derivative to a formal power expansion in Gevrey-1 class is well-defined, so that the median resummed form naturally returns ${\cal S}_{{\rm med},0} [f] \overset {\hbar \rightarrow 0_+} \sim f$.
Conversely, one can express the discontinuity in ${\cal S}_{0_\pm}[f]$ using the median resummation as
\be
   {\cal S}_{0_\pm}[f] &=& {\cal S}_{{\rm med},0} \circ {\frak S}_0^{\pm 1/2} [f] =  {\cal S}_{{\rm med},0} [f^{0_\mp}] \overset {\hbar \rightarrow 0_+} \sim f^{0_{\mp}}. \label{S0pm_f}
\ee
In addition, the procedure (\ref{eq:rel_fpm_f}) can be interpreted to enable to construct $f$ from $f^{0_+}$ and/or $f^{0_-}$ by the Stokes automorphism, ${\frak S}_0^{\pm 1/2}$, without passing through Borel resummation.

It is useful to introduce another Stokes automorphism, $\widehat{\frak S}_{\theta}$, that acts to a Borel resummed form and satisfies
\be
   {\cal S}_{{\rm med}, \theta} \circ {\frak S}^\nu_{\theta} = \widehat{\frak S}_{\theta}^\nu \circ {\cal S}_{{\rm med}, \theta}. \label{eq:gen_Stokes}
\ee
Choosing $\nu=\pm 1/2$ in the l.h.s. is equivalent to ${\cal S}_{\theta+0_\pm}$.
In this sense, this generalized Stokes automorphism, $\widehat{\frak S}^{\nu}_{\theta}$, can be regarded to be a continuous transform to Borel resummed forms, and the Borel resummation, ${\cal S}_{\theta+0_\pm}$, is a special case of Eq.(\ref{eq:gen_Stokes}).

A generalization of the Borel resummation to an $\ell$-th non-perturbative sector such that $f^{(\ell)}(\hbar) \sim e^{- \frac{\ell S}{\hbar}} c^{(\ell)}_n \hbar^{n+\ell \beta}$ with ${\rm Re}[S]>0$ and $\beta \notin {\mathbb Z}_{<0}$ is also defined by modifications of the Borel transform and Laplace integral as
\be
{\cal B}[f^{(\ell)}](\xi) = \sum_{n \in {\mathbb N}_0} \frac{c^{(\ell)}_n}{\Gamma(n + \ell \beta)}  (\xi - \ell S)^{n+\ell \beta -1} = f^{(\ell)}_B(\xi), \qquad 
{\cal L}_\theta [f^{(\ell)}_B](\hbar) := \int_{\ell S}^{\infty e^{i \theta}} d \xi \,e^{-\frac{\xi}{\hbar}} f^{(\ell)}_B(\xi).
\ee
The definition of the Stokes automorphism and alien derivative is obtained in the similar way to the case of a formal power expansion explained above.

Depending on problems, one obtains $f^{0_\pm}$ prior to knowing a specific form of $f$.
EWKB is the case.
In such a case, although taking Borel resummation ${\cal S}_{0_\pm}$ to $f^{0_\pm}$ formally solves the problem, it is in practice too tough to find their exact resummed forms as functions except in special cases.
Instead of that, we take another strategy for consideration of the ABS conjecture, that is, constructing a formal transseries $f$ from $f^{0_\pm}$ by Stokes automorphism using Eq.(\ref{eq:rel_fpm_f}).

Through this paper, we usually deal with formal transseries except where specifically noted.
The hat symbol, $\widehat{f}$, denotes the median resummed form of $f$, i.e., $\widehat{f} = {\cal S}_{{\rm med}}[f]$.
\\ \par 
In the below, we show some examples for the procedure of Borel resummation.
In these examples, Borel (median) resummed forms can be exactly obtained as functions.

\subsubsection{Example 1: $c_n = A S^{-n} n!$}
We consider  $c_n = A S^{-n} n!$ with $S \in {\mathbb R}_{>0}$ in Eq.(\ref{eq:def_f_asym}).
Acting Borel transform defined in Eq.(\ref{eq:_def_Borel_Lap}) to the formal power series gives
\be
   f_B := {\cal B}[f] = \frac{AS}{(S-\xi)^2}, \label{eq:exfB}
\ee
and one can immediately see that $f$ is Borel non-summable along $\theta = 0$.
Taking the Laplace integral with $\theta = 0_\pm$ leads to
\be
   {\cal S}_{0_\pm}[f] &=&  {\cal L}_{0_\pm}[f_B] = \frac{ AS e^{-\frac{S}{\hbar}}}{\hbar}  \left[  \text{Ei}\left(\frac{S}{\hbar} \right) \pm  \pi i \right] -A \nl
   &\overset {\hbar \rightarrow 0_+} \sim& \sum_{n \in {\mathbb N}} A S^{-n} n! \hbar^n \pm \pi i \frac{A S e^{-\frac{S}{\hbar}}}{\hbar}, \label{eq:ex_Sf}
\ee
where ${\rm Ei}(x)$ is the exponential integral.
Apparently, this result has a discontinuity at $\arg(\xi)=0$, that is $\pm \pi i \frac{A S e^{-\frac{S}{\hbar}}}{\hbar}$.
From Eq.(\ref{eq:exfB}), one can see that a set of singular points consists of only the double pole at $\xi =S$, i.e. $ \Gamma(\theta=0) = \{ S \}$, and thus the alien derivative is easily obtained by a residue integration around $\xi = S$ clockwisely, as
\be
\bul{\Delta}_S[f] = - \oint_{\xi = S} d\xi \, e^{-\frac{\xi}{\hbar}} f_B(\xi) = 2 \pi i \frac{A S e^{-\frac{S}{\hbar}}}{\hbar}, \qquad (\bul{\Delta}_S)^{n>1}[f] = 0.
\ee
From this result and Eq.(\ref{eq:Stokes_alien}), one can obtain action of the Stokes automorphism to $f$ as
\be
&& {\frak S}^{\nu}_0 [f] = f + 2 \pi \nu i \frac{A S e^{-\frac{S}{\hbar}}}{\hbar},
\ee
and Eq.(\ref{eq:rel_fpm_f}) gives $f^{0_\pm}$ as
\be
f^{0_\pm} =  {\frak S}^{\mp 1/2}_0[f]  =  f  \mp \pi  i \frac{A S e^{-\frac{S}{\hbar}}}{\hbar}.
\ee
Taking Borel resummation to $f^{0_\pm}$ yields
\be
\widehat{f} &=& {\cal S}_{0_\pm}[f^{0_\pm}] =  \frac{ AS e^{-\frac{S}{\hbar}}}{\hbar}  \text{Ei}\left(\frac{S}{\hbar} \right) -A \overset {\hbar \rightarrow 0_+} \sim \sum_{n \in {\mathbb N}} A S^{-n} n! \hbar^n.
\ee
This result could correctly reproduce $c_n = A S^{-n} n!$ and remove the discontinuity, $\pm \pi i \frac{AS e^{-\frac{S}{\hbar}}}{\hbar}$, appeared in Eq.(\ref{eq:ex_Sf}).
In addition, acting $\widehat{\frak S}_{0}^\nu$ to $\widehat{f}$ is written as
\be
\widehat{\frak S}_{0}^\nu[\widehat{f}] = \widehat{f} + 2 \pi \nu i \frac{A S e^{-\frac{S}{\hbar}}}{\hbar},
\ee
and taking $\nu = \pm 1/2$ corresponds to ${\cal S}_{0_\pm}[f]$ in Eq.(\ref{eq:ex_Sf}).

\subsubsection{Example 2: Zero-dimensional ${\cal PT}$ symmetric toy model}
Another example is the zero-dimensional model considered in Ref.~\cite{Ai:2022csx}, which is given by
\be
&& Z_{\cal PT} = \int_{\gamma_{\cal PT}} dx \, \exp \left[ - x^2 + g x^4\right], \qquad g \in {\mathbb R}_{>0}, \label{eq:ZPT_toy} \\
&& \gamma_{\cal PT} :=  s e^{+\frac{\pi}{4} i } \theta(-s) + s e^{-\frac{\pi}{4} i } \theta(+s), \qquad s \in {\mathbb R},
\ee
where $\theta(s)$ is the step function.
The integration in Eq.(\ref{eq:ZPT_toy}) is analytically performable, and one can obtain the exact solution as
\be
\widehat{Z}_{\cal PT} = \frac{\pi e^{-\frac{1}{8g}}}{4 \sqrt{g}} \left[ I_{-\frac{1}{4}}\left( \frac{1}{8g} \right) + I_{+\frac{1}{4}}\left( \frac{1}{8g} \right)  \right], \label{eq:exPT_toy}
\ee
where $I_{\nu}(x)$ is the modified Bessel function of the first kind.
We apply the above methods to a formal transseries of the exact solution expanded by $g$.
This example might be helpful to understand the whole story in this paper and Fig.~\ref{fig:summary_PT}.

Expanding Eq.(\ref{eq:exPT_toy}) gives
\be
\widehat{Z}_{\cal PT} \sim Z_{\cal PT} = \sqrt{\pi} \sum_{n \in {\mathbb N}_0} \frac{\left( \frac{1}{4} \right)_n\left( \frac{3}{4}\right)_n}{n!} (4 g)^n \quad \mbox{as} \quad g \rightarrow 0_+,
\ee
where $(a)_n := \Gamma(a+n)/\Gamma(a)$ is the Pochhammer symbol.
Here, we define $ {\cal J}_0 := g Z_{\cal PT}$ and deal with ${\cal J}_0$.
The coefficients of ${\cal J}_0$ are given by
\be
   {\cal J}_0 = \sum_{n \in {\mathbb N}} c_n g^n, \qquad c_{n} = \sqrt{\pi} \frac{\left( \frac{1}{4} \right)_{n-1} \left( \frac{3}{4} \right)_{n-1 }}{(n-1)!} 4^{n-1},
\ee
and Borel transform (\ref{eq:_def_Borel_Lap}) gives
\be
{\cal J}_{0,B} := {\cal B}[{\cal J}_{0}] = \frac{2 K\left(\frac{4 \sqrt{\xi}}{2 \sqrt{\xi}+1}\right)}{\sqrt{\pi } \sqrt{2 \sqrt{\xi}+1}},
\ee
where $K(x)$ is the elliptic integral.
This has a singular point at $\xi = \frac{1}{4}$, i.e. $\Gamma(\theta = 0) = \{ \frac{1}{4} \}$.
Taking Borel resummation yields
\be
{\cal S}_{0_\pm} [{\cal J}_0]
   &=&  \frac{\pi e^{-\frac{1}{8g}} \sqrt{g}}{4} \left[ I_{-\frac{1}{4}}\left( \frac{1}{8g} \right) + I_{+\frac{1}{4}}\left( \frac{1}{8g} \right) \pm i \frac{\sqrt{2}}{\pi}  K_{\frac{1}{4}}\left( \frac{1}{8g} \right) \right] \nl
   &=& \widehat{\cal J}_0 + \widehat{\cal J}_\pm,  \label{eq:exPT_toy_Spm}
\ee
where $\widehat{\cal J}_0 := g \widehat{Z}_{\cal PT}$, and $\widehat{\cal J}_\pm :=  \pm i \frac{e^{-\frac{1}{8g}} \sqrt{2 g}}{4} K_{\frac{1}{4}} \left( \frac{1}{8g}  \right)$ with the modified Bessel function of the second kind, $K_{\nu}(x)$.
Formal transseries of $\widehat{\cal J}_\pm$, denoted by ${\cal J}_\pm$, can be obtained by the asymptotic expansion of $K_{\frac{1}{4}}\left( \frac{1}{8 g}\right)$.
Notice that ${\cal J}_+ + {\cal J}_- = 0$ and ${\cal J}_+- {\cal J}_- = \pm 2 {\cal J}_\pm$.
From the results, one can find actions of alien derivatives and Stokes automorphism to ${\cal J}_0$ and ${\cal J}_\pm$ as
\be
&& (\bul{\Delta}_{\frac{1}{4}})[{\cal J}_0] =  \int_{+\infty + i 0_-}^{+\infty + i 0_+} d \xi \, e^{-\frac{\xi}{g}} {\cal J}_{0,B}(\xi) = {\cal J}_+ - {\cal J}_-, \qquad (\bul{\Delta}_{\frac{1}{4}})^{n>1}[{\cal J}_0] =0, \\
&& (\bul{\Delta}_{\theta=0})^{n \in {\mathbb N}}[{\cal J}_\pm] =0,
\ee
where $\int_{+\infty + i 0_-}^{+\infty + i 0_+} d \xi$ denotes integration along a Hankel contour going around the singular point at $\xi = \frac{1}{4}$ clockwisely, and
\be
&& {\frak S}_0^{\nu} [{\cal J}_0] = {\cal J}_0 + \nu ( {\cal J}_+ - {\cal J}_- ), \qquad  {\frak S}_0^{\nu} [{\cal J}_\pm] = {\cal J}_\pm, \label{eq:St_toy} \\
&& \widehat{\frak S}_0^{\nu} [\widehat{\cal J}_0] = \widehat{\cal J}_0 + \nu ( \widehat{\cal J}_+ - \widehat{\cal J}_- ), \qquad    \widehat{\frak S}_0^{\nu} [\widehat{\cal J}_\pm] = \widehat{\cal J}_\pm. \label{eq:hatSt_toy} 
\ee
Hence, ${\cal J}_0^{0_\pm}$ which satisfy ${\cal S}_{0_+}[{\cal J}_0^{0_+}] = {\cal S}_{0_-}[{\cal J}_0^{0_-}]$ are obtained as
\be
&& {\cal J}_0^{0_\pm} = {\frak S}_0^{\mp 1/2} [{\cal J}_0] = {\cal J}_0  \mp \frac{1}{2} ( {\cal J}_+ - {\cal J}_- ) = {\cal J}_0  - {\cal J}_\pm = {\cal J}_0 + {\cal J}_\mp, \\
&& {\cal S}_{0_\pm}[{\cal J}_0^{0_\pm}] = {\cal S}_{{\rm med},0}[{\cal J}_0] = \widehat{\cal J}_0.
\ee
Similarly, we consider analytic continuation of the Hermitian model given by 
\be
\widehat{Z}_{\cal H} := \int_{-\infty}^{+\infty} dx \, \exp \left[ - x^2  - \lambda  x^4\right] =  \frac{e^{\frac{1}{8 \lambda}}}{2 \sqrt{\lambda}} K_{\frac{1}{4}}\left(\frac{1}{8 \lambda}\right), \qquad \lambda \in {\mathbb R}_{>0}, \label{eq:ZH_toy} 
\ee
and define ${\cal J}_{\rm AC}^{\arg(\lambda) = \pm \pi} := g Z_{\cal H}^{\arg(\lambda) = \pm \pi}$.
By taking $\lambda = e^{\pm \pi i } g$ with $g \in {\mathbb R}_{>0}$, those can be expressed by
\be
   {\cal J}_{\rm AC}^{\arg(\lambda) = \pm \pi} &=& {\cal J}_0 + {\cal J}_\mp = {\cal J}_0 \mp \frac{1}{2} ( {\cal J}_+ - {\cal J}_-), \\
\widehat{\cal J}_{\rm AC}^{\arg(\lambda) = \pm \pi} 
&=& \frac{ \pi e^{-\frac{1}{8 g}} \sqrt{g}}{4} \left[ I_{\frac{1}{4}}\left(\frac{1}{8 g}\right) +  I_{-\frac{1}{4}}\left(\frac{1}{8 g}\right) \mp i \frac{\sqrt{2}}{\pi}  K_{\frac{1}{4}}\left(\frac{1}{8 g}\right) \right] \nl
&=& \widehat{\cal J}_0 + \widehat{\cal J}_\mp = \widehat{\cal J}_0 \mp \frac{1}{2} ( \widehat{\cal J}_+ - \widehat{\cal J}_-).
\ee
From Eqs.(\ref{eq:St_toy})(\ref{eq:hatSt_toy}), one can find that
\be
&&  {\frak S}_{0}^{\pm 1/2}[{\cal J}_{0}] = {\cal J}_{\rm AC}^{\arg(\lambda) = \mp \pi}, \qquad   {\frak S}_{0}^{\pm 1/2}[{\cal J}_{\rm AC}^{\arg(\lambda) = \pm \pi}] = {\cal J}_0, \nl
&& {\frak S}_{0}^{\pm 1}[{\cal J}_{\rm AC}^{\arg(\lambda) = \pm \pi}] = {\cal J}_{\rm AC}^{\arg(\lambda) = \mp \pi}, \\ \nl
&&  \widehat{\frak S}_{0}^{\pm 1/2}[\widehat{\cal J}_{0}] = \widehat{\cal J}_{\rm AC}^{\arg(\lambda) = \mp \pi}, \qquad \widehat{\frak S}_{0}^{\pm 1/2}[\widehat{\cal J}_{\rm AC}^{\arg(\lambda) = \pm \pi}] = \widehat{\cal J}_0, \nl
&& \widehat{\frak S}_{0}^{\pm 1}[\widehat{\cal J}_{\rm AC}^{\arg(\lambda) = \pm \pi}] = \widehat{\cal J}_{\rm AC}^{\arg(\lambda) = \mp \pi}, \\ \nl
&&    {\cal S}_{0_\pm}[{\cal J}_{0}] = \widehat{\cal J}_{\rm AC}^{\arg(\lambda) = \mp \pi}, \qquad \ \ \  \, {\cal S}_{0_\pm}[{\cal J}_{\rm AC}^{\arg(\lambda) = \pm \pi}] = \widehat{\cal J}_0.
\ee
Therefore, the zero-dimensional ${\cal PT}$ symmetric model satisfies the relations in Fig.~\ref{fig:summary_PT}.

It is remarkable to mention that this picture directly corresponds to Lefschetz thimble decomposition.
${\cal S}_{0_\pm}[{\cal J}_{0}]$ and ${\cal S}_{0_{\pm}}[{\cal J}_{\pm}]$ (divided by $g$) are nothing but thimble-integrations from a perturbative saddle at $x=0$ and non-perturbative saddles at $x= \pm \frac{1}{\sqrt{2 g}}$, respectively.
See Ref.~\cite{Ai:2022csx} in detail.

\subsection{Exact WKB analysis} \label{sec:EWKB}
We review EWKB and explain the procedure to obtain energy quantization conditions (QCs).
See, for example, Refs.~\cite{Voros1983,Silverstone,Schafke1,DDP2,DP1,Kawai1,AKT1,Iwaki1} in detail.

We consider the following Schr\"{o}dinger equation given by a potential $V(x)$:
\be
   && {\cal L} = - \hbar^2 \pd_x^2 + V(x) - E, \qquad {\cal L} \psi(x) = 0, \qquad (x \in {\mathbb C}, \ \   E, \hbar \in {\mathbb R}_{>0}) \label{eq:sch_eq_gen} 
\ee
where $E$ is an energy and $\psi$ is a wavefunction.
The variable, $x$, is normally taken as a real value, but we extend it to be complex-valued for analytic continuation of the wavefunction.
For simplicity, we assume that the potential $V$ is independent on $\hbar$ and a polynomial of $x$ bounded in the asymptotic limit, $x \rightarrow \pm \infty$, i.e., the wavefunction $\psi$ is normalizable along the real axis.
A main difference from the standard WKB is to take into account of its all orders with respect to $\hbar$.
We construct the wavefunction by preparing an ansatz expanded by $\hbar$, which is given by
\be
&& \psi_a (x,\hbar)  = \sigma(\hbar) \exp \left[ \int_{a}^{x} dx^\prime \, S(x^\prime,\hbar) \right], \qquad (x \in {\mathbb C}) \label{eq:psi_ansatz} \\
&& S(x,\hbar) = \sum_{n \in {\mathbb N}_0} S_{n-1}(x) \hbar^{n-1} \quad \mbox{as} \quad \hbar \rightarrow 0_+,
\ee
where $\sigma(\hbar)$ is an integration constant generally depending on $\hbar$, and $a$ is a normalization point.
Substituting $\psi$ into the Schr\"{o}dinger equation (\ref{eq:sch_eq_gen}) leads to Riccati equation in terms of $S(x,\hbar)$ as
\be
S(x,\hbar)^2 + \pd_x S(x,\hbar) = \hbar^{-2} Q(x), \qquad Q(x):= V(x) - E. \label{eq:Riccati}
\ee
From Riccati equation (\ref{eq:Riccati}),  $S(x,\hbar)$ is recursively obtained order by order, as
\be
&& S_{-1}(x) = \pm \sqrt{Q(x)}, \qquad S_{0}(x) = - \frac{\pd_x \log Q(x)}{4}, \nl
&& S_{+1}(x) = \pm \frac{1}{8 \sqrt{Q(x)}} \left[ \pd_x^2 \log Q(x) - \frac{(\pd_x \log Q(x))^2}{4} \right], \qquad \cdots.
\ee
Since Riccati equation is quadratic, we have two independent solutions which enable to be distinguished by the sign, $\pm$, in $S(x,\hbar)$.
One can also find from Eq.(\ref{eq:Riccati}) that $S(x,\hbar)$ can be decomposed into two parts; the one part consists of terms with the sign $\pm$ determined by choice of $S_{-1}(x) = \pm \sqrt{Q(x)}$ and the other part contains terms independent on the sign of $S_{-1}(x)$.
By denoting $S_{\rm od}(x,\hbar)$ and $S_{\rm ev}(x,\hbar)$ for the former and latter parts, respectively,  one can express the total $S(x,\hbar)$ as $S(x,\hbar) = \pm S_{\rm od}(x,\hbar) + S_{\rm ev}(x,\hbar)$, and $S_{\rm ev}(x,\hbar)$ can be written down by $S_{\rm od}(x,\hbar)$ as\footnote{
  If $Q(x)$ includes $\hbar^\alpha$ with $\alpha \in {\mathbb N}$, then $S_{\rm odd}(x,\hbar)$ contains both odd and even powers of the $\hbar$-expansion.
  The degenerate Weber-type Stokes graph in Appendix~\ref{sec:der_G} is the case.
}
\be
&& S_{\rm od}(x,\hbar) = \sum_{n \in {\mathbb N}_0} S_{2n-1}(x) \hbar^{2n-1}, \qquad S_{-1}(x) = \sqrt{Q(x)}, \\
&& S_{\rm ev}(x,\hbar) = \sum_{n \in {\mathbb N}_0} S_{2n}(x) \hbar^{2n}  = - \frac{1}{2} \pd_x \log S_{\rm od}(x,\hbar).
\ee
Therefore, the wavefunction (\ref{eq:psi_ansatz}) is given by
\be
\psi_{a\pm}(x, \hbar) &=&  \frac{\sigma_{\pm}(\hbar)}{\sqrt{S_{\rm od}(x,\hbar)}} \exp \left[ \pm \int_{a}^{x} dx^\prime \, S_{\rm od}(x^\prime,\hbar) \right] \nl
 &=& \sigma_{\pm}(\hbar) \exp \left[ \pm \frac{1}{\hbar} \int_{a}^{x} dx^\prime \, S_{{\rm od},-1}(x^\prime) \right] \sum_{n \in {\mathbb N}_0} \psi_{a \pm,n}(x) \hbar^{n+\frac{1}{2}}.
 \label{eq:psi_apm}
\ee
Without the loss of generality, we take $\sigma_{\pm}(\hbar)=1$.
The Borel transform and Laplace integral for the wavefunctions are defined as
\be
&& {\cal B}[\psi_{a\pm}](x, \xi) := \frac{\psi_{a \pm,n}(x)}{\Gamma(n+\frac{1}{2})}(\xi \pm \xi_0(x))^{n - \frac{1}{2}} = \psi_{B,a\pm}(x, \xi), \qquad \xi_0(x) := \sqrt{Q_0(x)}, \\
&& {\cal L}_{\theta}[\psi_{B,a\pm}](x, \hbar) := \int^{\infty e^{i \theta} }_{\mp \xi_0} d \xi \, e^{-\frac{\xi}{\hbar}} \psi_{B,a\pm}(x,\xi),
\ee
and the Borel resummation is a combination of the two operations, i.e. ${\cal S}_\theta = {\cal L}_\theta \circ {\cal B}$.

\begin{figure}[tbp]
  \begin{center}
    \begin{tabular}{cc}
      \begin{minipage}{0.5\hsize}
        \begin{center}
          \includegraphics[clip, width=75mm]{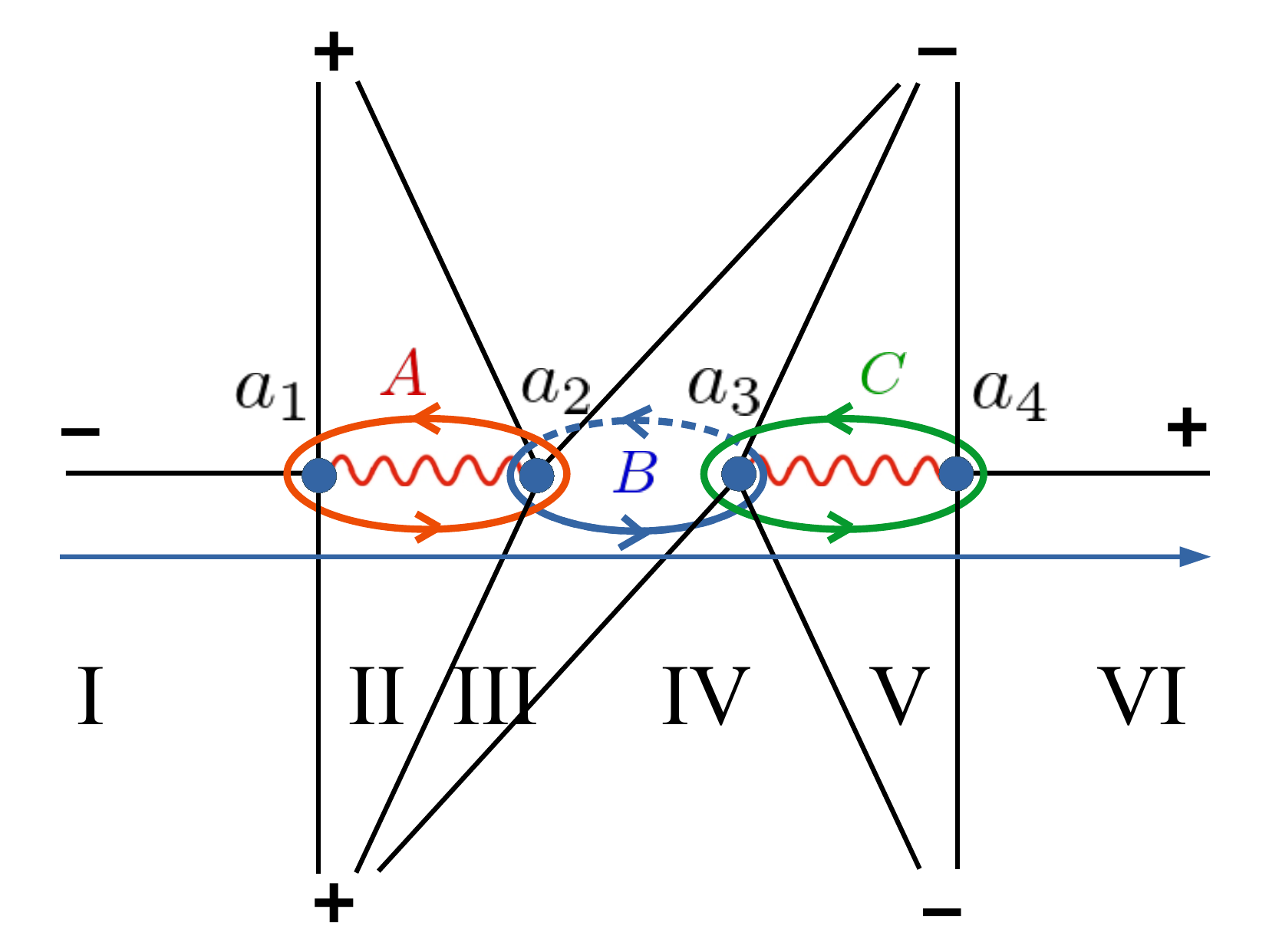}
          \hspace{1.6cm} (a) $\arg(\hbar) = 0_+$
        \end{center}
      \end{minipage}
      \begin{minipage}{0.5\hsize}
        \begin{center}
          \includegraphics[clip, width=75mm]{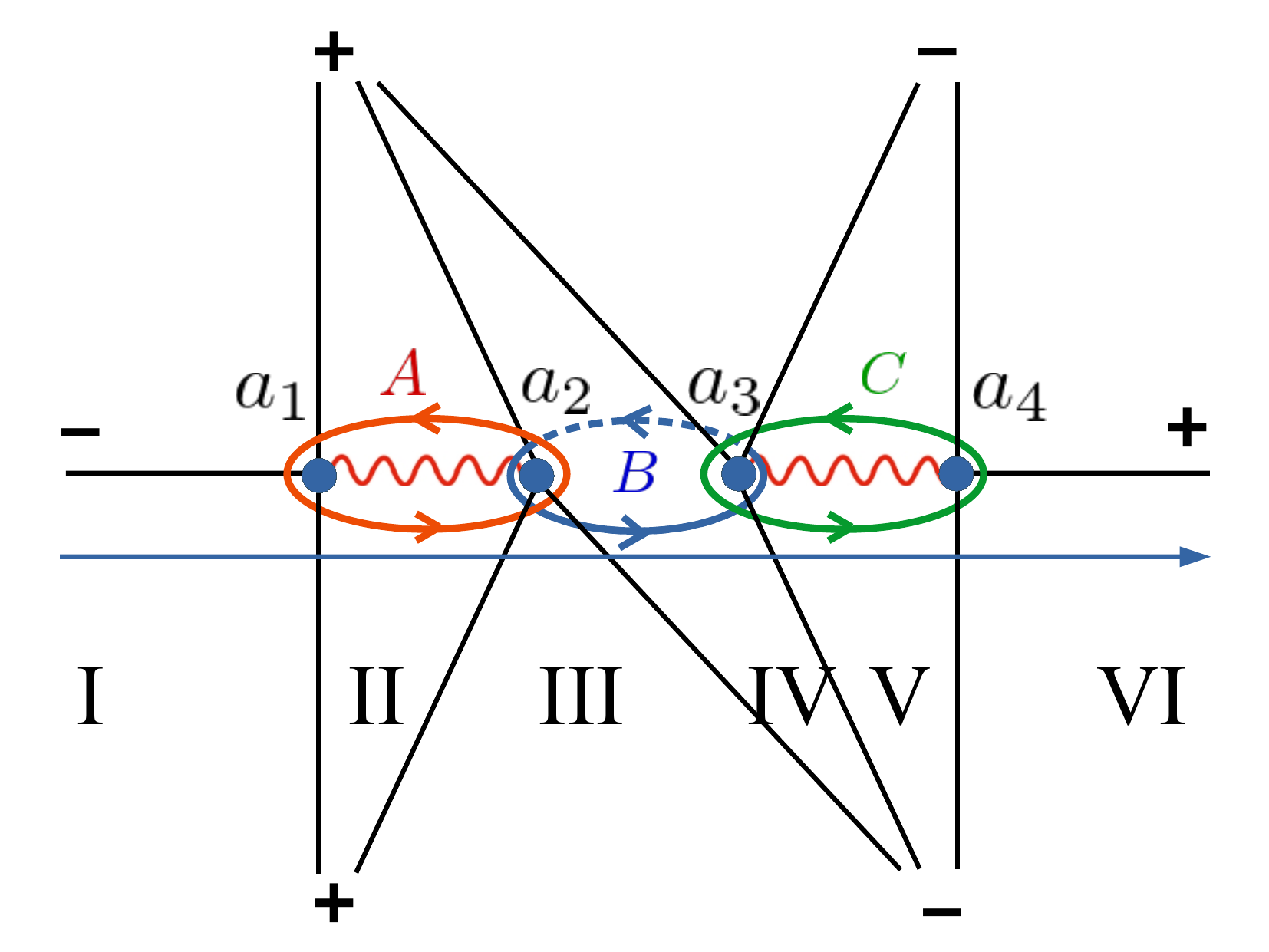}
          \hspace{1.6cm} (b) $\arg(\hbar) = 0_-$
        \end{center}
      \end{minipage}      
    \end{tabular} 
    \caption{Example of Stokes graph, a double-well potential.
      Turning points are denoted by blue dots, $a_{1,\cdots,4}$.
      The black solid and red wave lines mean Stokes lines and branch-cuts, respectively.
      The label, $+$ (resp. $-$), denotes an asymptotic domain where
      ${\rm Re} \left[\int_{a_j}^{x} dx^\prime \, S_{{\rm od},-1}(x^\prime)\right] \rightarrow + \infty$ (resp. $-\infty$).
      We perform analytic continuation along the blue line slightly below the real axis from $x  = -\infty$ to $x  = +\infty$.
      In these cycles, only the $B$-cycle is non-perturbative, and ${\rm C}_{\rm NP,\theta = 0} = \{ B \}$.
      The other cycles, $A$ and $C$, intersect with $B$ once, so that action of the Stokes automorphism to them is non-trivial and related to $B$.
      In this case, the intersection number is given by $\langle A,B \rangle = -\langle C,B \rangle = -1$.
    }
    \label{fig:Stokes_exam}
  \end{center}
\end{figure}

After formulating the wavefunction, we then draw \textit{Stokes graph}, that is defined by $\int dx \, S_{{\rm od},-1}(x)$ of $x \in {\mathbb C}$ and provides a structure of Borel summability of the wavefunction.
A Stokes graph normally consists of three kinds of objects: turning points, Stokes lines, and branch-cuts.
Turning points are defined from $Q(x)$ as
\be
{\rm TP} := \{ x \in {\mathbb C} \, | \, Q(x) = 0 \}. \label{eq:def_TPs}
\ee
These are used as a normalization point, denoted by $a$ in Eq.(\ref{eq:psi_apm}), in the wavefunction.
A Stokes line is defined as a line on the complex $x$-plane emerging from a turning point and satisfying the condition that
\be
{\rm Im} \left[ \frac{1}{\hbar} \int_{a_j}^x dx^\prime \, S_{{\rm od},-1}(x^\prime) \right] = 0, \qquad a_j \in {\rm TP}.
\ee
Along the lines, ${\rm Re} \left[\int_{a_j}^{x} dx^\prime \, S_{{\rm od},-1}(x^\prime)\right]$ is monotonically increasing (resp. decreasing), so that we add labels, $+$ (resp. $-$), to an asymptotic domain where ${\rm Re} \left[\int_{a_j}^{x} dx^\prime \, S_{{\rm od},-1}(x^\prime)\right] \rightarrow + \infty$ (resp. $-\infty$).
In addition, without some exceptions, branch-cuts normally have a role of swapping the independent solutions, $\psi_\pm$. 
Fig.~\ref{fig:Stokes_exam} is an example of Stokes graph in the case of a double-well potential, which would be also helpful to follow the below explanations.

\begin{figure}[tbp]
 \centering
 \includegraphics[width=100mm]{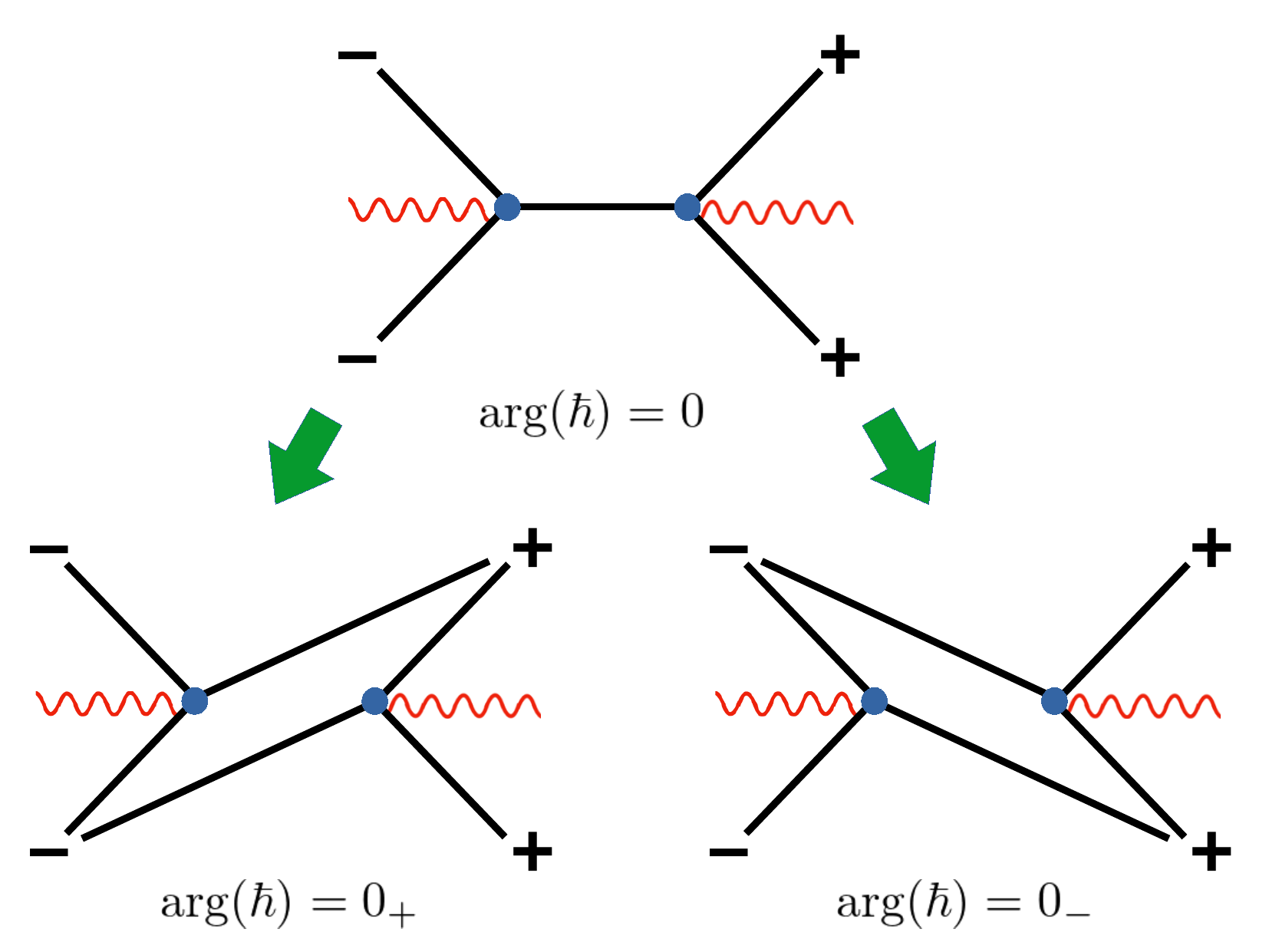}
 \caption{
   Example of degeneracy of two Stokes lines.
   When $\arg(\hbar) = 0$, two Stokes lines emerging from turning points (blue dots) on the one side go into the other side and degenerate.
   This degeneracy can be resolved by adding an infinitesimal complex phase to $\hbar$ as $\arg(\hbar) = 0_\pm$.
}
\label{fig:stokes_deg}
\end{figure}

When the wavefunction is Borel non-summable, there exist two possibilities for the reason: \textit{movable singularity} and \textit{fixed singularity}.
A movable singularity means a singularity on the Borel plane expressed by Stokes lines, and the wavefunction becomes Borel non-summable when it stays exactly on the lines.
Because of that, the wavefunctions defined on each the domain surrounded by the Stokes lines are generally discontinuous to each other.
A fixed singularity arises when a Stokes line reaches another turning point by taking certain values of $\hbar \in {\mathbb C}$, and then two Stokes lines normally degenerate.
This situation is called \textit{Stokes phenomenon}.
In such a case, the wavefunction is Borel non-summable on the complex $x$-plane entirely, but it can be resolved by adding an infinitesimal complex phase to $\hbar$ or parameters in $Q(x)$.
A example is drawn in Fig.~\ref{fig:stokes_deg}.
Below, we define the angle of the Laplace integral by $\hbar$, as $\theta = \arg(\hbar)$.

\begin{figure}[tbp]
 \centering
 \includegraphics[width=80mm]{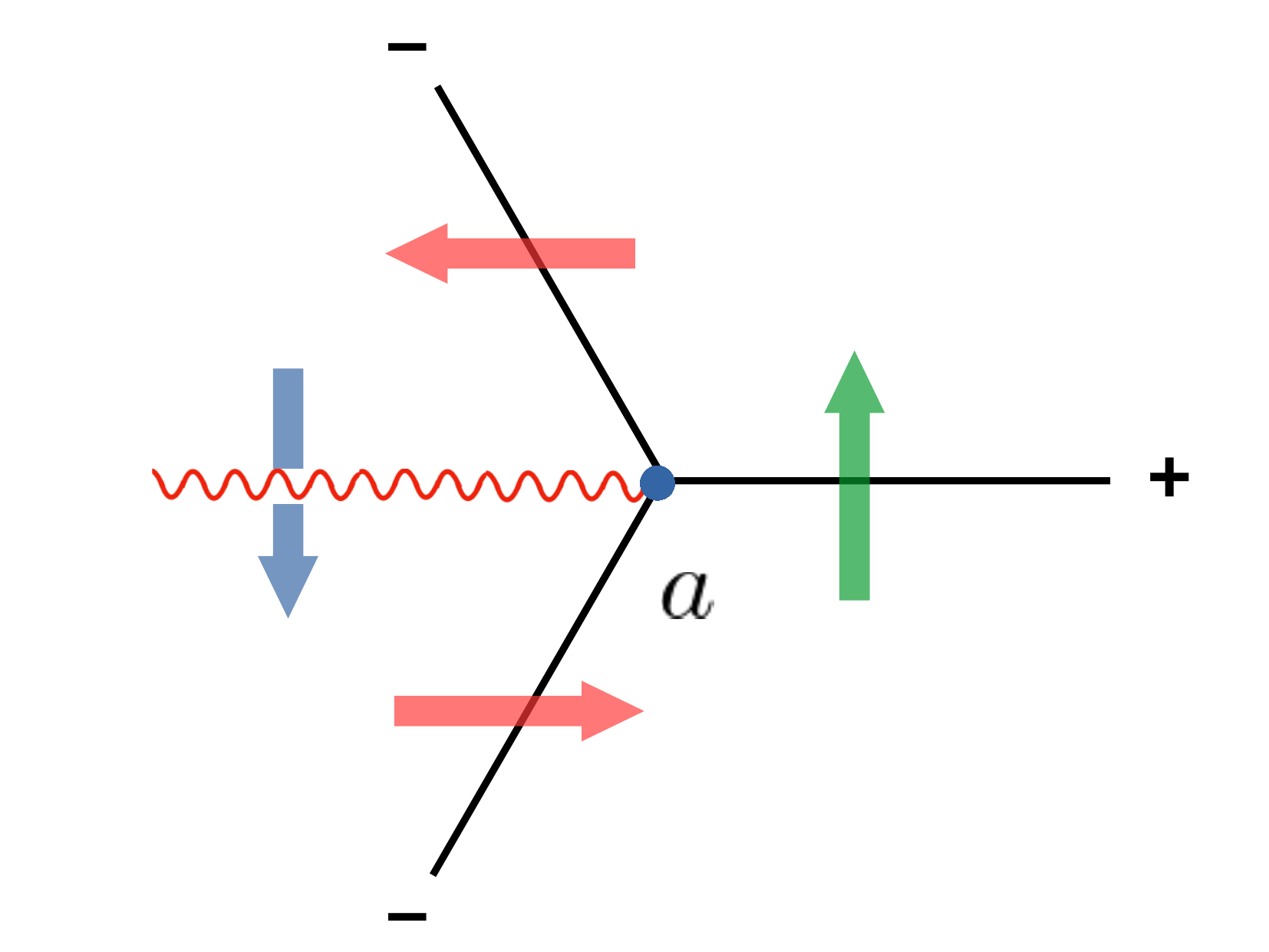}
 \caption{
   Airy-type Stokes graph defined by a simple turning point, $a$ (blue dot).
   The colored arrows crossing the Stokes lines and the branch-cut are expressed by $M_+$ (green), $M_-$ (red), and $T$ (blue) in Eq.(\ref{eq:M_T_mat}), respectively.
}
\label{fig:airy_connection}
\end{figure}

Our main task in EWKB is to obtain a QC by performing analytic continuation along a certain path on the complex $x$-plane.
For the computation, one has to connect the wavefunctions on each the domain separated by a Stokes line.
Here, we express the wavefunction by a vector form using two independent solutions of Riccati equation (\ref{eq:Riccati}), $\psi_\pm$, as
\be
\psi_a =
\begin{pmatrix}
  \psi_{a+} \\
  \psi_{a-}
\end{pmatrix}. \label{eq:psi_vec}
\ee
We suppose that $\psi^{\rm I}$ and $\psi^{\rm II}$ are wavefunctions on two domains separated by a Stokes line, ${\rm I}$ and ${\rm II}$, respectively.
A connection matrix crossing the Stokes line at $x = x_*$ is defined such that their Borel resummed wavefunctions become continuous at $x=x_*$, i.e.,
\be
&& {\cal S}_{\theta}[\psi^{\rm I}(x_* + 0_-)] = {\cal S}_{\theta}[\psi^{\rm I}(x_* + 0_+)], \label{eq:Spsi1_SpsiII} \\
&& \psi^{\rm I}(x_* + 0_+) := M^{{\rm I} \rightarrow {\rm II}} \psi^{\rm II} (x_* + 0_+),
\label{eq:psiI_MpsiII}
\ee
where $M^{\rm I \rightarrow II}$ is the connection matrix, and $x_* + 0_-$ and $x_* + 0_+$ belong to the domain ${\rm I}$ and ${\rm II}$, respectively.
In this paper, we mainly deal with Airy-type Stokes graph, shown in Fig.~\ref{fig:airy_connection}.
Anti-clockwisely crossing Stokes lines and branch-cut emerging from a turning point, $a$, can be expressed by the connection matrices $M_\pm$ and the branch-cut matrix $T$ given by
\be
&& M_+ =
\begin{pmatrix}
  1 & & i \\
  0 & & 1
\end{pmatrix}, \qquad 
M_- =
\begin{pmatrix}
  1 & & 0 \\
  i & & 1
\end{pmatrix}, \qquad
T =
\begin{pmatrix}
  0 &  -i \\
  -i  & 0
\end{pmatrix}, \label{eq:M_T_mat}
\ee
where the subscript of $M_\pm$ indicates the label of asymptotic behavior attached to the Stokes line we are crossing.
In addition, their inverse matrices correspond to crossing them clockwisely.
Such a connection formula is \textit{locally} formulated around a turning point, so that one has to change the normalization point depending on which turning point the Stokes line emerges from.
The normalization matrix changing a normalization point from $a_k$ to $a_j$ is given by
\be
N_{a_j,a_k} :=
\begin{pmatrix}
  e^{+\int_{a_k}^{a_j} dx \, S_{{\rm od}}(x,\hbar)}  &  \\
   &  e^{-\int_{a_k}^{a_j} dx \, S_{{\rm od}}(x,\hbar)}
\end{pmatrix} = N_{a_k,a_j}^{-1}, \qquad a_j,a_k \in {\rm TP}. \label{eq:norm_mat}
\ee
Acting those matrices (\ref{eq:M_T_mat})(\ref{eq:norm_mat}) to the wavefunction one by one along a path of analytic continuation yields a monodromy matrix.
When degeneracies are induced at $\theta = 0$ like Fig.~\ref{fig:stokes_deg}, one has to resolve the degeneracies by taking $\theta = 0_\pm$, and thus the resulting monodromy matrix depends on the complex phase.
For example, for a double-well potential shown in Fig.~\ref{fig:Stokes_exam}, those are given by
\be
&& {\cal S}_{0_\pm} [\psi_{a_1}^{0_\pm {\rm I}}] = {\cal S}_{0_\pm}[{\cal M}^{0_\pm} \psi_{a_1}^{0_\pm {\rm VI}}], \qquad \nl
&& {\cal M}^{0_+} = M_+ N_{a_1,a_2} M_+ N_{a_2,a_3} M_+ M_- N_{a_3,a_4} M_- N_{a_4,a_1},\\
&& {\cal M}^{0_-} = M_+ N_{a_1,a_2} M_+ M_- N_{a_2,a_3} M_- N_{a_3,a_4} M_- N_{a_4,a_1},
\ee
where ${\cal S}_{0_+}[\psi_a^{0_+ \bullet}] = {\cal S}_{0_-}[\psi_a^{0_-\bullet}]$\footnote{
  The wavefunctions, $\psi_a^{0_\pm}$, also have a discontinuity caused by a fixed singularity at $\arg(\hbar) = 0$ and can be constructed from $\psi_a$ in Eq.(\ref{eq:psi_vec}).
  Since this effect is consequently irrelevant to QCs and their solutions, we do not argue $\psi^{0_\pm}_a$.
  See Ref.~\cite{Takei2} for the construction.
}.
Normalizability to the wavefunction in the asymptotic limit, $x \rightarrow \pm \infty$, requires being zero to a corresponding component of the monodromy matrix, that gives a QC denoted by ${\frak D}^{0_\pm}$:
\be
&& {\cal M}^{0_\pm}_{jk} = {\frak D}^{0_\pm} = 0,
\ee
where $(j,k)$ is determined by normalizability of the wavefunction as
\be
&& (j,k) =
   \begin{cases} 
     (1,1) & \mbox{if} \ \  \psi_{-}^{0_\pm{\rm I}}(-\infty) = \psi_{+}^{0_\pm{\rm I}}(+\infty)  = 0 \\
     (1,2) & \mbox{if} \ \  \psi_{-}^{0_\pm{\rm I}}(-\infty) = \psi_{-}^{0_\pm{\rm I}}(+\infty)  = 0 \\
     (2,1) & \mbox{if} \ \  \psi_{+}^{0_\pm{\rm I}}(-\infty) = \psi_{+}^{0_\pm{\rm I}}(+\infty)  = 0 \\
     (2,2) & \mbox{if} \ \  \psi_{+}^{0_\pm{\rm I}}(-\infty) = \psi_{-}^{0_\pm{\rm I}}(+\infty)  = 0 
   \end{cases}. \label{eq:j_k}
\ee
The QCs are functions of the energy $E$, so that solving ${\frak D}^{0_\pm}(E) = 0$ gives a energy spectrum, but containing a discontinuity at $\theta = 0$.

Even if the resulting QCs are Borel non-summable at $\theta=0$ such that ${\frak D}^{0_+} \ne {\frak D}^{0_-}$, their Borel resummed forms are equivalent to each other, i.e.,  ${\cal S}_{0_+}[{\frak D}^{0_+}] \propto {\cal S}_{0_-}[{\frak D}^{0_-}]$.
This fact implies existence of Stokes automorphism that offers a relation between ${\frak D}^{0_+}$ and ${\frak D}^{0_-}$.
This Stokes automorphism is known as \textit{Delabaere-Dillinger-Pham (DDP) formula}~\cite{DDP2,DP1} in the context of EWKB.
Generally, QCs consist of \textit{Voros symbols} (cycles) defined as a contour integration of $S_{{\rm od}}$ going around two turning points such that $A=e^{a_{ij}}$, where
\be
a_{ij} := \oint_{a_i}^{a_j} dx \, S_{\rm od}(x,\hbar) 
= -a_{ji}, \qquad a_{ij} + a_{jk} = a_{ik}, \qquad (a_i,a_j,a_k \in {\rm TP}) \label{eq:def_aij}
\ee
and the DDP formula tells us perturbative/non-perturbative relations, i.e., resurgent relations, among cycles.
For cycles constructed only by simple turning points, e.g. in the case of Fig.~\ref{fig:Stokes_exam}, the DDP formula can be expressed by
\be
&& {\cal S}_{\theta+0_+}[A_j] = {\cal S}_{\theta+0_-}[A_j] \prod_{B_k \in {\rm C}_{{\rm NP},\theta}} ( 1 +  {\cal S}_{\theta+0_-} [B_k])^{\langle A_j,B_k \rangle}, \label{eq:def_DDP_A} \\
&& {\cal S}_{\theta+0_+}[B_k] = {\cal S}_{\theta+0_-}[B_k], \qquad  B_k \in {\rm C}_{{\rm NP},\theta}, \label{eq:def_DDP_B} 
\ee
or identically,
\be
&& {\frak S}_{\theta}^{\nu = 1} [A_j] = A_j \prod_{B_k \in {\rm C}_{{\rm NP},\theta}} ( 1 +  B_k)^{\langle A_j, B_k \rangle}, \label{eq:DDP_A_gen} \\
&& {\frak S}_{\theta}^{\nu=1}[B_k] = B_k, \qquad  B_k \in {\rm C}_{{\rm NP},\theta}, \label{eq:DDP_B_gen} 
\ee
where ${\rm C}_{{\rm NP},\theta}$ is a set of purely non-perturbative cycles that correspond to a degenerated Stokes line induced by a Stokes phenomenon at $\arg(\hbar) = \theta$, and orientation of the cycles in ${\rm C}_{{\rm NP},\theta}$ is fixed such that $ \lim_{\hbar \rightarrow 0_+}B_{k}(\hbar) = 0$.
The other cycles labeled by $j \in {\mathbb N}$, $A_{j}$, can be arbitrarily taken as far as being transversal to $B_k \in {\rm C}_{{\rm NP},\theta}$, but those are usually given as a set of cycles going around two turning points that enable to be connected by a branch-cut.
In addition, $\langle A,B \rangle$ denotes the intersection number between two cycles, $A$ and $B$,  determined by their orientation as
\be
\langle  \rightarrow , \uparrow \rangle = \langle  \leftarrow , \downarrow \rangle = +1, \qquad \langle  \rightarrow , \downarrow \rangle = \langle  \leftarrow , \uparrow \rangle = -1.
\ee
See also Ref.~\cite{Iwaki1} for the derivation.
Notice that ${\rm C}_{{\rm NP},\theta}$ depends on the value of $\theta$, and thus ${\rm C}_{{\rm NP},\theta} = \emptyset$ when no Stokes phenomenon happens on the Stokes graph at $\theta$.
The generalization to a one-parameter Stokes automorphism with $\nu \in {\mathbb R}$, denoted by ${\frak S}^{\nu}_{\theta}$, is available by replacing the intersection number in Eq.(\ref{eq:DDP_A_gen}) as $\langle A_j,B_k \rangle \rightarrow \langle A_j, B_k \rangle \times \nu$.
Eq.(\ref{eq:DDP_B_gen}) does not change for any $\nu$.
From Eqs.(\ref{eq:DDP_A_gen})(\ref{eq:DDP_B_gen}), the median resummed forms are obtained as
\be
&& \widehat{A}_j = {\cal S}_{\theta+0_\pm}[A_j^{\theta + 0_\pm}], \qquad A^{\theta+0_\pm}_j := {\frak S}^{\mp 1/2}_{\theta}[A_j] = A_j \prod_{B_k \in {\rm C}_{{\rm NP},\theta}} ( 1 +  B_k)^{\mp \langle A_j, B_k \rangle/2}, \\
&& \widehat{B}_k = {\cal S}_{\theta+0_\pm}[B_k^{\theta+ 0_\pm}], \qquad B_k^{\theta+ 0_\pm} = B_k,
\ee
where $\widehat{f} = {\cal S}_{\rm med,\theta}[f]$.
From the properties of Borel resummation and Stoke automorphism in Eqs.(\ref{eq:S_homo})(\ref{eq:G_homo})(\ref{eq:Stokes_prop_add}), the QCs satisfy
\be
&& {\frak S}^{+1/2}_{\theta}[{\frak D}^{\theta + 0_+}(A,B)] \propto {\frak S}^{-1/2}_{\theta}[{\frak D}^{\theta + 0_-}(A,B)]  \ \ =: {\frak D}^{\theta}(A,B), \label{eq:D_no_sing} \\
&& {\frak D}^{\theta}(A^{\theta+0_\pm},B^{\theta+0_\pm}) \propto {\frak D}^{\theta+0_\pm}(A,B).
\ee
Apparently, ${\cal S}_{\theta + 0_+}[{\frak D}^{\theta}(A^{\theta+0_+},B^{\theta+0_+})] \propto {\cal S}_{\theta + 0_-}[{\frak D}^{\theta}(A^{\theta+0_-},B^{\theta+0_-})] = {\frak D}^{\theta}(\widehat{A},\widehat{B})$.
This is a simple way to make sure of the relation that
\be
{\cal S}_{{\rm med},\theta} [{\frak D}^{\theta}(A,B)] \overset {\hbar \rightarrow 0_+} \sim {\frak D}^{\theta}(A,B).
\ee

It is quite crucial to remind that the energy, $E$, is a free-parameter in the DDP formula in Eqs.(\ref{eq:def_DDP_A})-(\ref{eq:DDP_B_gen}) and determined by solving the QC, ${\frak D}^\theta = 0$ (or $\widehat{\frak D}^{\theta} = 0$).
In this sense, the DDP formula of cycles is not relevant to the energy solution \textit{directly}.
However, one can formulate a one-parameter Stokes automorphism for a formal transseries of the energy solution by combining the DDP formula for cycles with the QC.
This issue would be discussed in Sec.~\ref{sec:PT_sym_mass} and Appendix~\ref{sec:DDP_energy}.

The complex phase $\theta$ to resolve a degeneracy of Stokes lines is normally introduced by $\arg(\hbar)$, but the other parameters in a given theory can be used if it induces the similar effect.
In our analysis, we use a coupling constant, $\lambda$ (resp. $g$), as $\theta = \arg(\lambda)$ (resp. $\theta = \arg(g)$).

\section{Quartic potential with a quadratic term} \label{sec:with_mass_term}
In this section, we apply EWKB to a quartic potential with $\omega > 0$.
In Sec.~\ref{sec:warm_up},  we firstly analyze the Hermitian potential as a warm-up.
In Sec.~\ref{sec:PT_sym_mass}, we then consider the negative coupling potential.
In Sec.~\ref{sec:reform_ABS}, from transseries solutions of the ${\cal PT}$ and the AC QCs obtained in Sec.~\ref{sec:PT_sym_mass}, 
we reformulate the ABS conjecture by Borel resummation theory.

\subsection{Hermitian potential: $V=\omega^2  x^2 + \lambda x^4$} \label{sec:warm_up}
We consider the Hermitian potential.
In Sec.~\ref{sec:warm_up_solve_QC}, we demonstrate the procedure that we repeatedly perform in the later analyses.
In Sec.~\ref{sec:warm_up_analytic_QC}, we also discuss analytic continuation of the QCs which will be helpful for analysis of the negative coupling potential and considerations in Sec.~\ref{sec:impossibility_PT_AC}.

\subsubsection{Quantization condition and energy spectrum} \label{sec:warm_up_solve_QC}

\begin{figure}[tbp]
  \begin{center}
    \begin{tabular}{cc}
      \begin{minipage}{0.5\hsize}
        \begin{center}
          \includegraphics[clip, width=75mm]{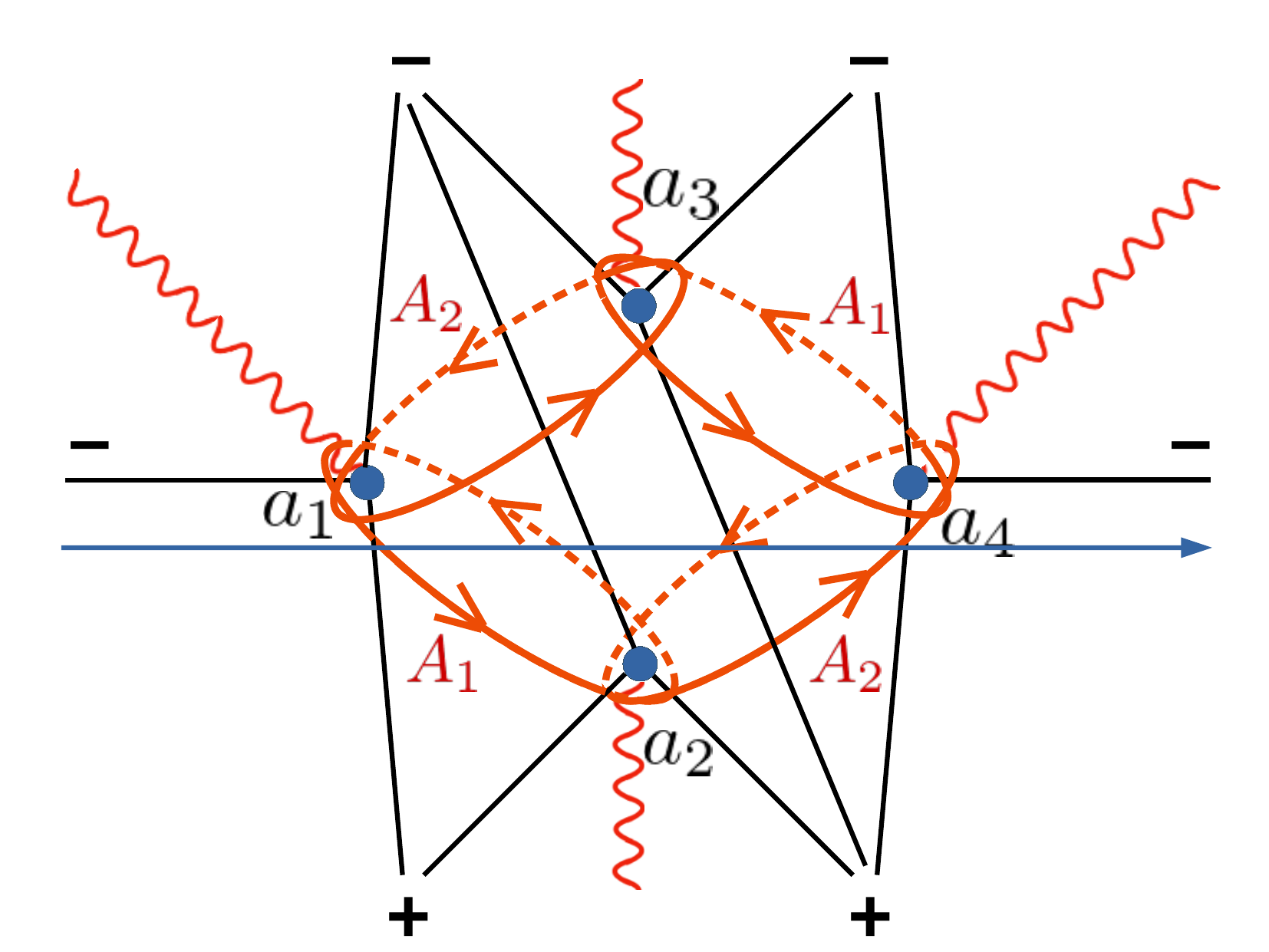} 
          \hspace{1.6cm} (a) $\arg(\lambda) = 0_+$
        \end{center}
      \end{minipage}
      \begin{minipage}{0.5\hsize}
        \begin{center}
          \includegraphics[clip, width=75mm]{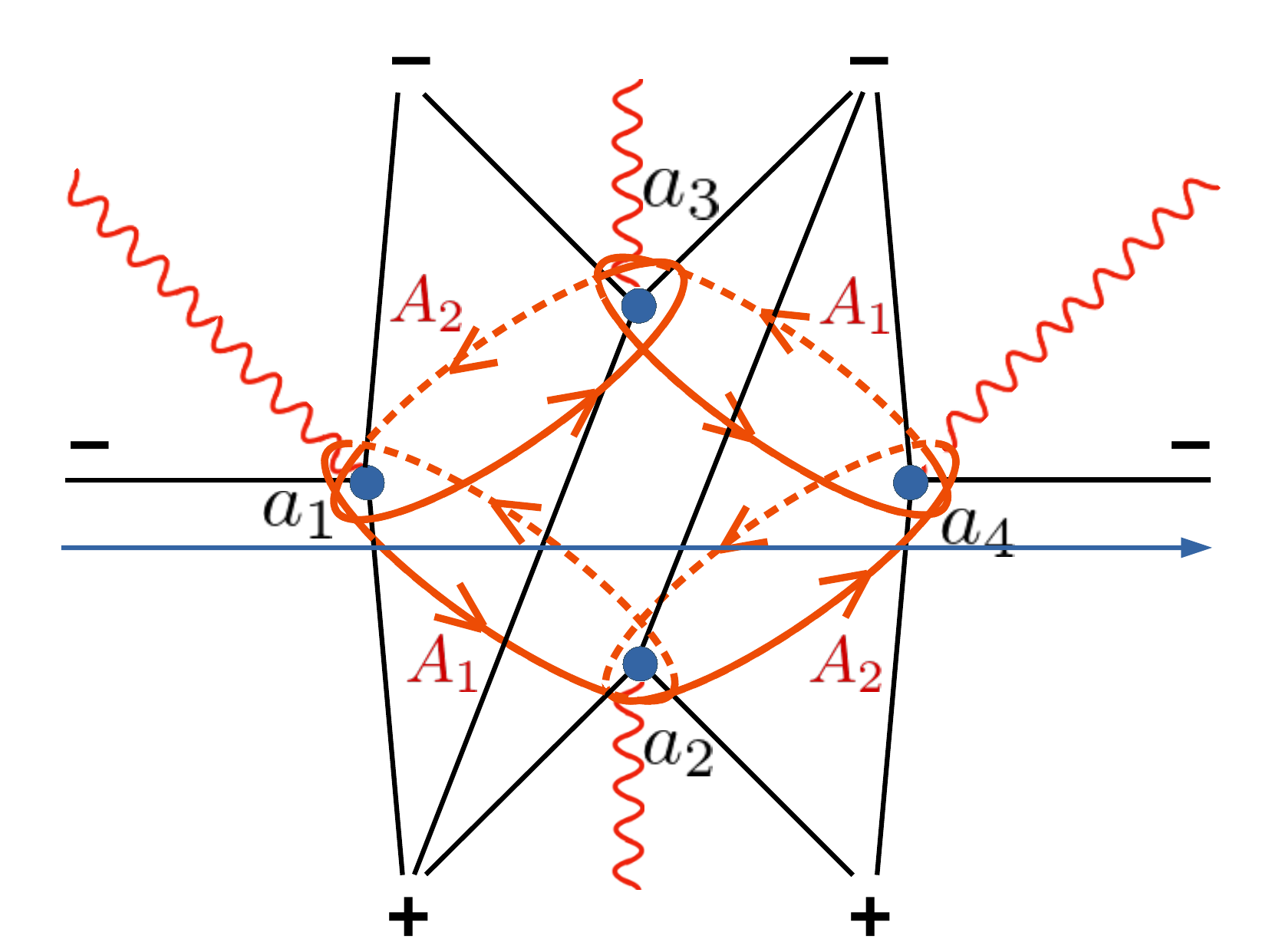} 
          \hspace{1.6cm} (b) $\arg(\lambda) = 0_-$
        \end{center}
      \end{minipage}      
    \end{tabular} 
    \caption{Stokes graph of the Hermitian potential with $\omega>0$.
      We take the analytic continuation along the path denoted the blue line to obtain the monodromy matrix.
    }
    \label{fig:stokes_warm}
  \end{center}
\end{figure}


We apply EWKB to a Schr\"{o}dinger equation of the Hermitian potential defined as
\be
&&   {\cal L} =  - \hbar^2 \pd_x^2 + \omega^2 x^2 + \lambda x^4 - E. \qquad {\cal L} \psi = 0, \label{eq:sc_mass_warm_Airy}
\ee
where $\omega, \lambda, E \in {\mathbb R}_{>0}$, and we take $E$ sufficiently small.
From Eq.(\ref{eq:def_TPs}), turning points are given by
\be
   {\rm TP} &=& \left\{ a_1 = -\sqrt{\frac{\sqrt{4 E \lambda +\omega ^4}-\omega ^2}{2 \lambda}}, \
   a_2 = - i \sqrt{\frac{\sqrt{4 E \lambda +\omega ^4} + \omega ^2}{2 \lambda}},   \right. \nl
   && \left. \ \ a_3 = + i \sqrt{\frac{\sqrt{4 E \lambda +\omega ^4} + \omega ^2}{2 \lambda}}, \ a_4 = +\sqrt{\frac{\sqrt{4 E \lambda +\omega ^4}-\omega ^2}{2 \lambda}} \right\}. \label{eq:TP_warm_mass}
\ee
Fig.~\ref{fig:stokes_warm} shows the Stokes graph drawn by Eq.(\ref{eq:sc_mass_warm_Airy}).
Because of ${\cal P}$ symmetry in the potential,  $V(-x) = V(x)$, one has two independent cycles, $A_{1,2}$, defined as 
\be
&& A_1 := e^{a_{12}} = e^{a_{34}}, \qquad A_2 := e^{a_{13}} = e^{a_{24}}, \label{eq:def_A12}
\ee
where $a_{jk}$ is defined by Eq.(\ref{eq:def_aij}).
Performing analytic continuation along a line slightly below the real axis generates a QC by connection matrices, as explained in Sec.~\ref{sec:EWKB}.
Since a Stokes phenomenon happens at $\arg(\lambda) = 0$, the monodromy matrix depends on the complex phase $\arg(\lambda) = 0_\pm$ and is obtained as
\be
&& {\cal M}^{0_+} = M_+ N_{a_1,a_2} M_-^{-1} N_{a_2,a_3} M_+ N_{a_3,a_4} M_+ N_{a_4,a_1}, \label{eq:M0p_warm} \\
&& {\cal M}^{0_-}  = M_+ N_{a_1,a_3} M_+ N_{a_3,a_2} M_-^{-1} N_{a_2,a_4} M_+ N_{a_4,a_1}. \label{eq:M0m_warm} 
\ee
Normalizability of the wavefunction determines the boundary condition and requires ${\cal M}^{0_\pm}_{12} = 0$ by Eq.(\ref{eq:j_k}).
As a result, one can write down the QCs represented by cycles as
\be
&& {\frak D}^{0_+}_{\cal H} \propto  1 +\frac{(2 + A_1)A_2}{1+A_1^{-1} A_2} = 1 +\frac{(2 + A_1)A_1D}{1+D}, \label{eq:qcondp_lam4_mass} \\
&& {\frak D}^{0_-}_{\cal H} 
\propto ( 1 + A_2)^2  + A_1 A_2 = ( 1 + A_1 D)^2  + A_1^2 D, \label{eq:qcondm_lam4_mass} 
\ee
where we defined $D=A_1^{-1} A_2$ exponentially damping without oscillation as a non-perturbative cycle, ${\rm C}_{{\rm NP},\arg(\lambda)=0} = \{ D\}$.

From Eqs.(\ref{eq:DDP_A_gen})-(\ref{eq:DDP_B_gen}), the DDP formula for $\arg(\lambda)=0$ is given by
\be
&&  {\frak S}_{0}^{\nu}[A_{j=1,2}] = A_j (1+D)^{\nu}, 
\qquad {\frak S}_{0}^{\nu}[D] = D. \qquad (\nu \in {\mathbb R}) \label{eq:DDP_A_warm}
\ee
The QC removed the discontinuity, ${\frak D}^{0}_{\cal H}$, is available by Eq.(\ref{eq:D_no_sing}) as ${\frak S}_0^{+1/2}[{\frak D}^{0_+}_{\cal H}] \propto {\frak S}_0^{-1/2}[{\frak D}^{0_-}_{\cal H}]= :{\frak D}^{0}_{\cal H}$, where
\be
   {\frak D}^{0}_{\cal H} \propto 1 + A_2 \left( A_1 + \frac{2}{\sqrt{1 + A_1^{-1} A_2}}\right)   = 1 + A_1 D \left( A_1 + \frac{2}{\sqrt{1 + D}}\right). \label{eq:qcond_lam4_mass}
\ee
   
It is worth to see that the median resummed form of ${\frak D}^0_{\cal H}$ has the same to Eq.(\ref{eq:qcond_lam4_mass}).
The discontinuous QCs in Eqs.(\ref{eq:qcondp_lam4_mass})(\ref{eq:qcondm_lam4_mass}) can be reexpressed as
\be
&&   {\frak D}^{0_\pm}_{\cal H} \propto  1 + A_2^{0_\pm}  \left( A_1^{0_\pm} + \frac{2}{\sqrt{1 + (A_1^{0_\pm})^{-1} A_2^{0_\pm}}}\right) =  1 + A_1^{0_\pm} D^{0_\pm} \left( A_1^{0_\pm} + \frac{2}{\sqrt{1 + D^{0_\pm}}}\right), \nl 
&& A_{j=1,2}^{0_\pm} = {\frak S}^{\mp 1/2}_0 [A_j] = A_j (1 + D)^{\mp 1/2}, 
\qquad D^{0_\pm} = D. \label{eq:A0pm_D0pm} 
\ee 
Since $\widehat{A}_j = {\cal S}_{{\rm med},0}[A_j] = {\cal S}_{0_\pm}[A_j^{0_\pm}]$ and $\widehat{D} = {\cal S}_{{\rm med},0}[D] = {\cal S}_{0_\pm}[D^{0_\pm}]$, one can immediately find that $\widehat{\frak D}^{0} _{\cal H}= {\cal S}_{{\rm med},0}[{\frak D}^{0}_{\cal H}] \propto {\cal S}_{0_\pm}[{\frak D}^{0_\pm}_{\cal H}]$.



\begin{figure}[tbp]
  \begin{center}
    \begin{tabular}{cc}
      \begin{minipage}{0.5\hsize}
        \begin{center}
          \includegraphics[clip, width=75mm]{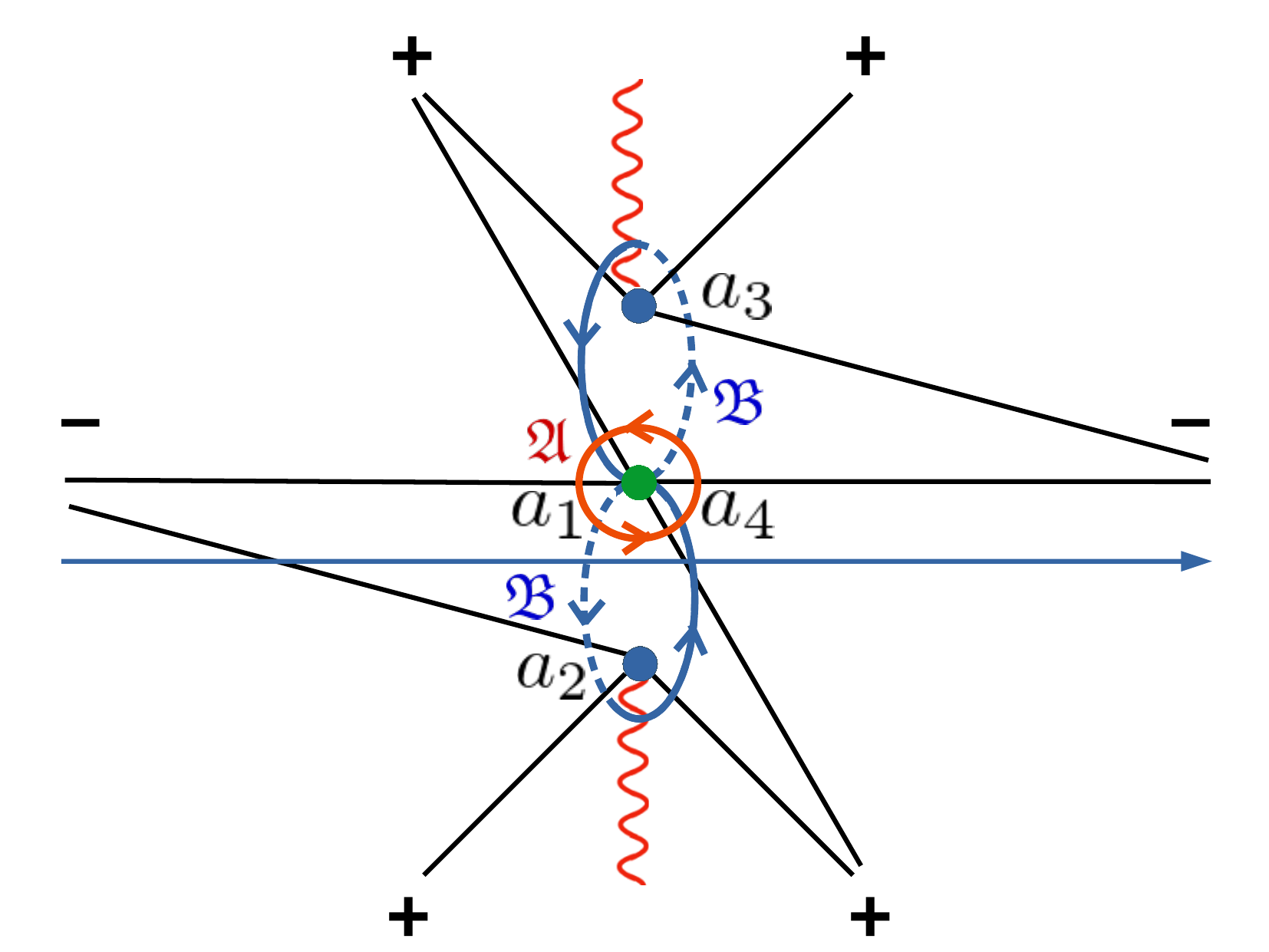}
          \hspace{1.6cm} (a) $\arg(\lambda)=0_+$
        \end{center}
      \end{minipage}
      \begin{minipage}{0.5\hsize}
        \begin{center}
          \includegraphics[clip, width=75mm]{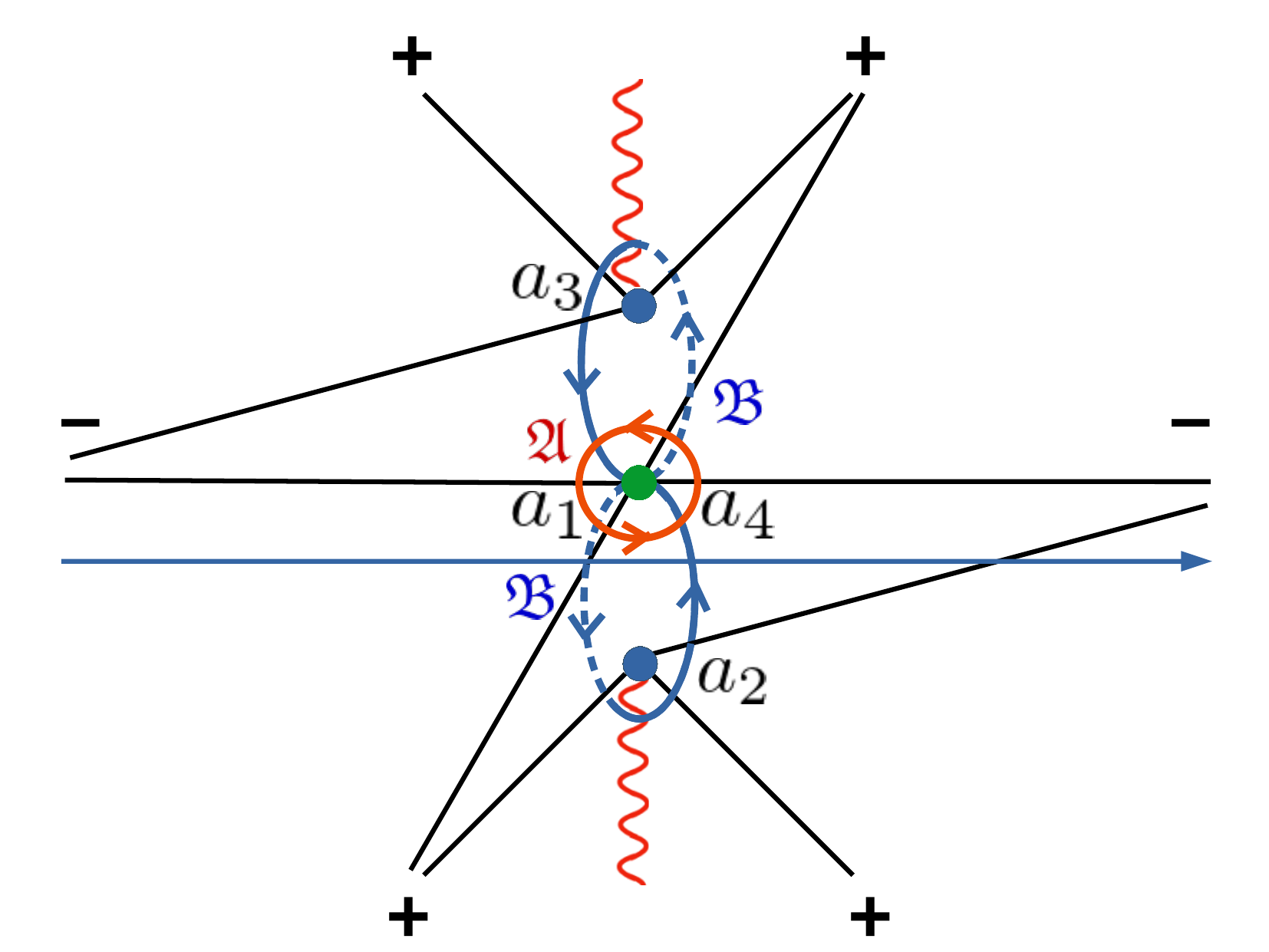}
          \hspace{1.6cm} (b) $\arg(\lambda)=0_-$
        \end{center}
      \end{minipage}      
    \end{tabular} 
    \caption{Stokes graph of the Hermitian potential with a quadratic term.
      Varying $E_0 \rightarrow 0_+$, the two simple turning points, $a_1$ and $a_4$, collide to each other and then become a double turning point (green dot) at $E_0 = 0$, where $E_0$ is the classical part of the energy.
    }
    \label{fig:stokes_warm_wb}
  \end{center}
\end{figure}

One can formally find the energy spectrum by solving the median resummed QC, but it is technically almost impossible to directly see details from it.
So that, our strategy is to find a transseries solution from ${\frak D}^0_{\cal H} = 0$ in Eq.(\ref{eq:qcond_lam4_mass}).
However, QCs formulated by the Airy-type Stokes graph usually have a problem such that the ordering of a double-expansion in terms of $E$ and $\hbar$ is generally non-commutative with each other.
Thus, the resulting transseries from the QC does not always correspond to a correct energy solution.
The main reason is that the energy solution is $O(\hbar)$ and exceeds the applicable limit or the assumption of the Airy-type connection formula.
In such a case, one needs to modify the Schr\"{o}dinger equation as
\be
&&   {\cal L} =  - \hbar^2 \pd_x^2 + \omega^2 x^2 + \lambda x^4 - \widetilde{E} \hbar, \qquad {\cal L} \psi = 0, \label{eq:sc_mass_warm}
\ee
where $E = E_0 + \widetilde{E} \hbar >0$ with $E_0 = 0$ and $\widetilde{E} = O(\hbar^0)$.
$E_0$ has an important role for the Stokes graph and is indeed a control parameter of it.
Taking $E_0 = 0$ causes a bifurcation such that two simple turning points corresponding to a harmonic oscillator collide to each other and become a double turning point.
In consequence, the set of turning points in Eq.(\ref{eq:TP_warm_mass}) changes to
\be
   {\rm TP} &=& \left\{ a_1 = a_4 = 0, \  a_2 = - i \frac{\omega}{\sqrt{\lambda}}, \ a_3 = +i \frac{\omega}{\sqrt{\lambda}} \right\}. 
\ee
The Stokes graph is shown in Fig.~\ref{fig:stokes_warm_wb}.
The local Stokes graph defined by a double turning point is called as \textit{degenerate Weber (DW)-type} Stokes graph, that gives another form of connection matrices~\cite{Takei3,DDP2}.
See also Appendix~\ref{sec:der_G}.
In spite of the difference between the Airy-type and the DW-type connection formulas,
the QCs and the DDP formula expressed by the Airy-type cycles can be directly translated into the DW-type~\cite{DP1,Zinn-Justin:2004qzw,Sueishi:2021xti}. 
For a simplified notation, by scaling dimensions in Eq.(\ref{eq:sc_mass_warm}) given by
\be
[x] = \frac{1}{2}, \qquad [\hbar] = 1, \qquad [\omega] = 0, \qquad [\lambda] = -1, \qquad [\widetilde{E}] = 0,
\ee
we rescale the parameters as
\be
  x \rightarrow \frac{\omega}{\sqrt{\lambda}} x, \qquad  \widetilde{E} \rightarrow \omega \widetilde{E}, \qquad \hbar \rightarrow \frac{ \omega^3}{\lambda} \hbar. \label{eq:resc_mass_warm}
\ee
Then, the Schr\"{o}dinger equation (\ref{eq:sc_mass_warm}) becomes dimensionless as
\be
 \frac{\lambda}{\omega^4}   {\cal L} \rightarrow {\cal L} =  - \hbar^2 \pd_x^2 + x^2 + x^4 - \widetilde{E} \hbar. \label{eq:sc_mass_warm_resc}
\ee
The QCs of the DW-type can be directly obtained by replacing $A_{1,2}$ as 
\be
&& A_1 \rightarrow {\frak A}^{1/2} {\frak B}, \qquad A_2 \rightarrow {\frak A}^{1/2} {\frak B}^{-1}, \\
&& {\frak A} = e^{- 2 \pi i F}, \qquad {\frak B} =  \frac{\sqrt{2 \pi} e^{-G}}{\Gamma(1/2 + F)} \left( \frac{\hbar}{2} \right)^{-F}, \label{eq:frakAB_warm}
\ee
where $F$ and $G$ are formal expansions of $\hbar$ and $\widetilde{E}$.
$F$ can be computed by a residue integration rotating clockwisely around the double turning point, but we do not argue $G$ in this subsection.
By this procedure, one can obtain the modified QC from Eq.(\ref{eq:qcond_lam4_mass}) as\footnote{
  Of course, one can obtain QCs by beginning with the DW connection formula.
  In this case, the QCs are given by
  \be
     {\frak D}^{0_+}_{\cal H} \propto {\frak D}^{0_-}_{\cal H} \propto {\frak B},
  \ee
  but the resulting QCs do not have dependence on $\arg(\lambda) = 0_\pm$.
  As we will see, this observation is not a contradiction to Eqs.(\ref{eq:qcond_lam4_mass})(\ref{eq:qcond_lam4_mass_deg}).
}
\be
   {\frak D}^{0}_{\cal H} \propto 1 + {\frak A} \left( 1 + 2 \frac{ {\frak A}^{-1/2} {\frak B}}{\sqrt{1 + {\frak B}^2}} \right).  \label{eq:qcond_lam4_mass_deg}
\ee

The next step is to solve the QC (\ref{eq:qcond_lam4_mass_deg}).
The perturbative part is given by ${\frak D}^0_{\cal H} \sim 1 + {\frak A}$ = 0, i.e.,
\be
F(\widetilde{E},\hbar) = \frac{q}{2}, \qquad q \in 2 {\mathbb Z} + 1, \label{eq:QC_warm_mass}
\ee
where $q$ is an energy level.
As described above, $F$ is obtained by the residue integration around the double turning point, as
\be   
   F(\widetilde{E};\hbar) =  -{\rm Res}_{x=0} S_{\rm od}(x,\widetilde{E};\hbar), \label{eq:F_def}
\ee
and the formal power expansion of $F$ is given by
\be
F(\widetilde{E},\hbar) &=&  - \frac{\widetilde{E}}{2} + \frac{3 \left(1 + \widetilde{E}^2 \right)}{16} \hbar - \frac{5 \left(17 \widetilde{E} + 7 \widetilde{E}^3 \right)}{128} \hbar^2 \nl
&& + \frac{105 \left(19 +50 \widetilde{E}^2 + 11 \widetilde{E}^4 \right)}{2048}\hbar^3 + O(\hbar^4).
\ee 
By setting the asymptotic form to $\widetilde{E}$ as $\widetilde{E}_{\cal H}(\hbar) \sim \sum_{n \in {\mathbb N}_0} e_{n} \hbar^n$, the coefficients $e_n$ can be obtained by solving the QC order by order.
Specifically, the first four coefficients are given by
\be
\widetilde{E}_{\cal H}(\hbar) &=& q + \frac{3 \left(q^2+1\right)}{8} 
\hbar-\frac{q \left(17 q^2+67\right)}{64}  \hbar^2  \nl
&& + \frac{3 \left(125 q^4+1138 q^2+513\right)}{1024} \hbar^3 + O(\hbar^4), 
\ee
where the energy level is constrained as $q \in 2{\mathbb N}_0 + 1$ to give a positive energy.
This also means that $F$ is a negative half-integer, $F \in -  {\mathbb N} +\frac{1}{2}$, and thus the non-perturbative effects do not appear because ${\frak B} = 0$ due to the gamma function in Eq.(\ref{eq:frakAB_warm}).
Therefore, the energy solution is Borel summable even though the DDP formula in Eq.(\ref{eq:DDP_A_warm}) looked non-trivial~\cite{Bender:1969si,DDP2,Bucciotti:2023trp}.

It might be meaningful to turn on the parameters by the scaling law (\ref{eq:resc_mass_warm}).
From $E_{\cal H} = \widetilde{E}_{\cal H} \hbar$, the energy solution can be written down as the following form:
\be
\frac{E_{\cal H}}{\omega \hbar}  &=& q + \frac{3\left(q^2+1\right)}{8}  \left( \frac{\lambda \hbar}{\omega^3} \right)  -\frac{q \left(17 q^2+67\right)}{64}  \left( \frac{\lambda \hbar}{\omega^3} \right)^2  \nl
&& + \frac{3 \left(125 q^4+1138 q^2+513\right)}{1024} \left( \frac{\lambda \hbar}{\omega^3} \right)^3  + O(\hbar^4).  \label{eq:E_H_mass}
\ee
One can see that 
taking $\omega \rightarrow 0$ induces divergence.
Therefore, the formal power series is ill-defined at $\omega = 0$.
It is important to note that the AC energies are not naively available from Eq.(\ref{eq:E_H_mass}) by replacing the coupling constant as $\lambda \rightarrow e^{\pm \pi i} g$ with $g \in {\mathbb R}_{>0}$ because in general an asymptotic form (and also transseries) of a function depends on the complex phase of its expansion parameter.
This fact also implies that reducing to transseries from Borel resummed forms and changing $\arg(\lambda)$ are generally not commutative with each other. 
In order to find the AC energies, there exist two ways: analytic continuation of the Hermitian QC to the negative coupling and directly constructing the AC QCs from the negative coupling potential.
These would be discussed in Secs.~\ref{sec:warm_up_analytic_QC} and \ref{sec:PT_sym_mass}, respectively, and indeed give the same result.

\subsubsection{Analytic continuation of the quantization condition} \label{sec:warm_up_analytic_QC}

\begin{figure}[tbp]
  \begin{center}
    \begin{tabular}{ccc}
      \begin{minipage}{0.33\hsize}
        \begin{center}
          \includegraphics[clip, width=50mm]{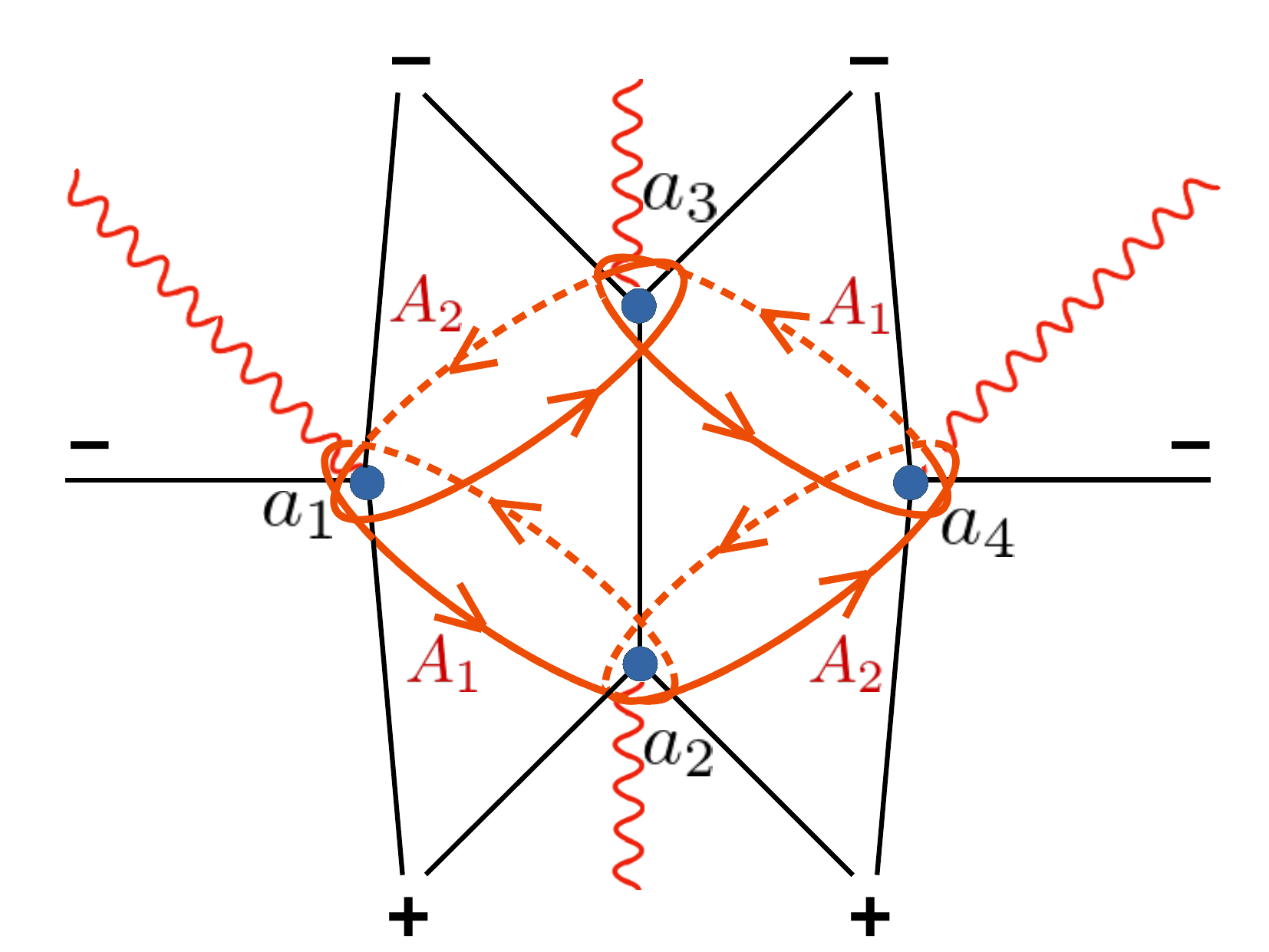}
          \hspace{1.6cm} (a) $\arg(\lambda)=0$
        \end{center}
      \end{minipage}
      \begin{minipage}{0.33\hsize}
        \begin{center}
          \includegraphics[clip, width=50mm]{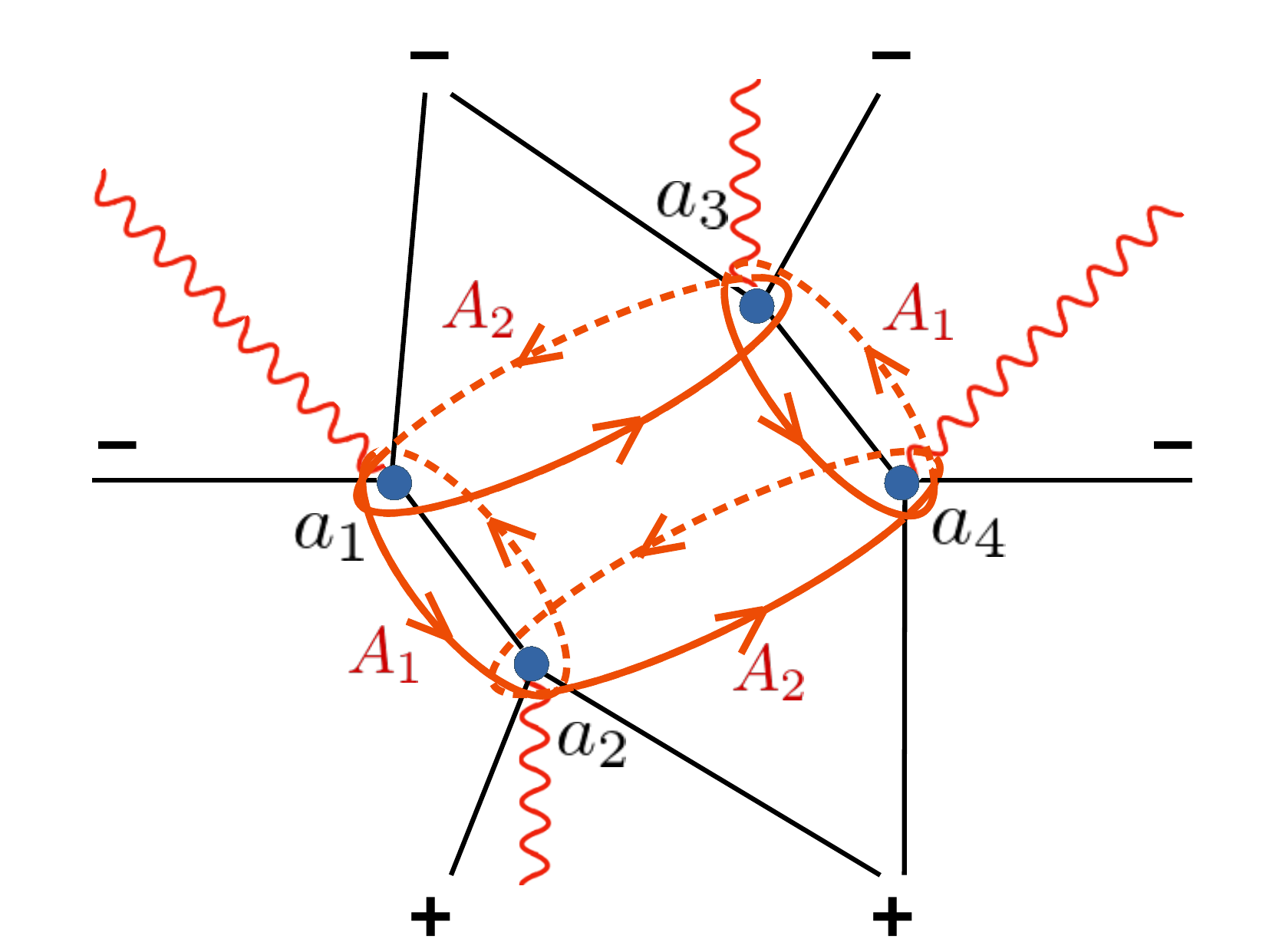}
          \hspace{1.6cm} (b) $\arg(\lambda)=\theta_*$
        \end{center}
      \end{minipage}
      \begin{minipage}{0.33\hsize}
        \begin{center}
          \includegraphics[clip, width=50mm]{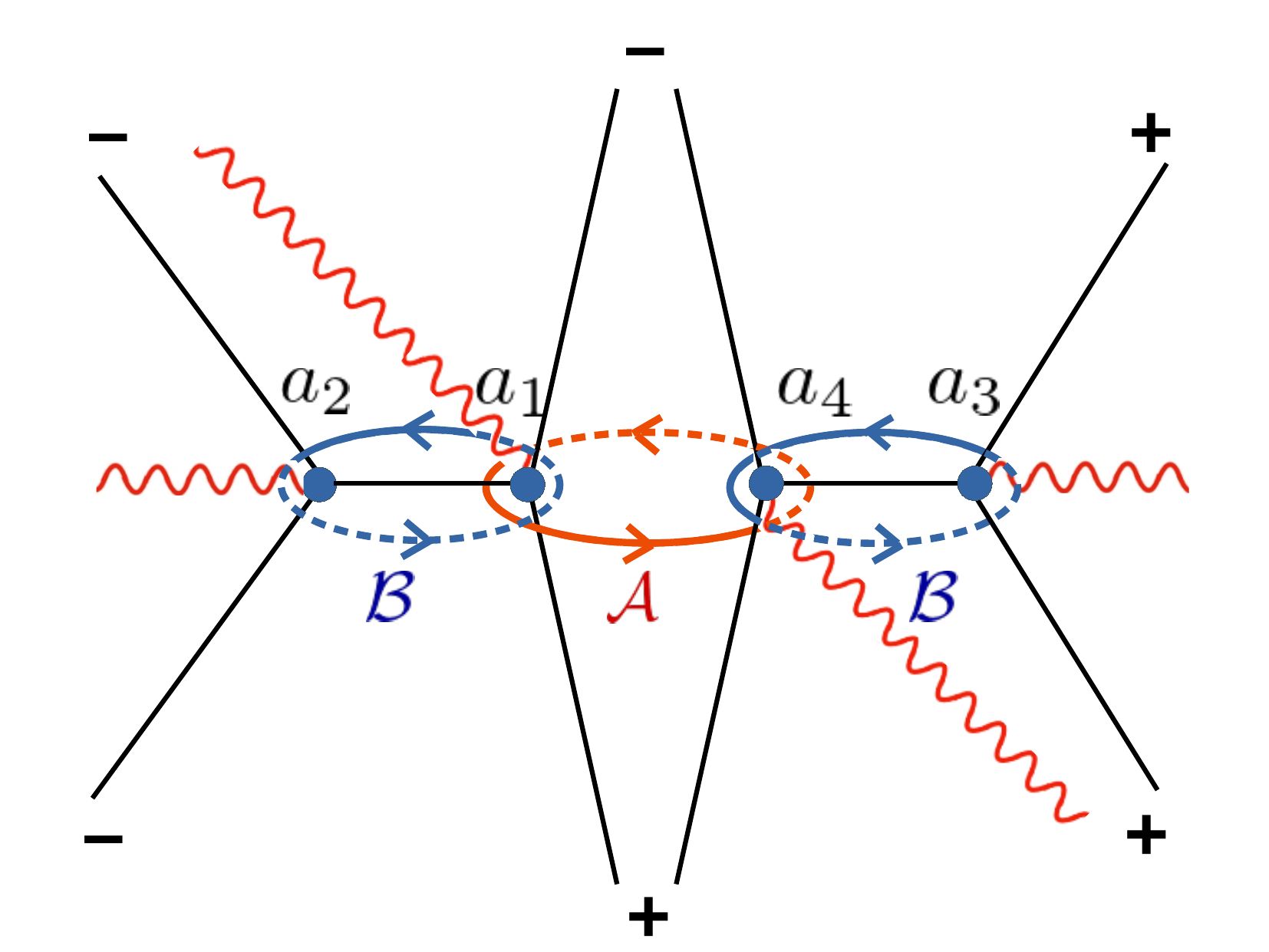}
          \hspace{1.6cm} (c) $\arg(\lambda)=+\pi$
        \end{center}
      \end{minipage}      
    \end{tabular} 
    \caption{
      Stokes graphs for $\arg(\lambda) = 0, \theta_* = \frac{3 \pi E |\lambda|}{4 \omega^4} + O(E^{3/2})$, and $+\pi$.
      The similar phenomena occur at $- \theta_*$ and $-\pi$ by varying $\arg(\lambda)$ from $0$ to $-\pi$.
    }
    \label{fig:stokes_warm_ac}
  \end{center}
\end{figure}

Before moving on the negative coupling case, it is worth to consider analytic continuation of the QCs by varying $\theta = \arg(\lambda)$ for a complex $\lambda$.
When QCs are expressed by a formal transseries, one generally needs to take into account of each Stokes phenomenon appearing during the analytic continuation.
Below, we consider the Airy-type cycles, but the similar consideration works for the DW-type thanks to the correspondence of cycle representations.

Varying $\arg(\lambda)$ from $0$ to $+ \pi$ (resp. $-\pi$)  
rotates the two turning points, $a_2$ and $a_3$, (resp. anti-)clockwisely, and they reach the real axis when $E_0>0$ is sufficiently small.
During the process, after resolving the degeneracy at $\arg(\lambda)=0$, Stokes phenomena happen twice at $\arg(\lambda) = \pm \frac{3 \pi E |\lambda|}{4 \omega^4} + O(E^{3/2})  =: \pm \theta_*$ and $\arg(\lambda) = \pm \pi$.
Fig.~\ref{fig:stokes_warm_ac} shows the Stokes graphs in which a Stokes phenomenon appears during the analytic continuation.
Each the phenomenon has its own DDP formula, and those can be expressed by using $A_{1,2}$ in Eq.(\ref{eq:def_A12}) as
\be
&&   {\frak S}_{+\theta_*}^{\nu} [A_1^{-1}] = A^{-1}_1, \qquad \qquad \quad \ \ {\frak S}_{+\theta_*}^{\nu} [A_2] = A_2 (1+A_1^{-1})^{2 \nu}, \nl
&&   {\frak S}_{+\pi}^{\nu} [{\cal A}] = {\cal A}(1 + {\cal B})^{-2 \nu},  \quad \quad \quad \, {\frak S}_{+\pi}^{\nu} [{\cal B}] = {\cal B}, \qquad ({\cal A} = A_1 A_2, \ {\cal B} = A_1)\\ \nl
&&   {\frak S}_{-\theta_*}^{\nu} [A_1] = A_1 (1+A_2)^{2 \nu}, \qquad \,  {\frak S}_{-\theta_*}^{\nu} [A_2] = A_2, \nl
&&   {\frak S}_{-\pi}^{\nu} [{\cal A}] = {\cal A}(1 + {\cal B})^{-2 \nu},  \quad  \quad \ \ \ \,  {\frak S}_{-\pi}^{\nu} [{\cal B}] = {\cal B}, \qquad ({\cal A} = A_1 A_2, \ {\cal B} = A_2^{-1})
\ee
where ${\rm C}_{{\rm NP},+\theta_*} = \{ A_1^{-1} \}$, ${\rm C}_{{\rm NP},-\theta_*} = \{ A_2 \}$, and ${\rm C}_{{\rm NP},\pm \pi} = \{{\cal B} \}$.
These give us analytic continuation of the QCs from $\arg(\lambda) = 0_+$ (resp. $0_-$) to $\arg(\lambda) = + \pi + 0_-$ (resp. $- \pi + 0_+$) as\footnote{
  In these equations, the symbol ``$0_{\pm}$'' is a different sense from the same symbol in the ABS conjecture.
  $\lambda = -g + i 0_\pm$ in the ABS conjecture corresponds to $\lambda = e^{ \pm \pi i } g$ in our notation.
  Our ``$0_{+}$'' (resp. ``$0_{-}$") means the limit from the right (resp. left) to the discontinuity on the Borel plane.
}
\be
&& {\frak D}^{+\theta_* + 0_+}_{\cal H} =  {\frak S}_{\theta_*}^{-1} [{\frak D}^{+\theta_* + 0_-}_{\cal H}], \quad \ \ \qquad {\frak D}^{+\pi + 0_+}_{\cal H} =  {\frak S}_{+\pi}^{-1} [{\frak D}^{+\pi + 0_-}_{\cal H}], \\
&& {\frak D}^{-\theta_* + 0_-}_{\cal H} =  {\frak S}_{-\theta_*}^{+1} [{\frak D}^{-\theta_* + 0_+}_{\cal H}], \qquad \quad \ {\frak D}^{-\pi + 0_-}_{\cal H} =  {\frak S}_{-\pi}^{+1} [{\frak D}^{-\pi + 0_+}_{\cal H}]. 
\ee
Notice that the cycles and the QCs are analytic under the change of $\arg(\lambda)$ until encountering the next Stokes phenomenon, i.e.,
\be
&& {\frak D}^{0_+}_{\cal H} \eqsym {\frak D}^{+\theta_* + 0_-}_{\cal H} \quad \quad \ \ \mbox{for} \quad 0 < \arg(\lambda) < \theta_*, \nl
&& {\frak D}^{+\theta_* + 0_+}_{\cal H} \eqsym {\frak D}^{+\pi + 0_-}_{\cal H} \quad \mbox{for} \quad \theta_* < \arg(\lambda) < \pi, \\ \nl
&& {\frak D}^{0_-}_{\cal H} \eqsym {\frak D}^{-\theta_* + 0_+}_{\cal H} \quad \quad \ \ \mbox{for} \quad - \theta_* < \arg(\lambda) < 0, \nl
&& {\frak D}^{-\theta_* + 0_-}_{\cal H} \eqsym {\frak D}^{-\pi + 0_+}_{\cal H} \quad \mbox{for} \quad - \pi < \arg(\lambda) < -\theta_*, 
\ee
where $\eqsym$ means that those cycle representations are analytic and keep the same symbolic forms in the domain.
Combining the above relations yields QCs at $\arg(\lambda) = +\pi + 0_{-}$ and $\arg(\lambda) = - \pi + 0_{+}$ as
\be
&& {\frak D}^{+ \pi + 0_-}_{\cal H} \eqsym {\frak S}_{+\theta_*}^{-1}[{\frak D}^{0_+}_{\cal H}] \propto 1 + {\cal A}, \label{eq:Dppi_zerom} \\
&& {\frak D}^{- \pi + 0_+}_{\cal H} \eqsym {\frak S}_{-\theta_*}^{+1}[{\frak D}^{0_-}_{\cal H}] \propto 1 + {\cal A}. \label{eq:Dmpi_zerop} 
\ee
As a result, by removing the discontinuity at $\arg(\lambda) = \pm \pi$ from ${\frak D}^{\pm \pi + 0_\mp}_{\cal H}$, one finds
\be
&&  {\frak D}^{+\pi}_{\cal H} = {\frak S}_{+\pi}^{-1/2} [{\frak D}^{+ \pi + 0_-}_{\cal H}] \propto 1 + {\cal A} (1 + {\cal B}), \label{eq:Dppi} \\
&&  {\frak D}^{-\pi}_{\cal H} =  {\frak S}_{-\pi}^{+1/2} [{\frak D}^{- \pi + 0_+}_{\cal H}] \propto 1 + \frac{{\cal A}}{1 + {\cal B}}. \label{eq:Dmpi} 
\ee
As one can readily see, the difference of the QCs by $\arg(\lambda) = \pm \pi$ appears in their non-perturbative part corresponding to ${\cal B}$ and is expected to eventually propagates to their energy solution.
It is notable that, as long as ${\cal B} \ne 0$, the energy solutions should be  complex values because ${\frak D}^{\pm \pi}_{\cal H}$ is not invariant under complex conjugate.
By denoting ${\cal C}$ as complex conjugate, since  $\log {\cal A} \in i {\mathbb R}$ and $\log {\cal B} \in {\mathbb R}$,
one can readily find that
\be
   {\cal C}[{\frak D}^{\pm \pi}_{\cal H}] \propto {\frak D}^{\mp \pi}_{\cal H}, \qquad {\cal C}[{\cal A}] = {\cal A}^{-1}, \qquad {\cal C}[{\cal B}] = {\cal B}. \label{eq:CD_D}
\ee
Complex conjugate is nothing but ${\cal T}$-transform, so that Eq.(\ref{eq:CD_D}) can be regarded that ${\cal T}$-transform swaps the QCs, ${\frak D}^{+\pi}_{\cal H} \leftrightarrow {\frak D}^{-\pi}_{\cal H}$.

In Sec.~\ref{sec:PT_sym_mass}, we would construct the ${\cal PT}$ and the AC QCs from the negative coupling potential.
As we will see, one can actually obtain the same forms to Eqs.(\ref{eq:Dppi})(\ref{eq:Dmpi}).


\subsection{Negative coupling potential: $V=\omega^2  x^2 - g x^4$} \label{sec:PT_sym_mass}

\begin{figure}[tbp]
  \begin{center}
    \begin{tabular}{cc}
      \begin{minipage}{0.5\hsize}
        \begin{center}
          \includegraphics[clip, width=75mm]{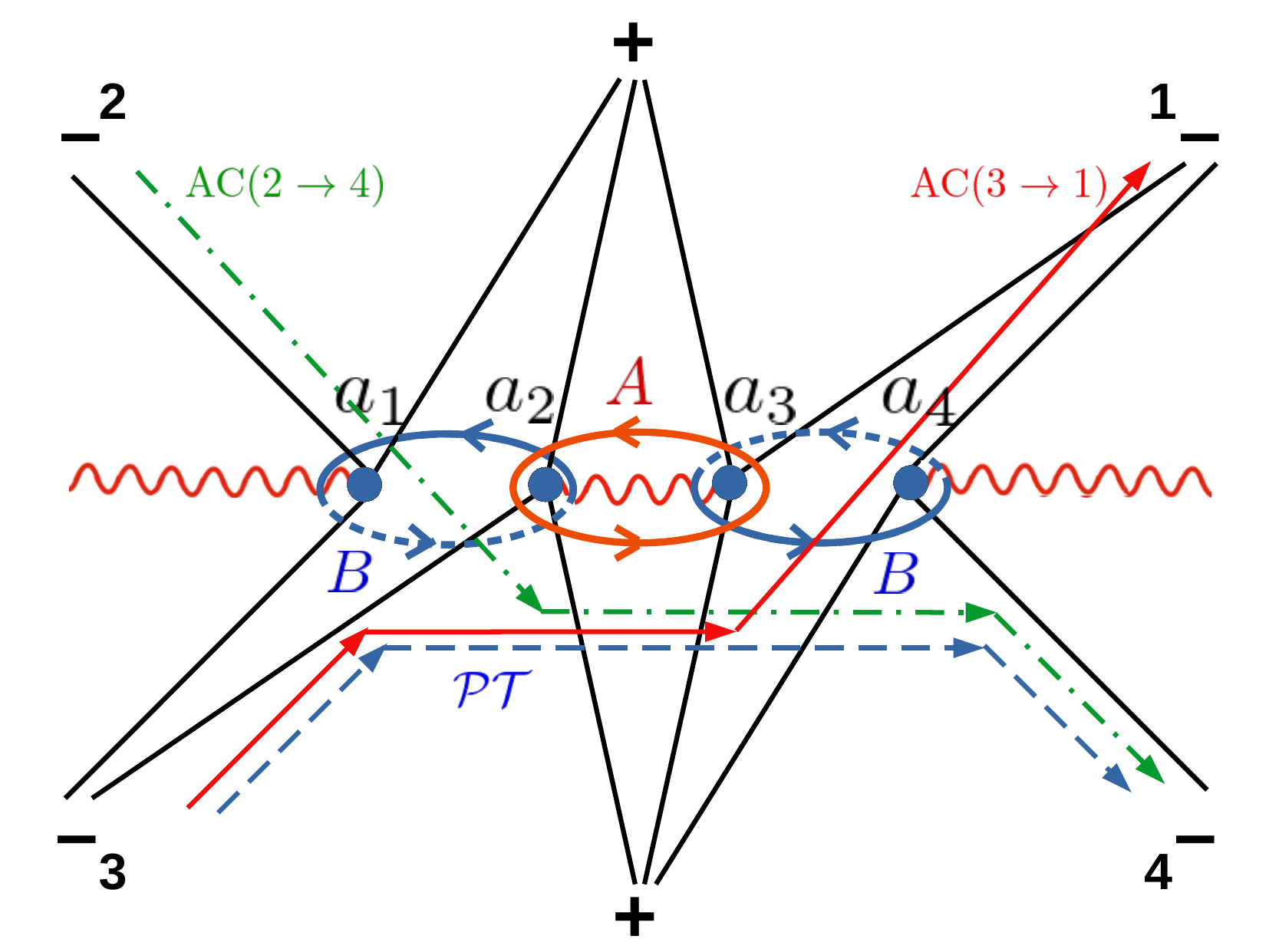}
          \hspace{1.6cm} (a) $\arg(g) = 0_+$
        \end{center}
      \end{minipage}
      \begin{minipage}{0.5\hsize}
        \begin{center}
          \includegraphics[clip, width=75mm]{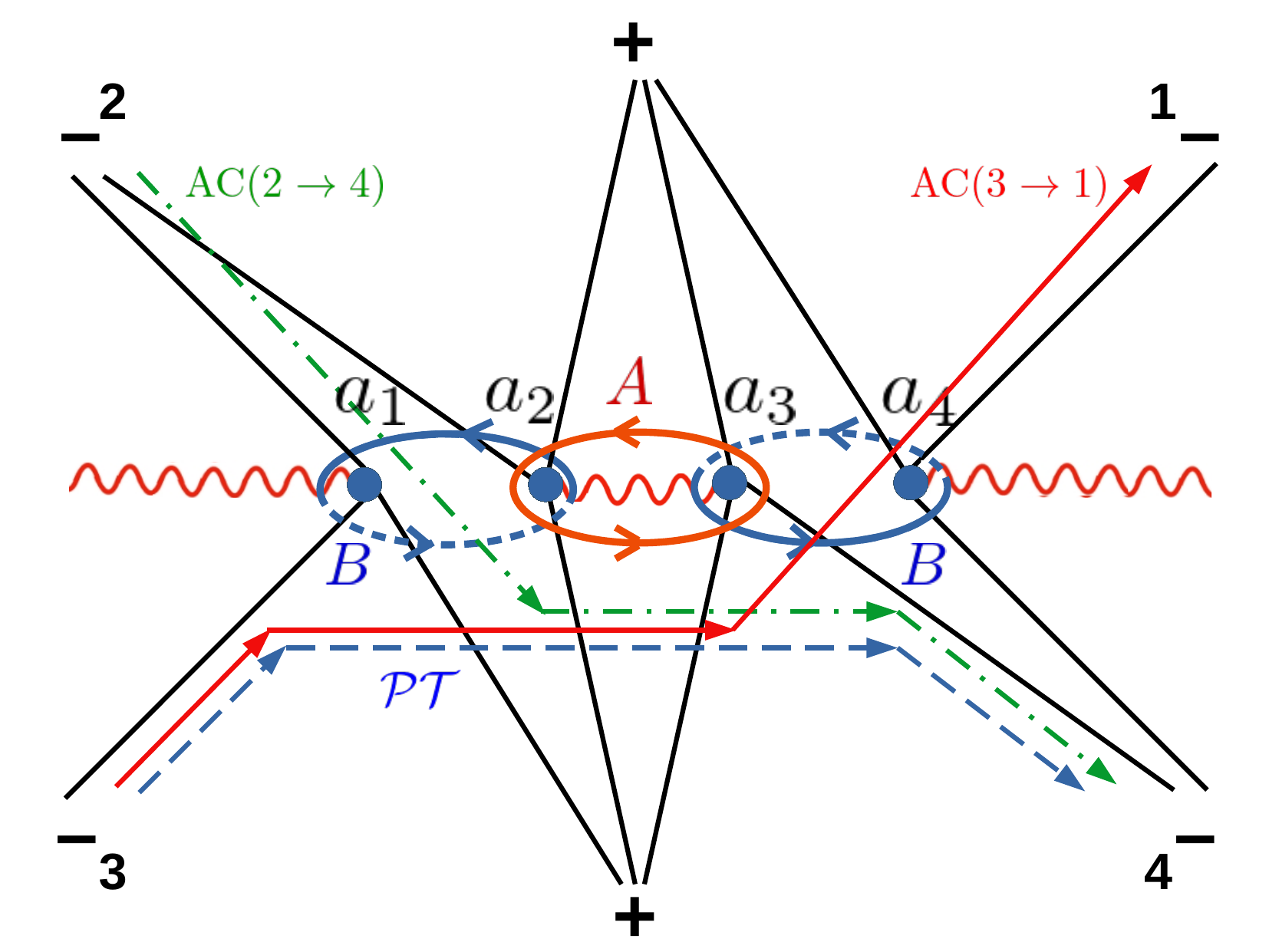}
          \hspace{1.6cm} (b) $\arg(g) = 0_-$
        \end{center}
      \end{minipage}      
    \end{tabular} 
    \caption{
      Stokes graph of the negative coupling potential with a quadratic term.
      The paths for the analytic continuation are denoted by colored lines, $\gamma_{3 \rightarrow 1}$ (red), $\gamma_{2 \rightarrow 4}$ (green), and $\gamma_{3 \rightarrow 4}$ (blue).
    }
    \label{fig:stokes_mass_PT}
  \end{center}
\end{figure}

We consider the negative coupling potential and derive transseries solutions of the ${\cal PT}$ and the ${\rm AC}$ energies.
The procedure of EWKB is almost parallel to Sec.~\ref{sec:warm_up_solve_QC}.

We begin with the following Schr\"{o}dinger equation:
\be
&& {\cal L} = - \hbar^2 \pd_x^2  + \omega^2 x^2 - g x^4 - E, \qquad {\cal L} \psi = 0,
\label{eq:sch_eq_no_mass_g} 
\ee
where $\omega, g, E, \in {\mathbb R}_{>0}$, and we take $E$ sufficient small.
From Eq.(\ref{eq:def_TPs}), turning points are given by
\be
   {\rm TP} &=& \left\{ a_2 = -  \sqrt{\frac{\omega^2 + \sqrt{\omega ^4 - 4 Eg}}{2 g}}, \ a_2 = -\sqrt{\frac{\omega^2-\sqrt{\omega ^4 - 4 E g}}{2 g}}, \right. \nl
   && \left. \ \ a_4 = +\sqrt{\frac{\omega^2 -\sqrt{\omega ^4 - E g}}{2 g}}, \  a_4 = +  \sqrt{\frac{\omega^2 + \sqrt{\omega ^4 - 4 Eg}}{2 g}} \right\}, \label{eq:TP_warm_mass_g}
\ee
and we define cycles as
\be
A = e^{a_{23}}, \qquad B = e^{a_{21}} = e^{a_{34}}.
\ee
Similar to the Hermitian case, a real positive-valued $g$ induces a Stokes phenomenon, so that we introduce an infinitesimal phase to $g$.
The Stokes graph is shown in Fig.~\ref{fig:stokes_mass_PT}.
In this case, the wavefunction at $x = \pm \infty$ is not normalizable because of oscillation, but there are six asymptotic domains to give a convergent (or divergent) wavefunction on the complex $x$-plane.
Hence, we take the following paths for analytic continuation:
\be
&& \gamma_{3 \rightarrow 1} : - e^{\frac{\pi}{4} i} \infty \ \rightarrow \ e^{\frac{\pi}{4} i} \infty,  \qquad 
\gamma_{2 \rightarrow 4} : - e^{-\frac{\pi}{4} i} \infty  \ \rightarrow \ e^{-\frac{\pi}{4}i} \infty,  \label{eq:path_AC}
\ee
for the AC energy, and
\be
\gamma_{3 \rightarrow 4} &:& - e^{\frac{\pi}{4} i} \infty \ \rightarrow \ e^{-\frac{\pi}{4} i} \infty,  \label{eq:path_PT}
\ee
for the ${\cal PT}$ energy\footnote{
The paths of analytic continuation are determined as continuous deformations of the asymptotic domains of the wavefunction on the complex $x$-plane from the real axis by varying a parameter in the potential.
For the AC energy, the deformation parameter is $\lambda$, and $\gamma_{3 \rightarrow 1/2 \rightarrow 4}$ are obtained by continuously changing the coupling as $\lambda \rightarrow e^{\mp \pi i} g$ with $g \in {\mathbb R}_{>0}$.
For the ${\cal PT}$ energy, in contrast, the deformation parameter is $\varepsilon$ in the potential given by $V_\varepsilon = g x^{2}(i x)^{\varepsilon}$ with $g \in {\mathbb R}_{>0}$, and continuously varying $\varepsilon = 0$ to $2$ gives $\gamma_{3 \rightarrow 4}$ from the real axis.
See, for example, Ref.~\cite{Bender:2019} and references within.
}
.
By taking the above paths, one can obtain the monodromy matrices for $\arg(g)=0_\pm$ as
\be
&& {\cal M}_{{\rm AC}(3 \rightarrow 1)}^{0_+} = N_{a_1,a_2} M_+ N_{a_2,a_3} M_+ M_- N_{a_3,a_1}, \\
&& {\cal M}_{{\rm AC}(3 \rightarrow 1)}^{0_-} = M_+ N_{a_1,a_2} M_+ N_{a_2,a_3} M_+ M_- N_{a_3,a_4} M_+^{-1} N_{a_4,a_1}, \\ \nl
&& {\cal M}_{{\rm AC}(2 \rightarrow 4)}^{0_+} = M_+^{-1} N_{a_1,a_2} M_- M_+ N_{a_2,a_3} M_+ N_{a_3,a_4} M_+ N_{a_4,a_1}, \\
&& {\cal M}_{{\rm AC}(2 \rightarrow 4)}^{0_-} = N_{a_1,a_2} M_- M_+ N_{a_2,a_3} M_+ N_{a_3,a_1}, \\ \nl
&& {\cal M}_{\cal PT}^{0_+} = N_{a_1,a_2} M_+ N_{a_2,a_3} M_+ N_{a_3,a_4}M_+ N_{a_4,a_1}, \\
&& {\cal M}_{\cal PT}^{0_-} = M_+ N_{a_1,a_2} M_+ N_{a_2,a_3} M_+ N_{a_3,a_1}.
\ee
By using Eq.(\ref{eq:j_k}), normalizability of the wavefunction determines the boundary condition as ${\cal M}^{0_\pm}_{12} = 0$ for both the ${\cal PT}$ and the AC matrices, and one can consequently find the QCs for $\arg(g)=0_\pm$ as
\be
&& {\frak D}_{{\rm AC}(3 \rightarrow 1)}^{0_+} \propto 1 + A, \qquad \qquad \quad \ \ \, {\frak D}_{{\rm AC}(3 \rightarrow 1)}^{0_-} \propto 1 + \frac{A}{(1+B)^2}, \\
&& {\frak D}_{{\rm AC}(2 \rightarrow 4)}^{0_+} \propto 1 + A(1+B)^{2}, \qquad {\frak D}_{{\rm AC}(2 \rightarrow 4)}^{0_-} \propto 1 + A, \\
&& {\frak D}_{\cal PT}^{0_+} \propto 1 + A(1+B), \qquad \qquad \ \,   {\frak D}_{\cal PT}^{0_-} \propto 1 + \frac{A}{1+B},
\ee
where ${\rm C}_{{\rm NP},\arg(g)=0} = \{ B \}$.
From Eqs.(\ref{eq:DDP_A_gen})-(\ref{eq:DDP_B_gen}), the DDP formula is also obtained by counting intersection numbers between the cycles as
\be
 {\frak S}_0^{\nu}[A] = A (1+B)^{-2 \nu}, \qquad  {\frak S}_0^{\nu}[B] = B. \label{eq:DDP_mass_g}
\ee
Thus, the QCs removed the discontinuity are obtained from Eq.(\ref{eq:D_no_sing}) as 
\be
&& {\frak D}^0_{{\rm AC}(3 \rightarrow 1)} \propto 1 + \frac{A}{1+B}, \qquad {\frak D}^0_{{\rm AC}(2 \rightarrow 4)} \propto 1 + A(1+B), \label{eq:DAC} \\
&& {\frak D}^0_{\cal PT} \propto 1 + A. \label{eq:DPT} 
\ee
It is remarkable that we could reproduce the QCs in Eqs.(\ref{eq:Dppi})-(\ref{eq:Dmpi}), that correspond to ${\frak D}^{\arg(\lambda) = +\pi}_{\cal H} \propto {\frak D}_{{\rm AC} (2 \rightarrow 4)}^{\arg(g) = 0}$ and ${\frak D}^{\arg(\lambda) = -\pi}_{\cal H} \propto {\frak D}_{{\rm AC} (3 \rightarrow 1)}^{\arg(g) = 0}$.
It is also notable that the path-dependence of the analytic continuation, given by Eq.(\ref{eq:path_AC}), can be seen as $A (1+B)^{\mp 1}$ in Eq.(\ref{eq:DAC}).
The difference consequently should appear in non-perturbative parts of the energy solution, as we pointed out in Sec.~\ref{sec:warm_up_analytic_QC}.
In contrast, the ${\cal PT}$ QC does not have such a non-perturbative effect, and the energy solution should contain only a perturbative part.
However, it does \textit{not} mean that the ${\cal PT}$ energy solution should be Borel summable.
It is indeed Borel non-summable unlike the Hermitian case, as we can see later.

\begin{figure}[tbp]
  \begin{center}
    \begin{tabular}{cc}
      \begin{minipage}{0.5\hsize}
        \begin{center}
          \includegraphics[clip, width=75mm]{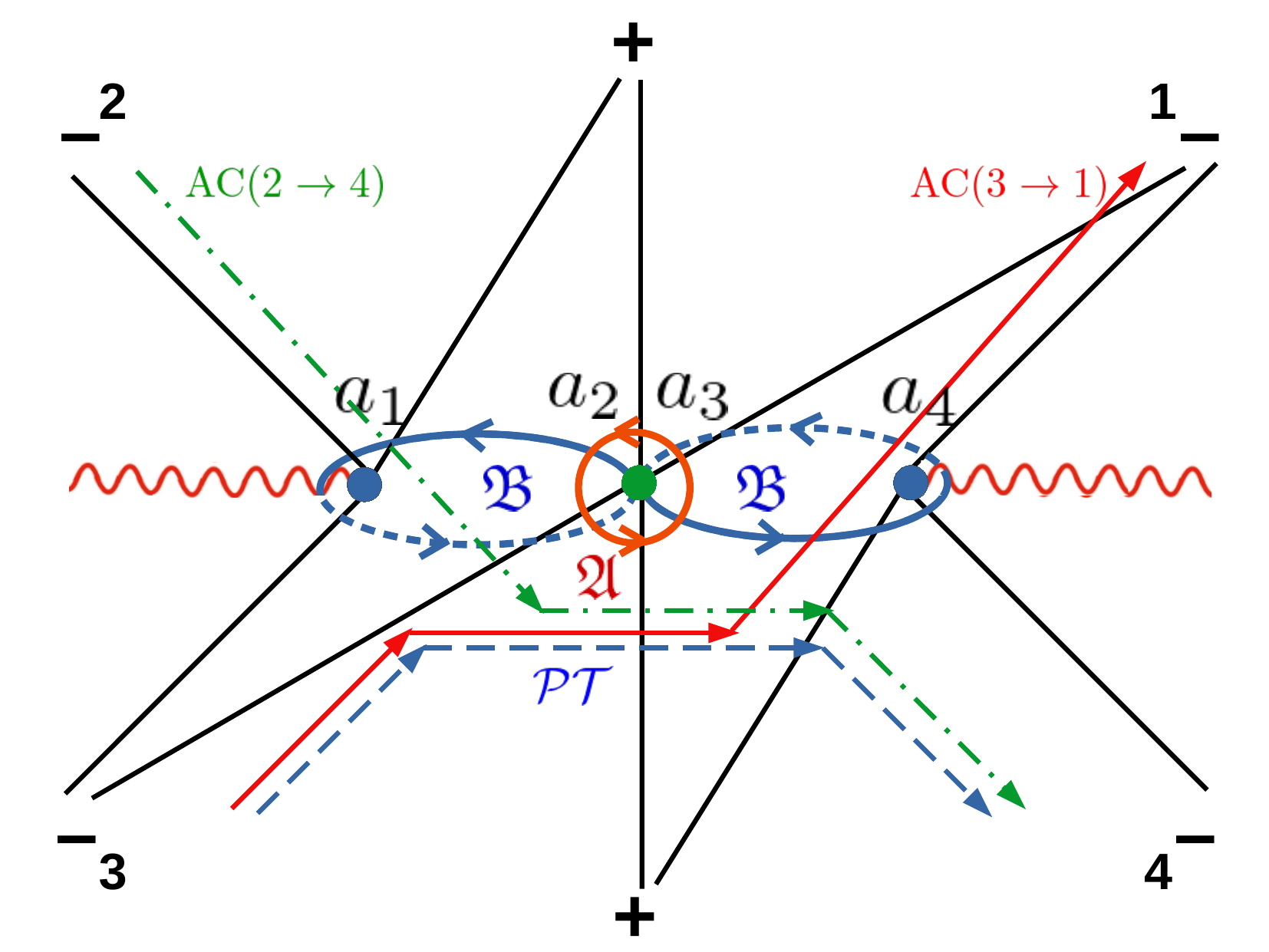}
          \hspace{1.6cm} (a) $\arg(g) = 0_+$
        \end{center}
      \end{minipage}
      \begin{minipage}{0.5\hsize}
        \begin{center}
          \includegraphics[clip, width=75mm]{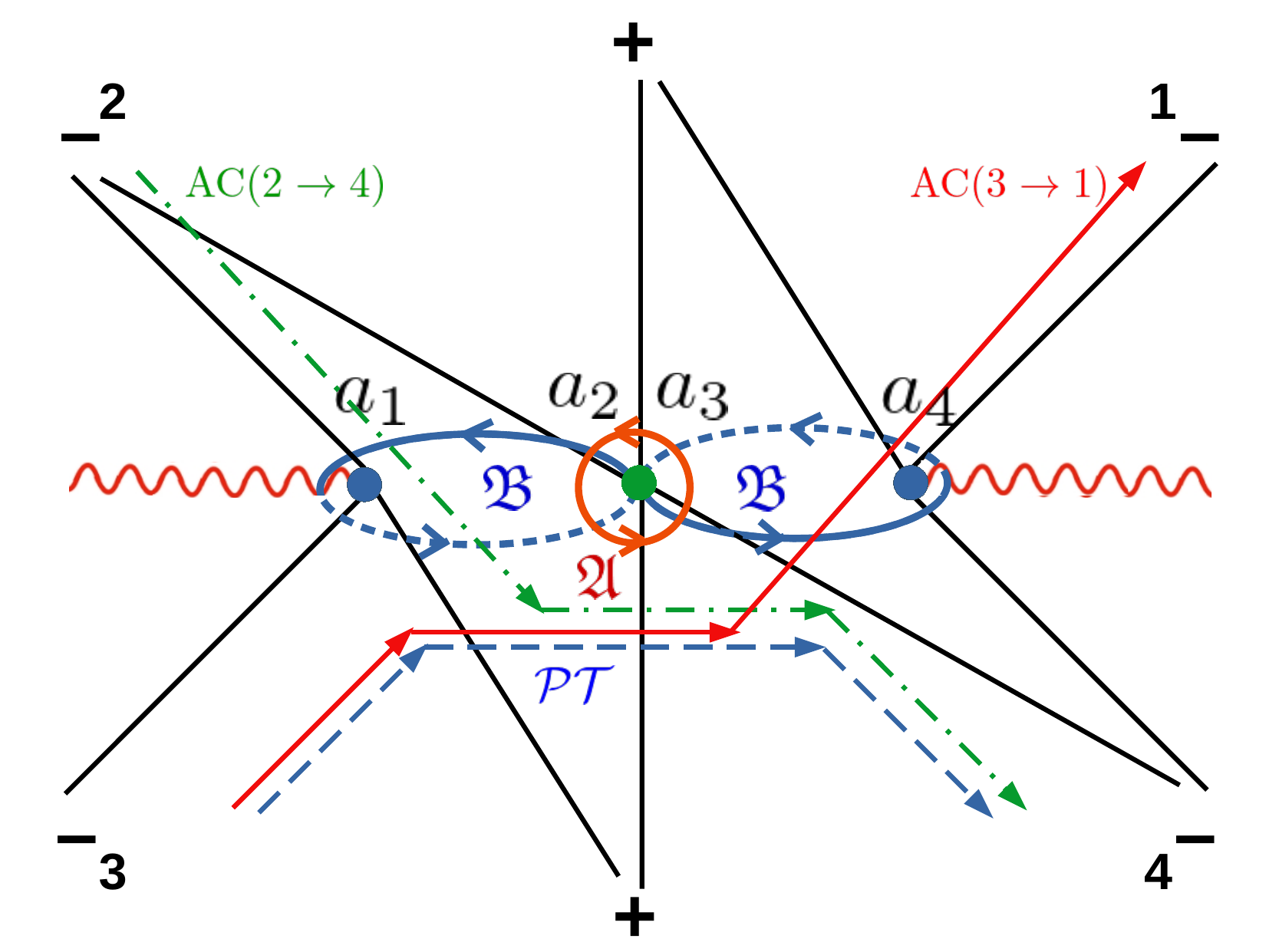}
          \hspace{1.6cm} (b) $\arg(g) = 0_-$
        \end{center}
      \end{minipage}      
    \end{tabular} 
    \caption{
      Stokes graph with $E_0 = 0$ corresponding to Fig.~\ref{fig:stokes_mass_PT}, where $E_0$ is the classical part of the energy.
      By taking $E_0 \rightarrow 0_+$, two simple turning points, $a_2$ and $a_3$, collide to each other and then become a double turning point (green dot) at $E_0 = 0$.
    }
    \label{fig:stokes_mass_PT_wb}
  \end{center}
\end{figure}

Let us solve the QCs (\ref{eq:DAC})(\ref{eq:DPT}) and obtain their formal transseries solutions.
As we have done in Sec.~\ref{sec:warm_up_solve_QC}, it is useful to use the DW-type by taking $E_0 \rightarrow 0_+$, where $E_0$ is a classical part of the energy.
The Stokes graph is shown in Fig.~\ref{fig:stokes_mass_PT_wb}.
For a simplified notation, we rescale variables as
\be
x \rightarrow \frac{\omega}{\sqrt{g}} x, \qquad  \widetilde{E} \rightarrow \omega \widetilde{E}, \qquad \hbar \rightarrow \frac{ \omega^3}{g} \hbar,
\label{eq:resc_mass_g}
\ee
where $E = E_0 +  \widetilde{E}\hbar>0$, and the Schr\"{o}dinger equation (\ref{eq:sch_eq_no_mass_g}) becomes dimensionless as
\be
&& \frac{g}{\omega^4}   {\cal L} \rightarrow {\cal L} =  - \hbar^2 \pd_x^2 + x^2 - x^4 - \widetilde{E} \hbar. \label{eq:sc_mass_g_resc}
\ee
By this procedure, the $A$- and $B$-cycles are replaced with
\be
&& A \rightarrow {\frak A} = e^{- 2 \pi i F}, \qquad  B \rightarrow {\frak B} = \frac{\sqrt{2 \pi} e^{-G}}{\Gamma(1/2  - F)} \left(\frac{\hbar}{2} \right)^{F}, \label{eq:rep_mass}
\ee
where $F$ and $G$ are formal power series of $\hbar$ and $\widetilde{E}$.
One can easily obtain $F$ by Eq.(\ref{eq:F_def}), but computation of $G$ is more technical.
We summarize the derivation of $G$ in Appendix~\ref{sec:der_G}.
Notice that the overall sign to $F$ in ${\frak B}$ is flipped to Eq.(\ref{eq:frakAB_warm}) because this sign is related to the intersection number $\langle {\frak A},{\frak B} \rangle$ and takes the opposite to the Hermitian case.
By this procedure, the QCs (\ref{eq:DAC})(\ref{eq:DPT}) lead to
\be
&& {\rm AC}: \  \log \frac{{\frak A}^{\pm 1}}{1 + {\frak B}} = \pi i q, \qquad
 {\cal PT}: \ \log {\frak A} = \pi i q, \qquad q \in 2 {\mathbb Z} + 1, \label{eq:QC_mass_g}
\ee
where $q$ is an energy level, and $\pm 1$ in ${\rm AC}$ corresponds to ${\frak D}_{{\rm AC}(3 \rightarrow 1)}$ and ${\frak D}_{{\rm AC}(2 \rightarrow 4)}$, respectively.
According to Eq.(\ref{eq:F_def}) and Appendix~\ref{sec:der_G}, $F$ and $G$ can be specifically written down as
\be
&& F = -\frac{\widetilde{E}}{2} -\frac{3 (1 + \widetilde{E}^2)}{16} 
\hbar -\frac{5 \widetilde{E} ( 17  + 7 \widetilde{E}^2)}{128} \hbar^2 
-\frac{105 (19 +50 \widetilde{E}^2 +11 \widetilde{E}^4)}{2048} \hbar^3 + O(\hbar^4), \nl
&& G = \frac{2}{3 \hbar} + \frac{\widetilde{E}^2}{8}\hbar + \frac{\widetilde{E}(55 +23 \widetilde{E}^2)
}{128} \hbar ^2  + \frac{441 +5388 \widetilde{E}^2 + 1091 \widetilde{E}^4}{3072} \hbar^3 + O(\hbar^4). \label{eq:F_G}
\ee
Perturbative parts of the ${\cal PT}$ and the AC energy solutions are obtained from the same condition, $\log {\frak A} = \pi i q$.
Substituting the ansatz, $\widetilde{E}^{(0)} = \sum_{n \in {\mathbb N}_0} e^{(0)}_n \hbar^n$, into $F$ in the condition leads to
\be
\widetilde{E}^{(0)}
&=& q - \frac{3 \left(q^2+1\right)}{8}  \hbar-\frac{q \left(17 q^2+67\right)}{64} \hbar^2 - \frac{3 \left(125 q^4+1138 q^2+513\right)}{1024} \hbar^3 + O(\hbar^4), \label{eq:E0_mass} \nl
\ee
with $q \in 2{\mathbb N}_0+1$.
Since ${\frak D}_{\cal PT}$ does not contain non-perturbative contributions, Eq.(\ref{eq:E0_mass}) is the transseries solution of the ${\cal PT}$ energy.
Hence, the perturbative part of the AC energy equals to the ${\cal PT}$ energy:
\be
\widetilde{E}_{\cal PT} = \widetilde{E}^{(0)}_{{\rm AC}(3 \rightarrow 1)}= \widetilde{E}^{(0)}_{{\rm AC}(2 \rightarrow 4)} \in {\mathbb R}_{>0}.
\ee
In contrast, ${\frak D}_{\rm AC}$ has non-perturbative effects, and those can be computed by expanding the energy $\widetilde{E}$ around its perturbative part $\widetilde{E}^{(0)}$ in Eq.(\ref{eq:QC_mass_g}).
Precisely, by shifting the energy as $\widetilde{E} = \widetilde{E}^{(0)} + \delta \widetilde{E}$, the AC QCs in Eq.(\ref{eq:QC_mass_g}) can be written as
\be
\sum_{s \in {\mathbb N}} \left. \frac{\pd^s F}{\pd \widetilde{E}^s} \right|_{\widetilde{E}=\widetilde{E}^{(0)}} \frac{(\delta \widetilde{E})^s}{s!}
&=& \mp \frac{1}{2 \pi i} \sum_{s \in {\mathbb N}_0} \left. \frac{\pd^s \log( 1 +  {\frak B})}{\pd \widetilde{E}^s} \right|_{\widetilde{E}=\widetilde{E}^{(0)}} \frac{(\delta \widetilde{E})^s}{s!}, \label{eq:dF_B}
\ee
where $\delta \widetilde{E} = \sum_{\ell \in {\mathbb N}}  \widetilde{E}^{(\ell)}$, and $\widetilde{E}^{(\ell)} = e^{-\frac{2 \ell}{3 \hbar}} \times \left( \cdots \right)$ is a non-perturbative part originated by $\ell$ bions (instanton-antiinstanton pairs).
Solving Eq.(\ref{eq:dF_B}) yields transseries solutions of $\widetilde{E}^{(\ell)}_{{\rm AC}(3 \rightarrow 1)}$, and the first two non-perturbative parts $(\ell = 1,2)$ can be written down as (cf. \cite{Bender:1973rz})
\be
\widetilde{E}_{{\rm AC}(3 \rightarrow 1)}^{(1)} &=& -i \sigma \left[
  1 -\frac{q (q+6)}{8} \hbar +  \frac{q^4+q^3-102 q^2-43 q-134}{128} \hbar^2 \right. \nl
  &&  \left. -\frac{q \left(q^5-15 q^4-184 q^3+4371 q^2+2400 q+20484\right)}{3072} \hbar^3 + O(\hbar^4) \right], \label{eq:E1_mass} \\ \nl
\widetilde{E}_{{\rm AC}(3 \rightarrow 1)}^{(2)} &=& \sigma^2 \left[
  \frac{\zeta_+}{2}  + \frac{ 2q + 3}{8} \hbar - \frac{q(q + 3)}{8} \zeta_+ \hbar  \right. \nl
  &&   \left. - \frac{8 q^3 + 3 q^2 - 102 q - 43}{128} \hbar^2 + \frac{2 q^4+q^3-51 q^2-43 q-67}{128} \zeta_+ \hbar^2 \right. \nl
  &&  \left. +   \frac{12 q^5-75 q^4-368 q^3+2988 q^2+2400 q+5121}{1536} \hbar^3 \right. \nl 
  &&  \left. - \frac{q \left(2 q^5-15 q^4-92 q^3+996 q^2+1200 q+5121\right)}{1536}  \zeta_+ \hbar^3  + O(\hbar^4) \right], \label{eq:E2_mass} 
\ee
where $\sigma$ and $\zeta_\pm$ are defined as
\be
\sigma:= \sqrt{\frac{2}{\pi}} \frac{e^{-\frac{2}{3\hbar}}}{\Gamma(\frac{q+1}{2})} \left(\frac{\hbar}{2}\right)^{-\frac{q}{2}}, \qquad \zeta_\pm :=  \psi ^{(0)}\left(\frac{q+1}{2}\right)+\log \left(\frac{\hbar }{2}\right) \pm \pi i. \label{eq:sig_zeta}
\ee
One can directly obtain $\widetilde{E}_{{\rm AC}(2 \rightarrow 4)}$ by complex conjugate of $ \widetilde{E}_{{\rm AC}(3 \rightarrow 1)}$ as ${\cal C}[\widetilde{E}_{{\rm AC}(3 \rightarrow 1)}] = \widetilde{E}_{{\rm AC}(2 \rightarrow 4)}$, i.e. $i \rightarrow -i$ and $\zeta_+ \rightarrow \zeta_-$ in Eqs.(\ref{eq:E1_mass})(\ref{eq:E2_mass}).

The important fact is that the non-perturbative parts of the AC energy, $\widetilde{E}^{(\ell \in {\mathbb N})}_{\rm AC}$, are complex values, and the signs of their imaginary parts correspond to the path of analytic continuation.
In particular, $\widetilde{E}^{(1)}_{\rm AC}$ is pure imaginary, but $\widetilde{E}^{(\ell>1)}_{\rm AC}$ contains both real and imaginary parts.
This point is extremely crucial for the ABS conjecture;
when exceeding the semi-classical level of the first non-perturbative order, the ABS conjecture (\ref{eq:conj_Ai}) is violated.

\subsection{Reformulation of the ABS conjecture} \label{sec:reform_ABS}
We reformulate the ABS conjecture by Borel resummation theory using the above results.
Although $\widetilde{E}_{\rm AC}$ and $\widetilde{E}_{\cal PT}$ are not naively related to each other, one can formulate a direct relation between $\widetilde{E}_{\rm AC}$ and $\widetilde{E}_{\cal PT}$ using the DDP formula (\ref{eq:DDP_mass_g}) thanks to the quadratic term.

Let us see this fact.
From the DDP formula (\ref{eq:DDP_mass_g}) and the QCs (\ref{eq:DAC})(\ref{eq:DPT}), one finds that
\be
   {\frak D}_{\cal PT} = {\frak S}^{-1/2}_0[{\frak D}_{{\rm AC}(3 \rightarrow 1)}] 
   ={\frak S}^{+1/2}_0[{\frak D}_{{\rm AC}(2 \rightarrow 4)}]. \label{eq:DPT_G_DAC}
\ee
Since the QCs are directly related to the energy solutions, Eq.(\ref{eq:DPT_G_DAC}) implies existence of Stokes automorphism for the energy solutions such that\footnote{
  Remind that $E = \widetilde{E} \hbar$, and this redefinition does not change the transformation law of Stokes automorphism.
}
\be
E_{\cal PT} = {\frak S}_0^{-1/2}[E_{{\rm AC}(3 \rightarrow 1)}] 
= {\frak S}^{+1/2}_0[E_{{\rm AC}(2 \rightarrow 4)}]. \label{eq:EPT_DDP}
\ee
Notice that $E_{{\rm AC}(3 \rightarrow 1)}$ and $E_{{\rm AC}(2 \rightarrow 4)}$ also have to be related to each other as
\be
&&  E_{{\rm AC}(3 \rightarrow 1)} = {\frak S}^{+1}_0[E_{{\rm AC}(2 \rightarrow 4)}]. \label{eq:EAC_DDP}
\ee
Thus, our task is specifically formulating the Stokes automorphism for the energy solutions.
It  is actually performable by computation of alien calculus, and we summarize the derivation in Appendix~\ref{sec:DDP_energy}.
By the results in Eqs.(\ref{eq:DEPT_res1})(\ref{eq:DEPT_res2}), one can reproduce the non-perturbative parts of $\widetilde{E}_{\rm AC}$ in Eqs.(\ref{eq:E1_mass})(\ref{eq:E2_mass}) from its perturbative part in Eq.(\ref{eq:E0_mass}), i.e. $\widetilde{E}_{\cal PT}$, by\footnote{
  The subscript, ``0'', in the alien derivative denotes $\arg(g)=0$, not locations of the singular points.
  The alien derivative contains non-perturbative effects from all the singular points along the real axis on the Borel plane.
  See Appendix~\ref{sec:DDP_energy}.
}
\be
{\frak S}_0^{\nu} [\widetilde{E}_{\cal PT}] &=& \left( 1 +\nu \bul{\Delta}_0  + \frac{( \nu \bul{\Delta}_0)^2}{2} + O(\nu^3)\right) [\widetilde{E}_{\cal PT}] \nl
&=&
\begin{cases}
  \widetilde{E}_{{\rm AC}(3 \rightarrow 1)} & \quad \mbox{for} \quad \nu= + \frac{1}{2} \\
  \widetilde{E}_{{\rm AC}(2 \rightarrow 4)} & \quad \mbox{for} \quad \nu= - \frac{1}{2}
\end{cases}. \label{eq:SEPT}
\ee
This result insists that Stokes automorphism for the energy solutions is actually constructable by alien calculus without any contradictions to Eqs.(\ref{eq:E0_mass})(\ref{eq:E1_mass})(\ref{eq:E2_mass}).
Notice that showing Eq.(\ref{eq:EAC_DDP}) is trivial by using the property of Stokes automorphism in Eq.(\ref{eq:Stokes_prop_add}).
We should remind that analytic continuation of the Hermitian QC with $\arg(\lambda) = \pm \pi $ and the AC QCs with $\arg(g) = 0$ are equivalent as ${\frak D}^{\arg(\lambda) = +\pi}_{\cal H}\propto {\frak D}_{{\rm AC} (2 \rightarrow 4)}^{\arg(g) = 0}$ and ${\frak D}^{\arg(\lambda) = -\pi}_{\cal H} \propto {\frak D}_{{\rm AC} (3 \rightarrow 1)}^{\arg(g) = 0}$.
Hence, the energy solutions should hold the same equivalence, i.e. $E^{\arg(\lambda) = +\pi}_{\cal H} = E_{{\rm AC} (2 \rightarrow 4)}^{\arg(g) = 0}$ and $E^{\arg(\lambda) = -\pi}_{\cal H} = E_{{\rm AC} (3 \rightarrow 1)}^{\arg(g) = 0}$.
One can also provide the same results for the Borel resummed forms using Eq.(\ref{eq:gen_Stokes}).
From Eqs.(\ref{eq:EPT_DDP})(\ref{eq:EAC_DDP}), those are given by
\be
&& \widehat{E}_{\cal PT} = {\cal S}_{0_-} [E_{{\rm AC}(3 \rightarrow 1)}] = \widehat{{\frak S}}_{0}^{-1/2} [\widehat{E}_{{\rm AC}(3 \rightarrow 1)}] \nl
&& \qquad  \,\,  = {\cal S}_{0_+}[E_{{\rm AC}(2 \rightarrow 4)}] = \widehat{{\frak S}}_{0}^{+1/2} [\widehat{E}_{{\rm AC}(2 \rightarrow 4)}], \label{eq:ABS_hatEPT} \\ \nl
&& \widehat{E}_{{\rm AC}(3 \rightarrow 1)} = {\cal S}_{0_+}[E_{\cal PT}] = \widehat{\frak S}^{+1/2}_{0}[\widehat{E}_{\cal PT}], \nl
&& \qquad \qquad \, \,  = {\cal S}_{0_+} \circ {\frak S}^{+1/2}_{0} [E_{{\rm AC}(2 \rightarrow 4)}] = \widehat{\frak S}^{+1}_{0}[\widehat{E}_{{\rm AC}(2 \rightarrow 4)}], \\ \nl
&& \widehat{E}_{{\rm AC}(2 \rightarrow 4)} = {\cal S}_{0_-}[E_{\cal PT}] = \widehat{\frak S}^{-1/2}_{0}[\widehat{E}_{\cal PT}] \nl
&& \qquad \qquad \, \,  = {\cal S}_{0_-} \circ {\frak S}^{-1/2}_{0} [E_{{\rm AC}(3 \rightarrow 1)}] = \widehat{\frak S}^{-1}_{0}[\widehat{E}_{{\rm AC}(3 \rightarrow 1)}], 
\ee
where $\widehat{E} = {\cal S}_{{\rm med},0}[E]$.
Therefore, the modified ABS conjecture for  the energy solutions are reformulated by Stokes automorphism and Borel resummation, as Eqs.(\ref{eq:EPT_DDP})(\ref{eq:ABS_hatEPT}).

One can directly map the results of the energies to  Euclidean partition functions by using the definition,
\be
Z = \sum_{q \in 2 {\mathbb N}_0 +1} \exp \left[ - \beta E_{q} \right],
\ee
where $q$ denotes the energy level.
Formal transseries of the ${\cal PT}$ and the AC partition functions are related as
\be
&& Z_{\cal PT} =  {\frak S}^{-1/2}_0[Z_{{\rm AC}(3 \rightarrow 1)}] =  {\frak S}^{+1/2}_0[Z_{{\rm AC}(2 \rightarrow 4)}], \label{eq:ZPT_ZAC}
\ee
where
\be
   {\frak S}_{0}^{\nu} [Z] = \sum_{q \in 2 {\mathbb N}_0 +1} \exp \left[ - \beta {\frak S}_{0}^{\nu}[E_{q}] \right].
\ee
The Borel resummed forms are expressed by
\be
&& \widehat{Z}_{\cal PT} = {\cal S}_{0_-} [Z_{{\rm AC}(3 \rightarrow 1)}] = \widehat{{\frak S}}_{0}^{-1/2} [\widehat{Z}_{{\rm AC}(3 \rightarrow 1)}] \nl
&& \qquad  \,\,  = {\cal S}_{0_+}[Z_{{\rm AC}(2 \rightarrow 4)}] = \widehat{{\frak S}}_{0}^{+1/2} [\widehat{Z}_{{\rm AC}(2 \rightarrow 4)}]. \label{eq:ABS_hatZPT}
\ee

Finally, we briefly see the energy solutions from the viewpoint of the continuous deformation by Stokes automorphism.
The QCs in Eqs.(\ref{eq:DAC})(\ref{eq:DPT}) can be unifiedly expressed by a real parameter, $\nu$, as
\be
&& {\frak D}^{\nu \in {\mathbb R}} := 1 + A (1+B)^{-2\nu} \propto
\begin{cases}
  {\frak D}_{{\cal PT}} & \quad \mbox{if} \quad \nu = 0 \\
  {\frak D}_{{\rm AC}(3 \rightarrow 1)} & \quad \mbox{if} \quad \nu = + \frac{1}{2} \\
  {\frak D}_{{\rm AC}(2 \rightarrow 4)} & \quad \mbox{if} \quad \nu =  - \frac{1}{2}
\end{cases}, \\
&& {\frak S}^{\nu \in {\mathbb R}}_{0}[{\frak D}^{\nu_0}] = {\frak D}^{\nu_0 +\nu}.
\ee
Denoting $\widetilde{E}(\nu_0)$ as the energy solution of ${\frak D}^{\nu_0}$, one can calculate the $\nu$-evolution from $\widetilde{E}(\nu_0)$ by modifying Eq.(\ref{eq:dF_B}) as
\be
&& \left. F \right|_{\widetilde{E} = \widetilde{E}(\nu_0)} = - \frac{q}{2} - \frac{\nu_0}{\pi i } \left. \log (1 + {\frak B}) \right|_{\widetilde{E} = \widetilde{E}(\nu_0)}, \\
&& \sum_{s \in {\mathbb N}} \left. \frac{\pd^s}{\pd \widetilde{E}^s} \left( F + \frac{\nu_0 }{\pi i} \log (1+{\frak B}) \right) \right|_{\widetilde{E} = \widetilde{E}(\nu_0)} \frac{(\delta \widetilde{E})^s}{s !} \nl
&& \qquad =   - \frac{\nu}{\pi i} \sum_{s\in {\mathbb N}_0} \left. \frac{\pd^s \log (1 + {\frak B})}{\pd \widetilde{E}^s} \right|_{\widetilde{E} = \widetilde{E}(\nu_0)} \frac{(\delta \widetilde{E})^{s}}{s!}. \label{eq:alien_Enu0}
\ee
One can express $\delta \widetilde{E}$ by the alien derivatives as
\be
\delta \widetilde{E} = \sum_{n \in {\mathbb N}} \frac{\nu^n}{n !} (\bul{\Delta}_0)^n[\widetilde{E}(\nu_0)], \label{eq:delE_alien}
\ee
and substituting Eq.(\ref{eq:delE_alien}) into Eq.(\ref{eq:alien_Enu0}) determines  $(\bul{\Delta}_0)^n[\widetilde{E}(\nu_0)]$ recursively.
Taking $\nu_0 = 0$ and $\nu = \pm 1/2$ reproduces Eq.(\ref{eq:SEPT}).
Therefore, the ${\cal PT}$ and the AC energy solutions are parameterized by $\nu$ and continuously connected by the one-parameter Stokes automorphism ${\frak S}^{\nu}_0$, as  ${\frak S}^{\nu}_0[E(\nu_0)] = E(\nu_0 + \nu)$.
In addition, these Borel resummed forms obey the following initial value problem:
\be
\frac{d \widehat{E}(\nu)}{d \nu} &=& \frac{d \widehat{\frak S}_{0}^{\nu} [ \widehat{E}(0)]}{d \nu}  
=  {\cal S}_{\rm med,0}   \circ \bul{\Delta}_0[E(\nu)] \nl
&=& - \frac{\hbar}{\pi i}   \left. \frac{\log(1+\widehat{\frak B})}{\frac{\pd }{\pd \widetilde{E}} \left( \widehat{F} + \frac{\nu}{\pi i} \log (1 + \widehat{\frak B}) \right)} \right|_{\widetilde{E} = \widehat{E}(\nu)/\hbar}, \nl
 \widehat{E}(\nu = 0) &=&  \widehat{E}_{\cal PT},
\ee
where $\widehat{\frak S}^{\nu}_0[\widehat{E}(\nu_0)] = \widehat{E}(\nu_0 + \nu)$.

We have now completed to make all the relations in Fig.~\ref{fig:summary_PT}.

\section{Pure quartic potential}  \label{sec:without_mass_term}
In this section, we consider the Schr\"{o}dinger equation given by the pure quartic potential, i.e. $\omega = 0$.
In this case, since no quadratic local minimum exists in the potential, one might imagine that EWKB expanding $\hbar$ is meaningless.
Indeed, the scaling law in the Hamiltonian gives a monomial with a rational exponent of $\hbar$ such that
\be
E = c(k) (\lambda \hbar^4)^{1/3}, \qquad c(k) \in {\mathbb R}_{>0}, \label{eq:Echbar4/3}
\ee
with an overall function, $c(k)$, depending on the energy level, $k$.
However, although the $\hbar$-expansion is meaningless for the potential, there is another parameter for a transseries solution of the energy, that is the (inverse) energy level, $k$~\cite{BPV,Bucciotti:2023trp}.
Hence, our main strategy for the pure quartic potential is to derive a formal transseries of $c(k)$ in Eq.(\ref{eq:Echbar4/3}).

We firstly consider EWKB of the Hermitian potential, $V(x) = \lambda x^4$, in Sec.~\ref{sec:Herm_sym_wo_mass}, and then analyze the negative coupling potential, $V(x) = -g x^4$, in Sec.~\ref{sec:PT_sym_wo_mass}.
From these results, we finally discuss (im)possibility of an alternative form of the ABS conjecture in Sec.~\ref{sec:impossibility_PT_AC}.

\subsection{Hermitian potential: $V = \lambda x^4$} \label{sec:Herm_sym_wo_mass}
The Schr\"{o}dinger equation with the Hermitian potential, $V(x) = \lambda x^{4}$, is defined as
\be
   {\cal L} = - \hbar^{2} \pd_x^2  + \lambda x^{4} - E, \qquad {\cal L} \psi = 0,
   \label{eq:sc_wo_mass_warm}
\ee
where $\lambda, E \in {\mathbb R}_{>0}$.
When naively applying EWKB to Eq.(\ref{eq:sc_wo_mass_warm}), a remarkable feature of the pure quartic potential can be readily seen from $\int dx \, S_{\rm od}$, which takes the form that
\be
&& \int dx \, S_{\rm od}(x,\hbar) = \Phi_{-1}(x) \eta^{-1} + \Phi_{+1}(x) \eta^{+1} + \cdots, \qquad \eta := \frac{\lambda^{1/4} \hbar}{E^{3/4}}, \label{eq:Sod_exp_eta}
\ee
where the coefficients, $\Phi_{n \in 2 {\mathbb N}_0-1}(x)\in {\mathbb C}$, are independent on  $\lambda$ and $E$ for all $n$.
This suggests that taking $\eta$ should be a more reasonable choice of the expansion parameter.
The Schr\"{o}dinger equation of $\eta$ is easily obtained by scaling dimensions of the parameters in Eq.(\ref{eq:sc_wo_mass_warm}), that are
\be
[x]  = \frac{1}{4}, \qquad    [\hbar] = \frac{3}{4}, \qquad [\lambda]  = 0, \qquad [E] = 1. \label{eq:scale_dim_wo_mass_warm}
\ee
Rescaling  $x \rightarrow \left( \frac{E}{\lambda} \right)^{1/4} x$ and the relation of $\eta$ with $\hbar$ in Eq.(\ref{eq:Sod_exp_eta}) lead to the dimensionless operator ${\cal L}$ as
\be
\frac{1}{E}   {\cal L} \rightarrow {\cal L} =  - \eta^2  \pd_x^2 + Q, \qquad Q = x^4 - 1. \label{eq:L_scaled}
\ee

Let us perform EWKB for the dimensionless equation (\ref{eq:L_scaled}).
The Stokes graph has the same topology as Fig.~\ref{fig:stokes_warm} consisting of turning points with different values given by
\be
   {\rm TP} = \left\{ a_1 = - 1, a_2 = -i,  a_3 = +i, a_4 = + 1 \right\}.
\ee   
Similarly, the Stokes phenomenon happens at $\arg(\lambda) = 0$.
Thanks to the same topology of the Stokes graph, the monodromy matrix, the QC, and the DDP formula have the same cycle representations to Eqs.(\ref{eq:M0p_warm})-(\ref{eq:A0pm_D0pm}).
The difference is only specific forms of $A_{j=1,2}$, and those are given by
\be
A_1 = e^{\phi (e^{- \pi i /4}\eta)}, \qquad  A_2 = e^{-\phi (e^{+\pi i /4}\eta)},
\ee
where $\phi=\phi(\eta)$ is a formal power expansion of $\eta$ as
\be
&& \phi(\eta) := \sum_{n \in {\mathbb N}_0} v_{2n-1} \eta^{2n-1}, \qquad v_{2n-1} \in {\mathbb R} \ \ \mbox{for all} \ \ n \in {\mathbb N}_0. \label{eq:phi}
\ee
Those coefficients, $v_{2n-1}$, are evaluated as
\be
&& v_{-1} = \frac{4 \sqrt{2}K(-1)}{3}, \quad v_{1} = -\frac{4 \sqrt{2\pi} \Gamma(7/4)}{9 \Gamma(-3/4)}, \quad v_{3} = -\frac{11 \sqrt{2\pi} \Gamma(5/4)}{384 \Gamma(3/4)}, \quad \cdots
\ee
where $K(x)$ denotes the complete elliptic integral of the first kind.
For solving the QC, it is technically useful to decompose the exponents of $A_{j=1,2}$ into the real and the imaginary parts.
By defining 
\be
\phi(e^{-\pi i/4} \eta) = \phi_{\rm R}(\eta) + i \phi_{\rm I}(\eta), \qquad \phi_{\rm R,I}(\eta) \in {\mathbb R},
\ee
the cycles are expressed by $\log A_1 = \phi_{\rm R} + i \phi_{\rm I}$ and $\log A_2 = -\phi_{\rm R} + i \phi_{\rm I}$, and thus one can write down the QC, which has the same cycle representation to Eq.(\ref{eq:qcond_lam4_mass}), as
\be
{\frak D}_{\cal H}^{\arg(\eta) = 0} \propto \cos \phi_{\rm I}  + \frac{e^{-\phi_{\rm R}}}{\sqrt{1 + e^{-2\phi_{\rm R}}}}. \label{eq:Qc_x4}
\ee

As we described above, $E$ appears only in $\eta$ as  coupling to $\hbar$. 
This fact implies that any ansatz expanded by $\hbar$ does not work for the energy solution.
The coupling constant, $\lambda$, is also the case.
Although any extra parameters explicitly do not exist in the potential, solving QCs generally gains another parameter to distinguish the energy state, that is an energy level.
For this reason, we use the (inverse) energy level for our ansatz.
Some properties of formal transseries with the $\kappa^{-1}$-expansion for $V = \lambda x^4$ are discussed in Ref.~\cite{Bucciotti:2023trp}.
Here, we prepare the following ansatz for $\eta$\footnote{
  The transmonomial, $\sigma$, is available after determining the leading order, $\phi_{\rm R} = \phi_{\rm I} =  \kappa$.
  By taking $\phi_{\rm R,I} = \kappa + \delta \phi_{\rm R,I}$, and the linearized Eq.(\ref{eq:Qc_x4}) in terms of  $\delta \phi_{\rm R,I}$ leads to
\be
-(-1)^{k} \delta \phi_{\rm I} + e^{- \kappa} + O(\delta \phi^2, e^{-\kappa} \delta \phi) = 0 \quad \Rightarrow \quad \delta \phi_{\rm I} = (-1)^{k}  e^{-\kappa}.
\ee
}: 
\be
&& \eta^{-1} = \frac{E^{3/4}}{\lambda^{1/4} \hbar} \sim \sum_{n \in {\mathbb N}_0} e^{(0)}_{2n-1}\kappa^{{1-2n}} + \sum_{\ell \in {\mathbb N}} \sum_{n \in {\mathbb N}_0} e^{(\ell)}_{n} \sigma^{\ell}  \kappa^{-n} \quad \mbox{as} \quad \kappa \rightarrow +\infty, \label{eq:ansatz_k} \\
&& \kappa = \kappa(k) = \pi  \left( k + \frac{1}{2} \right),  \qquad \sigma:= e^{- \kappa}, \qquad k \in {\mathbb N}_0, \qquad e^{(\ell)}_{n} \in {\mathbb R} \label{eq:kappa}.
\ee
Notice that we use the $\kappa^{-1}$-expansion for technical simplicity, but the $k^{-1}$-expansion is always available from it.
The coefficients in the perturbative part, $e^{(0)}_{2n-1}$, can be determined by the first term in Eq.(\ref{eq:Qc_x4}), i.e. $\cos \phi_{\rm I}(\eta) = 0$, and are obtained as
\be
e^{(0)}_{-1} &=& \frac{3}{4 K(-1)}, \qquad \qquad \ \  e^{(0)}_{1} = -\frac{4 \sqrt{\pi } \Gamma \left(7/4\right)}{9 \Gamma \left(-3/4\right)}, \nl
e^{(0)}_{3} &=& -\frac{11 \Gamma \left(5/4\right) [\Gamma \left(1/4\right)]^4}{6912 \sqrt{\pi } \Gamma \left(3/4\right)}-\frac{8 \sqrt{2 \pi } [\Gamma \left(11/4\right)]^2}{1323},\quad \cdots.
\ee
The second term in Eq.(\ref{eq:Qc_x4}) generates the non-perturbative parts, and the coefficients for the first two bion contributions, $e^{(\ell =1,2)}_{n}$, can be written down as
\be
&& e^{(1)}_0 = (-1)^k \frac{3 }{4 K(-1)}, \qquad \qquad \qquad
e^{(1)}_1 = (-1)^k \frac{8 \sqrt{\pi}  \Gamma \left(7/4\right)}{9 \Gamma \left(-3/4\right)}, \nl
&& e^{(1)}_2 = -(-1)^k \frac{2 \sqrt{2} (6-\pi) \pi ^{3/2} }{27 [\Gamma \left(-3/4\right)]^2}, \qquad
 e^{(1)}_3 =   (-1)^k\frac{\sqrt{2} (36-\pi) \pi ^{5/2} }{243 [\Gamma \left(-3/4 \right)]^2}, \quad \cdots, \\ \nl
 && e^{(2)}_0 = -\frac{3}{4 K(-1)}, \qquad \qquad \qquad \qquad e^{(2)}_1  = -\frac{16 \sqrt{\pi } \Gamma \left(7/4 \right)}{9 \Gamma \left(-3/4\right)}, \nl
 && e^{(2)}_2 =  \frac{4 \sqrt{2} (9-2 \pi ) \pi ^{3/2}}{27 [\Gamma \left(-3/4\right)]^2}, \qquad \qquad \quad e^{(2)}_3 = - \frac{4 \sqrt{2} (2 \pi (18 - \pi)- 27) \pi ^{3/2}}{243 [\Gamma \left(-3/4\right)]^2}, \quad \cdots, 
\ee
Taking $E_{\cal H}  = \eta^{-4/3} (\lambda \hbar^4)^{1/3}$ yields the energy solution, and $E^{(\ell=0,1,2)}_{\cal H}$ are obtained as (cf. \cite{BPV})
\be
\frac{E^{(0)}_{\cal H}}{(\lambda \hbar^{4})^{1/3}} &=& \kappa^{4/3} \left[ \frac{(3/4)^{4/3}}{[K(-1)]^{4/3}} -\frac{2^{10/3} \sqrt{\pi}   \Gamma \left(7/4\right)}{3^{8/3} [K(-1)]^{1/3} \Gamma \left(-3/4\right)} \kappa^{-2} + O(\kappa^{-4})\right], \label{eq:E0_lam_wo_mass} \\
\frac{E^{(1)}_{\cal H}}{(\lambda \hbar^{4})^{1/3}} &=& (-1)^k \sigma \kappa^{1/3}
\left[ \frac{(3/4)^{1/3} }{[K(-1)]^{4/3}} + \frac{2^{13/3} \sqrt{\pi }  \Gamma \left(7/4\right)}{3^{8/3} [K(-1)]^{1/3} \Gamma \left(-3/4\right)} \kappa^{-1} \right. \nl
&& \left. - \frac{4 \sqrt{\pi} \left(2^{5/6} \pi (6 - \pi) + 2^{4/3} \Gamma \left(7/4\right) \Gamma \left(-3/4\right)\right)}{3^{11/3} [K(-1)]^{1/3} [\Gamma \left(-3/4\right)]^2} \kappa^{-2} + O(\kappa^{-3}) \right], \label{eq:E1_lam_wo_mass} \\
\frac{E^{(2)}_{\cal H}}{(\lambda \hbar^{4})^{1/3}} &=& \sigma^2 \kappa^{1/3} \left[ -\frac{(3/4)^{1/3}}{[K(-1)]^{4/3}}  - \frac{128 \sqrt{\pi } [K(-1)] \Gamma \left(7/4\right) - 9 \Gamma \left(-3/4\right)}{3 \cdot 6^{5/3} [K(-1)]^{4/3} \Gamma \left(-3/4\right)} \kappa^{-1} \right. \nl
  && \left.  + \frac{2^{10/3} \sqrt{\pi} \left( \sqrt{2} \pi ( 9 - 2 \pi  ) + 3 \Gamma \left(7/4\right) \Gamma \left(-3/4\right)\right)}{3^{11/3} [K(-1)]^{1/3} [\Gamma \left(-3/4\right)]^2} \kappa^{-2} + O(\kappa^{-3})  \right]. \label{eq:E2_lam_wo_mass} 
\ee

In the above, we naively applied the ansatz (\ref{eq:ansatz_k}) for solving the QC, but need to clarify what we actually did from the aspect of EWKB of the $\kappa^{-1}$-expansion.
The point is  that the dimensionless operator (\ref{eq:sc_wo_mass_warm}) can be interpreted to be given by another potential, $V(x)=x^4-1$, and a \textit{fixed} energy already determined as zero\footnote{
  This ``zero-energy'' is not directly relevant to $E$ in Eq.(\ref{eq:sc_wo_mass_warm}), and it is just an analogy with a usual Schr\"{o}dinger equation.
}.
Here, let us replace $\eta$ with the formal expansion, $\eta^{-1} \sim \sum_{n \in {\mathbb N}_0} e^{(0)}_{2n-1} \kappa^{1-2 n}$ as $\kappa \rightarrow \infty$, in Eq.(\ref{eq:L_scaled})\footnote{
We omitted non-perturbative sectors, $\sigma^\ell = e^{-\ell \kappa}$, for simplicity, but the similar discussion works for the case including them.
}.
Then, we define a modified potential $\widetilde{Q}(\kappa) \sim \sum_{n \in {\mathbb N_0}} \widetilde{Q}_{n} \kappa^{-2n}$ as
\be
&& \eta^{-2} {\cal L} = - \pd_{x}^2  + \eta^{-2} Q, \nl
&& \eta^{-2} Q \sim \kappa^{2} \left[ \widetilde{Q}_0 + \widetilde{Q}_1 \kappa^{-2} + \widetilde{Q}_2 \kappa^{-4}  + O(\kappa^{-6})  \right], \label{eq:mod_pot_eta}
\ee
where
\be
\widetilde{Q}_{0} = (e^{(0)}_{-1})^2 Q, \qquad \widetilde{Q}_{1} = 2 e^{(0)}_{-1} e^{(0)}_{1}  Q, \qquad \widetilde{Q}_{2} = ( 2 e^{(0)}_{-1} e^{(0)}_{3} + ( e^{(0)}_{1} )^2 ) Q .
\ee
By analogy with a standard $\hbar$-expansion, this potential $\widetilde{Q}(\kappa)$ can be regarded as a \textit{quantum deformed potential} with respect to the $\kappa^{-1}$-expansion.
Assuming $e^{(0)}_{-1} \ne 0$, one can in principle draw a Stokes graph and construct a QC by using the Airy-type connection formula without details of the coefficients, $c_n$, because the Stokes graph depends only on $\arg(\kappa)$ and $\widetilde{Q}_0$.
After that, the coefficients are determined by solving the resulting QC.
In other words, EWKB of the $\kappa^{-1}$-expansion can be interpreted as an inverse problem; we determined the specific potential form of $\widetilde{Q}(\kappa)$ by solving the QC constrained only by $\widetilde{Q}_0$ and topology of its Stokes graph.

Unlike the case that $\omega>0$ discussed in Sec.~\ref{sec:with_mass_term}, one can directly obtain the AC energy solution from Eqs.(\ref{eq:E0_lam_wo_mass})-(\ref{eq:E2_lam_wo_mass}) by taking $\lambda = e^{\pm \pi i} g$ and fixing $\arg(\eta) = 0$ because the energy is a monomial with respect to $(\lambda \hbar^4)^{1/3}$, as is shown in Eq.(\ref{eq:Echbar4/3}).
By this procedure, one finds
\be
\frac{E_{\rm AC}(g)}{(g \hbar^4)^{1/3}} = e^{\pm \frac{\pi}{3} i} \frac{E_{\cal H}(\lambda)}{(\lambda \hbar^4)^{1/3}} \label{eq:sol_DAC_wo_mass_g},
\ee
where $E_{\cal H}$ is given by Eqs.(\ref{eq:E0_lam_wo_mass})-(\ref{eq:E2_lam_wo_mass}).
Notice that $E_{\rm AC}$ contains non-perturbative parts with respect to $\kappa^{-1}$.
We would further discuss the resulting AC energy in Sec.~\ref{sec:PT_sym_wo_mass} by directly beginning with the negative coupling potential.

\subsection{Negative coupling potential: $V = -g x^4$} \label{sec:PT_sym_wo_mass}

\begin{figure}[tbp]
  \begin{center}
    \begin{tabular}{cc}
      \begin{minipage}{0.5\hsize}
        \begin{center}
          \includegraphics[clip, width=75mm]{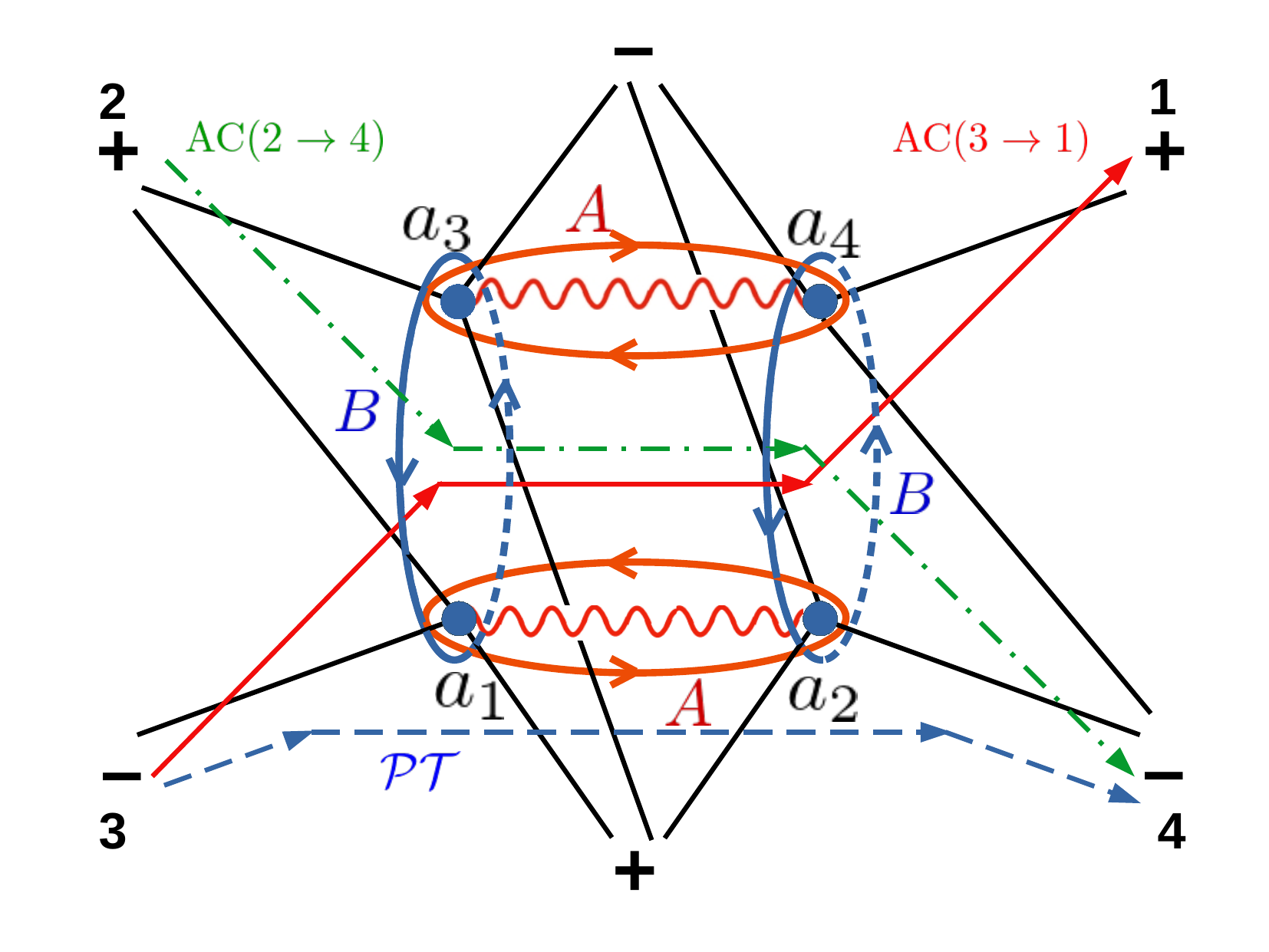}
          \hspace{1.6cm} (a) $\arg(g) = 0_+$
        \end{center}
      \end{minipage}
      \begin{minipage}{0.5\hsize}
        \begin{center}
          \includegraphics[clip, width=75mm]{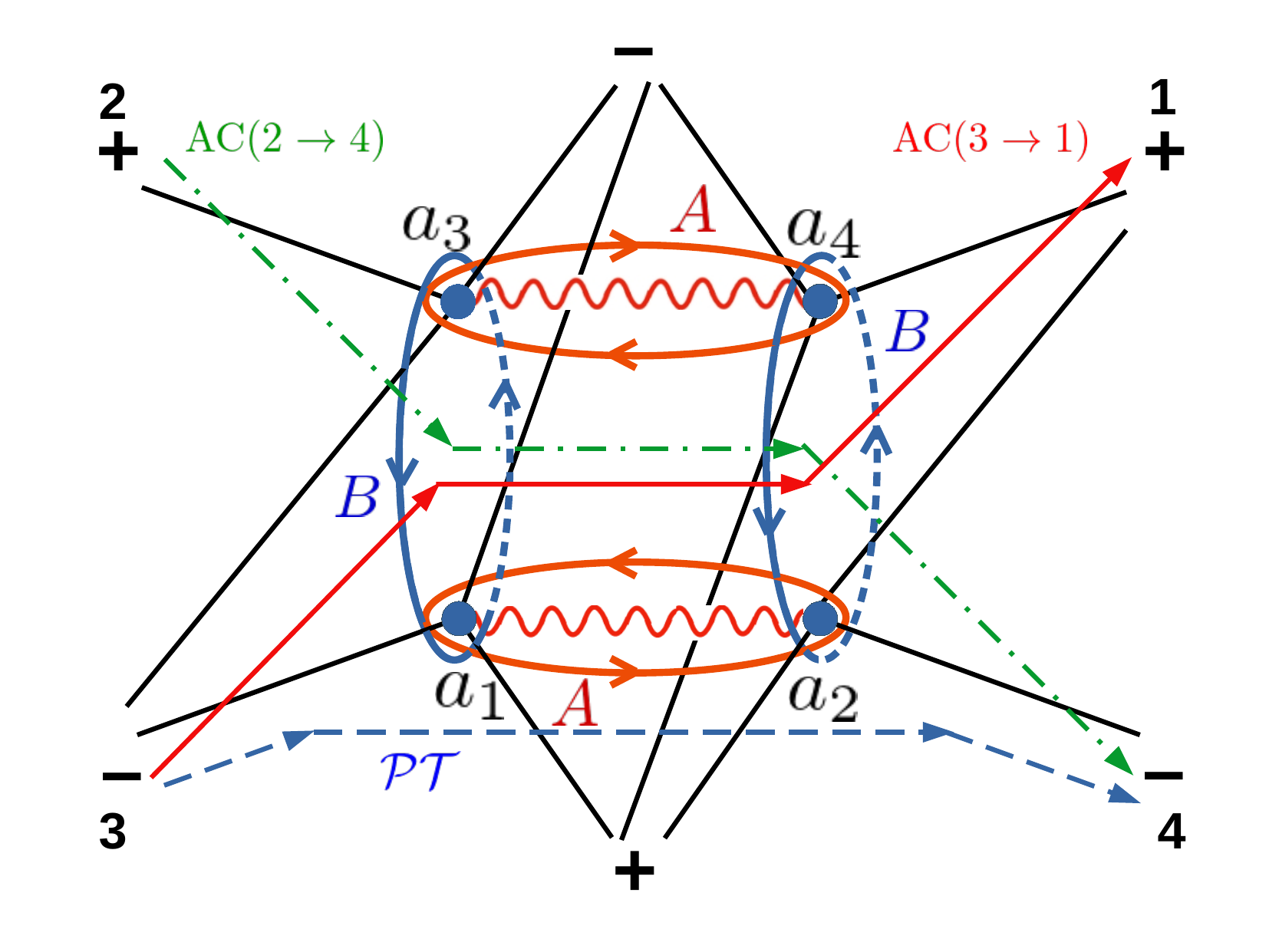}
          \hspace{1.6cm} (b) $\arg(g) = 0_-$
        \end{center}
      \end{minipage}      
    \end{tabular} 
    \caption{Stokes graph of the negative coupling potential without a quadratic term.
      The paths for analytic continuation are denoted by colored lines, $\gamma_{3 \rightarrow 1}$ (red), $\gamma_{2 \rightarrow 4}$ (green), and $\gamma_{3 \rightarrow 4}$ (blue).
    }
    \label{fig:stokes_nomass_PT}
  \end{center}
\end{figure}

We consider the negative coupling potential with $\omega = 0$ and derive transseries solutions of the ${\cal PT}$ and the ${\rm AC}$ energies.
Although the transseries solutions of the ${\rm AC}$ energies have been found in Eq.(\ref{eq:sol_DAC_wo_mass_g}), it would be worth to reconsider it from the negative coupling potential as a consistency check.
Furthermore, the DDP formula given in the negative coupling potential would be used in discussions in Sec.~\ref{sec:impossibility_PT_AC}.

The Schr\"{o}dinger equation is given by 
\be
   {\cal L} = - \hbar^{2} \pd_x^2 -g x^{4} - E, \qquad {\cal L} \psi = 0, \label{eq:sc_wo_mass_g}
\ee
where $g,E \in {\mathbb R}_{>0}$.   
By using the scaling law (\ref{eq:scale_dim_wo_mass_warm}) and rescaling $x \rightarrow \left( \frac{E}{g} \right)^{1/4} x$, one  obtains the dimensionless equation as
\be
\frac{1}{E}   {\cal L} \rightarrow {\cal L} =  - \widetilde{\eta}^2  \pd_x^2 + Q, \qquad Q = -x^4 - 1, \qquad \widetilde{\eta} := \frac{g^{1/4} \hbar}{E^{3/4}}.
\label{eq:L_scaled_wo_mass}
\ee
From Eq.(\ref{eq:def_TPs}), turning points are given by
\be
   {\rm TP} = \left\{ a_1 =  e^{-\frac{3\pi}{4} i },  a_2 = e^{-\frac{\pi}{4} i },  a_3 = e^{\frac{3\pi}{4} i }, a_4 =  e^{\frac{\pi}{4} i }  \right\},
\ee
and the Stokes graph is shown in Fig.~\ref{fig:stokes_nomass_PT}.
Performing analytic continuation taking the paths in Eqs.(\ref{eq:path_AC})(\ref{eq:path_PT}) yields the monodromy matrices, ${\cal M}^{\arg(\widetilde{\eta})= 0_\pm}$, as
\be
&& {\cal M}_{{\rm AC}(3 \rightarrow 1)}^{0_+} = M_-^{-1} M_+^{-1} N_{a_1,a_3}  M_- N_{a_3,a_2} M_-^{-1} N_{a_2,a_4} M_- M_+ N_{a_4,a_1}, \\
&& {\cal M}_{{\rm AC}(3 \rightarrow 1)}^{0_-} =  M_-^{-1} M_+^{-1} N_{a_1,a_4}  M_- M_+ N_{a_4,a_1}, \\ \nl
&& {\cal M}_{{\rm AC}(2 \rightarrow 4)}^{0_+} = N_{a_1,a_3} M_+ M_- N_{a_3,a_2} M_+^{-1} M_-^{-1} N_{a_2,a_1}, \\
&& {\cal M}_{{\rm AC}(2 \rightarrow 4)}^{0_-} = N_{a_1,a_3} M_+ M_- N_{a_3,a_1} M_+^{-1} N_{a_1,a_4} M_- N_{a_4,a_2}  M_+^{-1} M_-^{-1} N_{a_2,a_1}, \\ \nl
&& {\cal M}_{\cal PT}^{0_+} = M_+ N_{a_1,a_3} M_+ N_{a_3,a_2} M_+ N_{a_2,a_1}, \\
&& {\cal M}_{\cal PT}^{0_-} = M_+ N_{a_1,a_4} M_+ N_{a_4,a_2} M_+ N_{a_2,a_1}.
\ee
From Eq.(\ref{eq:j_k}), normalizability of the wavefunctions determines a matrix component to be ${\cal M}^{\arg(\widetilde{\eta})=0_\pm}_{ij}=0$.
For each the QC, those are given as ${\frak D}_{{\rm AC}(3 \rightarrow 1)}^{\arg(\widetilde{\eta})=0_\pm} = {\cal M}^{\arg(\widetilde{\eta})=0_\pm}_{{\rm AC}(3 \rightarrow 1), 11}$, ${\frak D}_{{\rm AC}(2 \rightarrow 4)}^{\arg(\widetilde{\eta})=0_\pm} = {\cal M}^{\arg(\widetilde{\eta})=0_\pm}_{{\rm AC}(2 \rightarrow 4), 22}$, and ${\frak D}_{\cal PT}^{\arg(\widetilde{\eta})=0_\pm} = {\cal M}^{\arg(\widetilde{\eta})=0_\pm}_{{\cal PT}, 12}$.
Specifically,
\be
&&   {\frak D}_{{\rm AC}(3 \rightarrow 1)}^{0_+} \propto 1 + \frac{A B}{(1+B)^2}, \qquad {\frak D}_{{\rm AC}(3 \rightarrow 1)}^{0_-} \propto 1 + A B, \label{eq:DAC31_wo_mass_g}\\
&&   {\frak D}_{{\rm AC}(2 \rightarrow 4)}^{0_+}  \propto 1 + A^{-1} B, \qquad \quad \, \, {\frak D}_{{\rm AC}(2 \rightarrow 4)}^{0_-} \propto 1 + \frac{A^{-1} B}{(1+B)^2}, \label{eq:DAC24_wo_mass_g}\\ 
&& {\frak D}_{\cal PT}^{0_+} \propto 1 + \frac{A}{1+B}, \qquad \qquad \quad \, \, {\frak D}_{\cal PT}^{0_-} \propto 1 + A(1+B),
\ee
where
\be
A = e^{a_{12}} = e^{a_{43}}, \qquad B = e^{a_{31}} = e^{a_{42}},
\ee
and ${\rm C}_{{\rm NP}, \arg(\widetilde{\eta}) = 0} = \{ B \}$.
From Eqs.(\ref{eq:DDP_A_gen})-(\ref{eq:DDP_B_gen}), one can obtain the DDP formula for $\arg(\widetilde{\eta})=0_\pm$ as
\be
{\frak S}^{\nu}_0[A] = A (1+B)^{2 \nu}, \qquad {\frak S}^{\nu}_0[B] = B, \qquad (\nu \in {\mathbb R}) \label{eq:DDP_no_mass_g}
\ee
and thus the QCs removed the discontinuity are derived by Eq.(\ref{eq:D_no_sing}) as
\be
&&  {\frak D}^{0}_{{\rm AC}(3 \rightarrow 1)} \propto 1 + \frac{A B}{1+B}, \qquad
   {\frak D}^{0}_{{\rm AC}(2 \rightarrow 4)} \propto 1 + \frac{A^{-1} B}{1+B}, \label{eq:D0_AC_wo_mass} \\
&& {\frak D}^{0}_{\cal PT} \propto  1 + A. \label{eq:D0_PT_wo_mass} 
\ee

\begin{figure}[tbp]
 \centering
 \includegraphics[width=80mm]{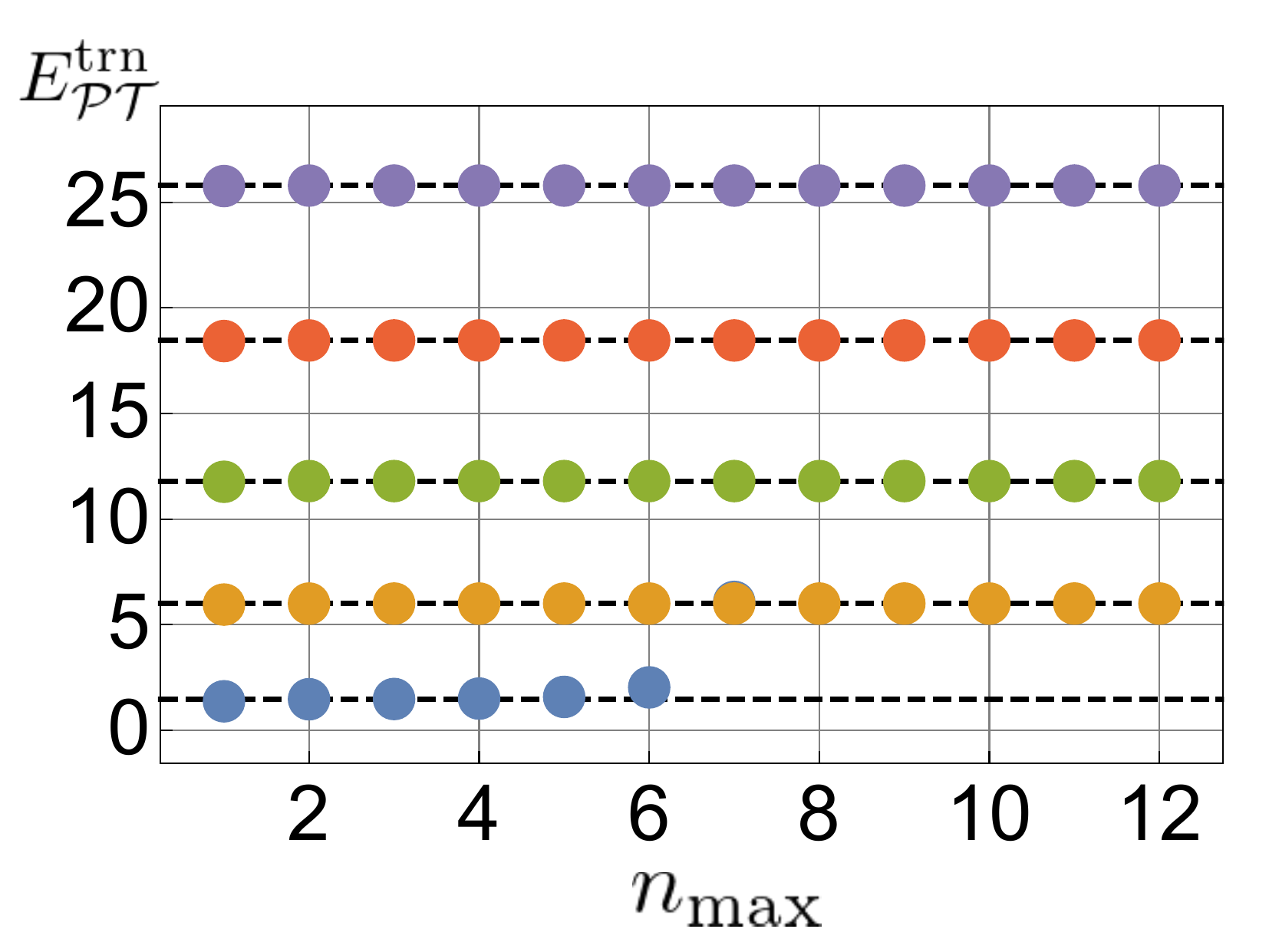}
 \caption{Truncated ${\cal PT}$ energy spectrum for first five energy levels, $k=0,\cdots,4$, evaluated by Eq.(\ref{eq:trnc_EPT}) with $g \hbar^4 = 1$.
   $n_{\rm max}$ in the horizontal axis denotes the truncation order of the truncated ${\cal PT}$ energy solution. 
   The black dashed lines are exact values from Ref.~\cite{Bender:2019}.   
 }
 \label{fig:PT_energy_womass}
\end{figure}

Then, we solve the QCs in Eqs.(\ref{eq:D0_AC_wo_mass})(\ref{eq:D0_PT_wo_mass}).
The cycles can be written as
\be
&& A = e^{i \widetilde{\phi}(\widetilde{\eta}) }, \qquad B = e^{-\phi \left(\widetilde{\eta} \right)}, 
\ee
where $\phi(\widetilde{\eta}) \in {\mathbb R}$ is given by Eq.(\ref{eq:phi}), and $\widetilde{\phi}(\widetilde{\eta})$ is defined from $\phi$ as
\be
\widetilde{\phi}(\widetilde{\eta}) := -i  \phi (\widetilde{\eta} e^{- \frac{\pi i}{2}} ) = \sum_{n \in {\mathbb N}_0} v_{2n-1} (-1)^n \widetilde{\eta}^{2n-1} \quad \in {\mathbb R}. \label{eq:phi_2til}
\ee
Thus, the QCs (\ref{eq:D0_AC_wo_mass})(\ref{eq:D0_PT_wo_mass}) can be rewritten as
\be
&& {\rm AC}: \  \pm i \widetilde{\phi} - \phi -  \log ( 1+e^{-\phi} ) = 2 i \kappa, \qquad
 {\cal PT}: \ \widetilde{\phi} = 2 \kappa, \label{eq:QC_wo_mass_g}
\ee
where the sign,  $\pm$ , in ${\rm AC}$ corresponds to ${\frak D}^{0}_{{\rm AC}(3 \rightarrow 1)}$ and ${\frak D}^{0}_{{\rm AC}(2 \rightarrow 4)}$, respectively, and $\kappa$ is defined by Eq.(\ref{eq:kappa}).
In the similar way to the analysis in Sec.~\ref{sec:Herm_sym_wo_mass}, we set the following ansatz to $\widetilde{\eta}$:
\be 
&& \widetilde{\eta}^{-1} = \frac{E^{3/4}}{g^{1/4} \hbar} \sim \sum_{n \in {\mathbb N}_0} \widetilde{e}^{(0)}_{2n-1}\kappa^{{1-2n}} + \sum_{\ell \in {\mathbb N}} \sum_{n \in {\mathbb N}_0} \widetilde{e}^{(\ell)}_{n} \sigma^{\ell}  \kappa^{-n} \quad \mbox{as} \quad \kappa \rightarrow +\infty, \label{eq:ansatz_tilk}
\ee
where $\sigma = e^{-\kappa}$.
One can easily solve ${\frak D}_{\cal PT}^{0} = 0$ and obtains
\be
&& \widetilde{e}^{(0)}_{-1} = \frac{3}{2 \sqrt{2} K(-1)}, \qquad \widetilde{e}^{(0)}_{1} -\frac{2 \sqrt{2 \pi } \Gamma \left(7/4 \right)}{9 \Gamma \left(-3/4\right)}, \nl
&& \widetilde{e}^{(0)}_{3} = \frac{3 \sqrt{2} \pi ^3 \Gamma \left(7/4 \right)-704 [\Gamma \left(5/4\right)]^7}{324 \pi ^{3/2} \Gamma \left(-3/4\right)}, \qquad \cdots
\ee
and $\widetilde{e}^{(\ell \in {\mathbb N})}_{n} = 0$ for all $n \in {\mathbb N}_0$.
Taking $E_{\cal PT} = \widetilde{\eta}^{-4/3} (g \hbar^4)^{1/3}$ yields
\be
\frac{E_{\cal PT}}{(g \hbar^{4})^{1/3}} &=& \kappa^{4/3} \left[
  \frac{3^{4/3}}{4 [K(-1)]^{4/3}}  -\frac{8 \sqrt{\pi } \Gamma \left(7/4\right)}{3^{8/3} [K(-1)]^{1/3} \Gamma \left(-3/4 \right)} \kappa^{-2} + O(\kappa^{-4}) \right]. \label{eq:EPT_no_mass}
\ee
Notice that the solution is real and Borel non-summable. 
In Fig.~\ref{fig:PT_energy_womass}, we show the truncated ${\cal PT}$ energy solutions defined as
\be
E_{{\cal PT}}^{\rm trn} = (g \hbar^4)^{1/3} \kappa^{4/3}  \sum_{n=0}^{n_{\rm max}-1} c_{n}\kappa^{-2n}, \label{eq:trnc_EPT}
\ee
where $c_{n \in {\mathbb N}_0}$ are coefficients obtained by our EWKB.
These plots are quite close to the exact values in $n_{\rm max} \lesssim 5$.
The lowest energy starts to deviate from the exact value around $n_{\rm max} \approx 6$ and then becomes divergent.
Since the energy spectrum with any energy level is a divergent series of $\kappa^{-1}$, all of them should be eventually divergent by taking $n_{\rm max} \rightarrow + \infty$.
Hence, one can interpret these plots as follows: only the lowest energy goes beyond the limitation of optimal truncation in $n_{\rm max} \gtrsim 6$, and the others are still stable as keeping almost the exact value up to, at least, $n_{\rm max} = 12$~\cite{boyd1999devil}.
Therefore, Fig.~\ref{fig:PT_energy_womass} insists that EWKB of the $\kappa^{-1}$-expansion works well.

\begin{figure}[tbp]
  \begin{center}
    \begin{tabular}{ccc}
      \begin{minipage}{0.33\hsize}
        \begin{center}
          \includegraphics[clip, width=50mm]{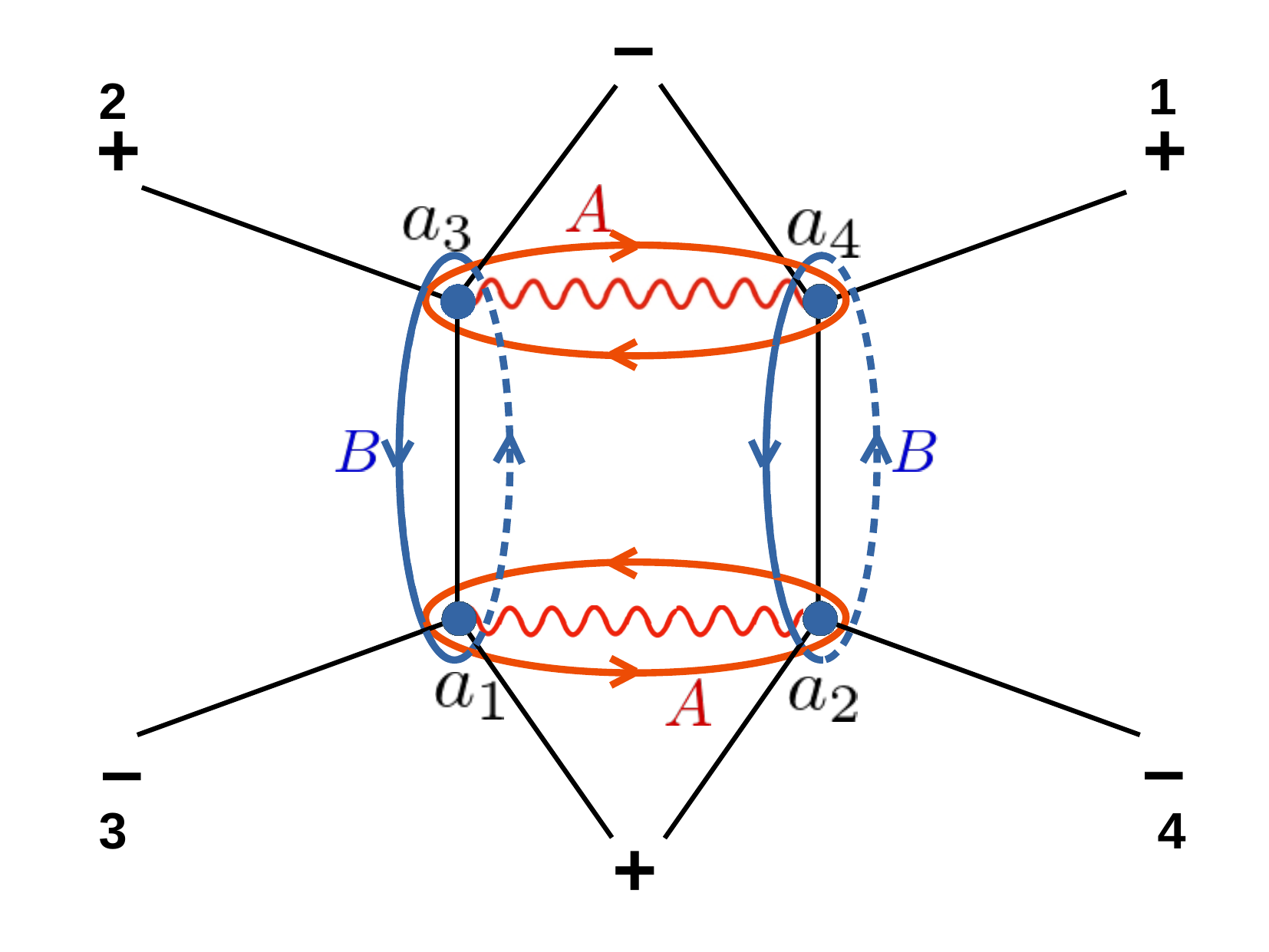}
          \hspace{1.6cm} (a) $\arg[\widetilde{Q}_0 \kappa^{2}] = 0$
        \end{center}
      \end{minipage}
      \begin{minipage}{0.33\hsize}
        \begin{center}
          \includegraphics[clip, width=50mm]{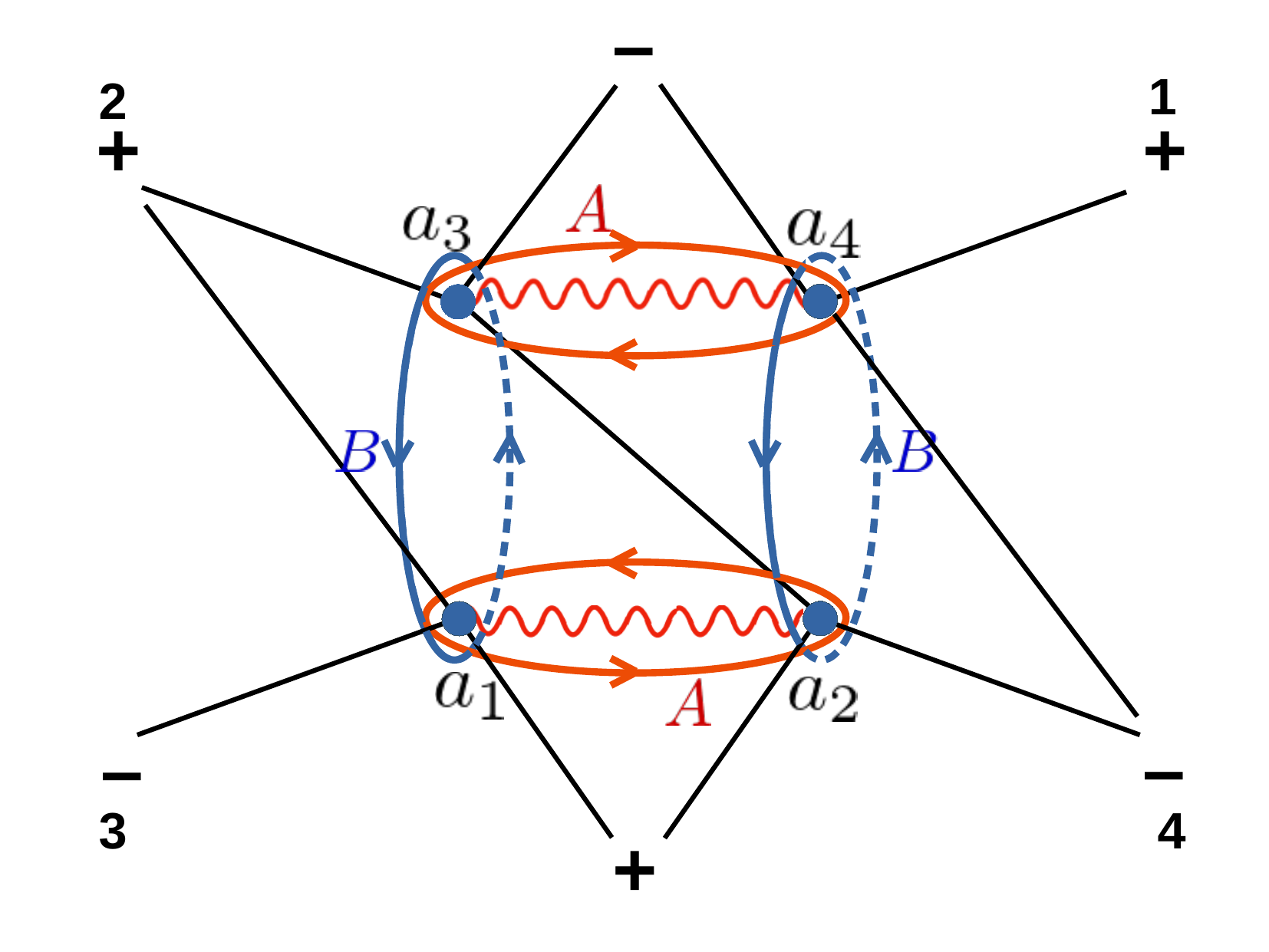}
          \hspace{1.6cm} (b) $\arg[\widetilde{Q}_0 \kappa^{2}] = -\frac{\pi}{2}$
        \end{center}
      \end{minipage}
      \begin{minipage}{0.33\hsize}
        \begin{center}
          \includegraphics[clip, width=50mm]{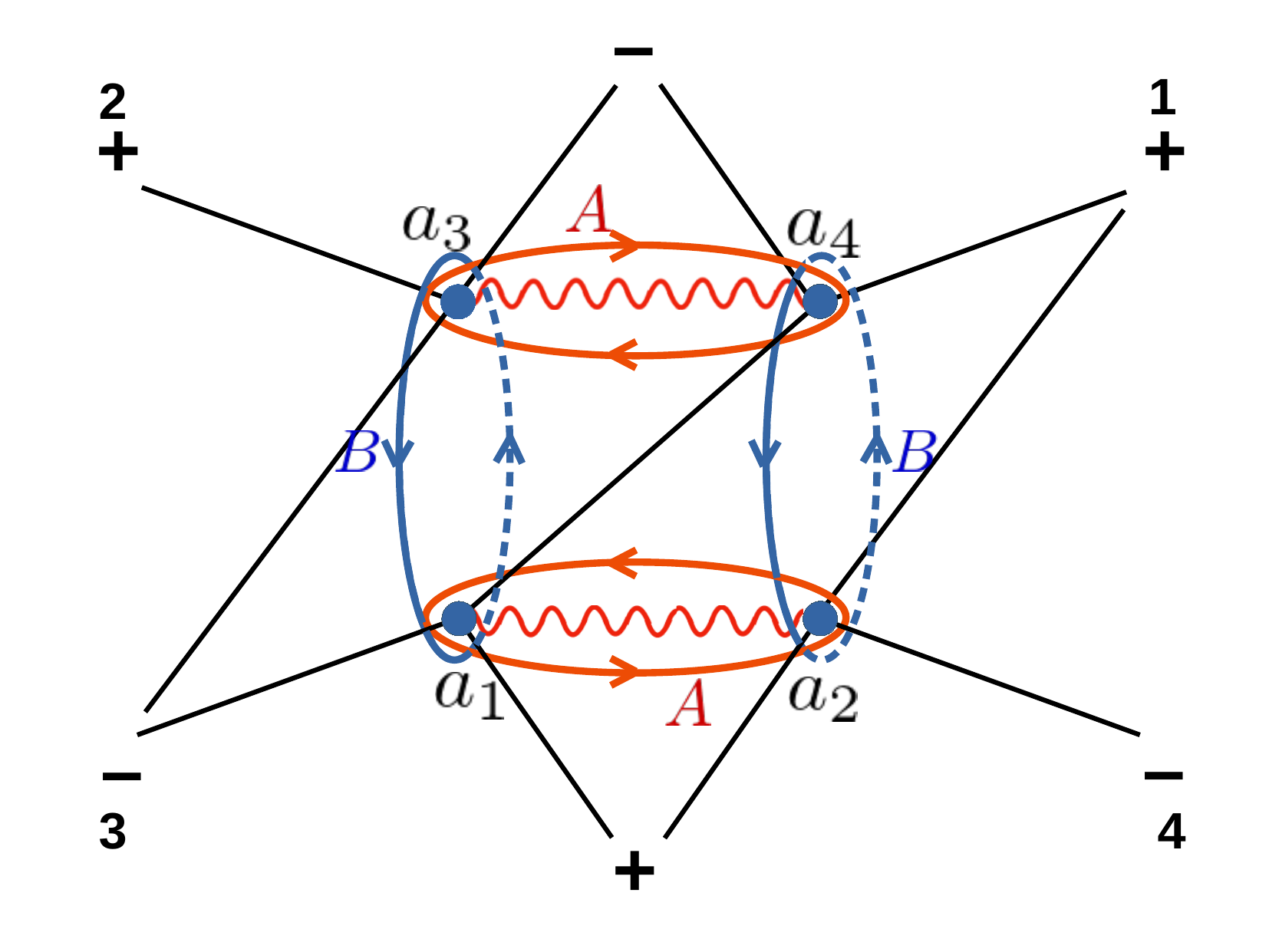}
          \hspace{1.6cm} (c) $\arg[\widetilde{Q}_0 \kappa^{2}] = +\frac{\pi}{2}$
        \end{center}
      \end{minipage}      
    \end{tabular} 
    \caption{
      Stokes graphs for $\arg[\widetilde{Q}_0 \kappa^{2}] = 0, \mp \frac{\pi}{2}$.
      The QCs in Eqs.(\ref{eq:D0_AC_wo_mass})(\ref{eq:D0_PT_wo_mass}) are constructed under the assumption that $\arg[\widetilde{Q}_0 \kappa^{2}] = 0$ shown in (a).
      The solution of ${\frak D}_{\rm AC}^{\arg(\widetilde{\eta}) = 0} = 0$ violates the assumption and reproduces either (b) or (c), where another Stokes phenomenon occurs, depending on the paths of analytic continuation, $\gamma_{3 \rightarrow 1}$ or $\gamma_{2 \rightarrow 4}$.}
    \label{fig:stokes_nomass_PT_sp}
  \end{center}
\end{figure}

We show the fact that no appropriate solution of ${\frak D}^{\arg(\widetilde{\eta})=0}_{\rm AC}$ exists. 
It can be seen by constructing the modified potential $\widetilde{Q}(\kappa)$ in Eq.(\ref{eq:mod_pot_eta}).
Since the AC QCs (\ref{eq:D0_AC_wo_mass}) have a special form such that the perturbative part is given by ${\frak D}_{\rm AC}^{\arg(\widetilde{\eta})=0} \sim1+ A^{\pm 1} B$, the leading order is a complex value, that is $\widetilde{e}_{-1}^{(0)} =  \frac{3 e^{\pm \pi i /4}}{4K(-1)}$.
This coefficient makes the leading order of $\widetilde{Q}$ complex-valued as $\widetilde{Q}_0(x) \kappa^{2} = Q(x) \left(\frac{3}{4K(-1)}\right)^2  (e^{\pm \pi i /4}\kappa)^{2}$.
Due to the phase in $\widetilde{Q}_0$, i.e. $e^{\pm \pi i/2}$, it does not give the same Stokes graph to Fig.~\ref{fig:stokes_nomass_PT_sp}(a), but (b) or (c). 
Since the DDP formulas are generally defined from each Stokes phenomenon, the QCs (\ref{eq:D0_AC_wo_mass}) formulated by the DDP formula corresponding to Fig.~\ref{fig:stokes_nomass_PT_sp}(a) violates the assumption that its solution must reproduce the same Stokes graph.
Therefore, the solution of ${\frak D}_{\rm AC}^{\arg(\widetilde{\eta})=0}=0$ in Eq.(\ref{eq:D0_AC_wo_mass}) should be rejected\footnote{
  We should also remind that $\kappa$ is now a real value.
}.
Instead of Fig.~\ref{fig:stokes_nomass_PT_sp}(a), one can begin with Fig.~\ref{fig:stokes_nomass_PT_sp}(b)(c) by taking $\arg(\widetilde{\eta})=\pm \frac{\pi}{4}$
and perform EWKB to obtain the AC energy. 
By this procedure, appropriate solutions are available from either (b) or (c) depending on the path of analytic continuation, as (b) for $\gamma_{3 \rightarrow 1}$ and (c) for $\gamma_{2 \rightarrow 4}$.
The Stokes graph and the details of the calculations are parallel to the analysis in Sec.~\ref{sec:Herm_sym_wo_mass}, and one can consequently find the same the Hermitian energy in Eqs.(\ref{eq:E0_lam_wo_mass})-(\ref{eq:E2_lam_wo_mass}) except an overall phase due to the phase rotation of the coupling constant, $\lambda = g e^{\pm \pi i}$:
\be
E^{\arg(\widetilde{\eta}) = +\pi/4}_{{\rm AC}(3 \rightarrow 1)}(g) = {\cal C} \left[ E^{\arg(\widetilde{\eta}) = -\pi/4}_{{\rm AC}(2 \rightarrow 4)}(g) \right] = E_{\cal H}(\lambda = g e^{+\pi i }), \qquad g \in {\mathbb R}_{>0},
\ee
where ${\cal C}$ is complex conjugate.
This is the same conclusion to Eq.(\ref{eq:sol_DAC_wo_mass_g}), and we should accept this solution as $E_{\rm AC}$.

\subsection{Impossibility of an alternative form of the ABS conjecture} \label{sec:impossibility_PT_AC}
Finally, we point out that the ${\cal PT}$ and the AC energies for the pure quartic potential are independent solutions on each other, meaning that those can not be related by Stokes automorphism and Borel resummation.

Let us see this fact.
In order to make such a relation, if there exists, those QCs need to be equivalent by DDP formula at a certain $\arg(\lambda)$ and/or $\arg(\kappa)$\footnote{
These two complex phases are distinguished for the ansatz in Eq.(\ref{eq:ansatz_k}).
  The former is an overall phase, and the latter affects the coefficients.
}.
The discussion in Sec.~\ref{sec:warm_up_analytic_QC} is helpful for the consideration.
Since no Stokes phenomenon occurs in 
$0 < |\arg(\eta)| < \pi/4$, the QC for $\arg(\eta) = 0_\pm$ in Eqs.(\ref{eq:qcondp_lam4_mass})(\ref{eq:qcondm_lam4_mass}) can be directly analytic-continued to the AC QCs for $\arg(\widetilde{\eta})=0_\mp$ in Eqs.(\ref{eq:DAC31_wo_mass_g})(\ref{eq:DAC24_wo_mass_g}) by replacing symbols $(A_1,A_2) \rightarrow (A,B)$ as
\be
&& {\frak D}_{{\rm AC}(3 \rightarrow 1)}^{\arg(\widetilde{\eta})=0_+} \propto \left. {\frak D}_{\cal H}^{\arg(\eta)=0_-}\right|_{(A_1,A_2) \rightarrow (A,B)}, \qquad {\frak D}_{{\rm AC}(2 \rightarrow 4)}^{\arg(\widetilde{\eta})=0_-} \propto \left. {\frak D}_{\cal H}^{\arg(\eta)=0_+}\right|_{(A_1,A_2) \rightarrow (A,B)}. \label{eq:rel_DAC_Dlam} \nl
\ee
This procedure is the same to rotating complex phases of $\lambda$ and $\kappa$ at once.
Thus, we now have the direct relation between ${\frak D}_{\cal H}^{\arg(\eta)=0_\pm}$ and ${\frak D}_{\rm AC}^{\arg(\widetilde{\eta})=0_\pm}$ given by Eq.(\ref{eq:rel_DAC_Dlam}).
However, no connection from ${\frak D}_{\rm AC}^{\arg(\widetilde{\eta})=0_\pm}$ to ${\frak D}_{\cal PT}^{\arg(\widetilde{\eta})=0}$ is constructable by the DDP formula in Eq.(\ref{eq:DDP_no_mass_g}).
Especially, eliminating the non-perturbative part, $B$, from ${\frak D}_{\rm AC}^{\arg(\widetilde{\eta})=0_\pm}$ is extremely tough, which implies that the perturbative/non-perturbative structures between the ${\cal PT}$ and the AC energies are not related to each other.
Notice that $E_{\cal PT}$ in Eq.(\ref{eq:EPT_no_mass}) is purely perturbative with respect to the $\kappa^{-1}$-expansion, but $E_{\rm AC}$ in Eq.(\ref{eq:sol_DAC_wo_mass_g}) is not.
Furthermore, even the coefficients in their perturbative part do not match with each other.
Therefore, the ABS conjecture is not satisfied, and those must be independent solutions.
No alternative form of the ABS conjecture can be reformulated by Stokes automorphism and Borel resummation.
In summary, the above consideration can be schematically expressed by
\be
&& {\frak D}_{\cal H}^{\arg(\eta)=0_-} \xrightarrow{\lambda = g e^{+\pi i }}   {\frak D}_{{\rm AC}(3 \rightarrow 1)}^{\arg(\widetilde{\eta})=+\frac{\pi}{4}+0_-} \xrightarrow{\arg(\kappa) = 0\rightarrow +\frac{\pi}{4} + 0_-}   {\frak D}_{{\rm AC}(3 \rightarrow 1)}^{\arg(\widetilde{\eta})=0_+} \xnrightarrow{{\frak S}^{\forall \nu \in {\mathbb R}}_{\arg(\widetilde{\eta})=0}} {\frak D}^{\arg(\widetilde{\eta})=0}_{\cal PT}, \nl
&&  {\frak D}_{\cal H}^{\arg(\eta)=0_+} \xrightarrow{\lambda = g e^{-\pi i }}   {\frak D}_{{\rm AC}(2 \rightarrow 4)}^{\arg(\widetilde{\eta})=-\frac{\pi}{4}+0_+} \xrightarrow{\arg(\kappa) = 0\rightarrow -\frac{\pi}{4} + 0_+}   {\frak D}_{{\rm AC}(2 \rightarrow 4)}^{\arg(\widetilde{\eta})=0_-} \xnrightarrow{{\frak S}^{\forall \nu \in {\mathbb R}}_{\arg(\widetilde{\eta})=0}} {\frak D}^{\arg(\widetilde{\eta})=0}_{\cal PT}.
\ee

Notice that the violation of the ABS conjecture for $\omega=0$ is essentially irrelevant to the values of $\beta$, $g$, and $\hbar$.
It is because our EWKB of the $\kappa^{-1}$-expansion does not depend on them, and their dependence appears as an overall factor of the energy only.
Our result is consistent with the observations in Ref.~\cite{Lawrence:2023woz}.

\section{Additional remarks} \label{sec:add_remark}

\subsection{Generalization to $V_{\cal PT} = \omega^2 x^{2} + g x^{2K} (i x)^{\varepsilon=2}$ with $K \in {\mathbb N}$} \label{sec:general_eps}
We would make comments on a generalization of the modified ABS conjecture to 
\be
V_{\cal PT} = \omega^2 x^{2} + g x^{2K} (i x)^{\varepsilon=2}, \qquad (\omega \in {\mathbb R}_{\ge 0}, \ g \in {\mathbb R}_{>0}. \ K \in {\mathbb N})
\ee
In such a case, the paths of analytic continuation in Eqs.(\ref{eq:path_AC})(\ref{eq:path_PT}) are modified and defined by values of $\varepsilon$ and $K$. 
(See, for example, Ref.\cite{Bender:2019} and references within.)
When $\varepsilon =2$, those are given by
\be
&& \gamma_{3 \rightarrow 1} : - e^{\alpha \pi i} \infty \ \rightarrow \ e^{ \alpha \pi i} \infty,  \qquad 
\gamma_{2 \rightarrow 4} : - e^{-\alpha \pi i} \infty  \ \rightarrow \ e^{-\alpha \pi i} \infty,  
\label{eq:path_AC_gen}
\ee
for the AC energy, and
\be
\gamma_{3 \rightarrow 4} &:& - e^{\alpha \pi i} \infty \ \rightarrow \ e^{-\alpha \pi i} \infty,  \label{eq:path_PT_gen}
\ee
for the ${\cal PT}$ energy, where $\alpha = \frac{1}{2 K + 2}$.
Along these paths, one can directly apply the same analyses in Secs.~\ref{sec:with_mass_term} and \ref{sec:without_mass_term}, and the similar results can be eventually found.

For $\omega >0$, similar to the case that $K = 1$, the ${\cal PT}$ energy is purely perturbative (but Borel non-summable), and the AC energy includes non-perturbative contributions in addition to the same perturbative ${\cal PT}$ energy.
The QCs and the DDP formula also take the same forms.
As a result, Eq.(\ref{eq:EPT_DDP}) is satisfied, and our modified ABS conjecture is unchanged for any $K \in {\mathbb N}$.

For $\omega = 0$,  the generalization of the energy in Eq.(\ref{eq:Echbar4/3}) to $K \in {\mathbb N}$ gives
\be
E = c(k) (\lambda \hbar^{2K+2})^{1/(K+2)}, \qquad c(k) \in {\mathbb R}_{>0}, \label{eq:Echbar_gen}
\ee
and the scaling law in Eq.(\ref{eq:scale_dim_wo_mass_warm}) is modified as
\be
[x]  = \frac{1}{2 K + 2}, \qquad    [\hbar] = \frac{K + 2}{2 K + 2}, \qquad [\lambda]  = 0, \qquad [E] = 1. \label{eq:scale_dim_wo_mass_warm_gen}
\ee
The ansatz in Eqs.(\ref{eq:ansatz_k})(\ref{eq:ansatz_tilk}) based on the $\kappa^{-1}$-expansion works by a slight modification of the dimensionless parameters, $\eta$ and $\widetilde{\eta}$, as
\be
&& \eta = \frac{\lambda^{1/(2K + 2)} \hbar}{E^{(K+2)/(2K+2)}}, \qquad \widetilde{\eta} = \frac{g^{1/(2K + 2)} \hbar}{E^{(K+2)/(2K+2)}}.
\ee
The ${\cal PT}$ energy is purely perturbative with respect to the $\kappa^{-1}$ expansion, and the AC energy contains non-perturbative parts not only the perturbative part like the case of $K=1$.
The relation between the Hermitian and the AC energies in Eq.(\ref{eq:sol_DAC_wo_mass_g}) is modified as
\be
\frac{E_{\rm AC}(g)}{(g \hbar^{2 K + 2})^{1/(K + 2)}} 
= e^{\pm \frac{\pi}{K+2} i} \frac{E_{\cal H}(\lambda)}{(\lambda \hbar^{2 K +2})^{1/(K+2)}} \label{eq:sol_DAC_wo_mass_g_gen}.
\ee
No relation between $E_{\rm AC}(g)$ and $E_{\cal PT}(g)$ is constructable by Stokes automorphism and Borel resummation.

\subsection{Spectral reality} \label{sec:comm_qcondition}
Observations from QCs and DDP formula are quite useful to know properties of a energy spectrum of a given theory.
In EWKB, spectral reality can be determined solely by the corresponding QC without specifically solving them, and showing spectral reality is equivalent to obtaining a real QC by analytic continuation.
For example, by denoting $a_A(E,\hbar) = -i \log A(E,\hbar) \in {\mathbb R}$, Eqs.(\ref{eq:DAC})(\ref{eq:DPT}) are reexpressed by
\be
&& {\frak D}^{0}_{{\rm AC}(3 \rightarrow 1)/(2 \rightarrow 4)} \propto \cos \left[ \frac{a_A}{2} \pm  \frac{i}{2} \log  (1+B)   \right], \qquad {\frak D}^{0}_{\cal PT} \propto \cos \frac{a_A}{2}. \label{eq:D_cos}
\ee
Since ${\frak D}^{0}_{\cal PT}$ is real, the ${\cal PT}$ energy solution should be real.
In addition, one can immediately see that the spectral reality of the above AC energy solutions are violated by the non-perturbative effect, $B$, because of the complex ${\frak D}^{0}_{\rm AC}$.

The methods used in our analysis based on the Airy-type and the DW-type connection formulas should work for ${\cal PT}$ symmetric polynomial potentials.
By taking appropriate paths of analytic continuation, one can make sure that Eq.(\ref{eq:VPT_def}) with $\omega \in {\mathbb R}_{> 0}$ and $\varepsilon \in {\mathbb N}$ provides the same cycle representation to Eq.(\ref{eq:D_cos}) for ${\frak D}^{0}_{\cal PT}$.
For more complicated potentials such as $V_{\cal PT}(x) = g (i x)^{\alpha} \log (i x)$, one has to formulate a connection formula beyond the Airy- and DW-type.
This would be an interesting problem as a future work.

\acknowledgments
We would thank Toshiaki~Fujimori, Tatsuhiro~Misumi, Muneto~Nitta, Norisuke~Sakai, Naohisa~Sueishi, and Hidetoshi~Taya for helpful discussion about exact WKB analysis. 
We would thank Marco~Serone for helpful discussion about connection formula of the degenerate Weber-type Stokes graph.
We would thank Mithat~\"{U}nsal for comments on a draft of this paper.
S.~K is supported by JSPS KAKENHI Grant No.~22H05118.

\appendix

\section{Derivation of $G$} \label{sec:der_G}

\begin{figure}[tbp]
 \centering
 \includegraphics[width=80mm]{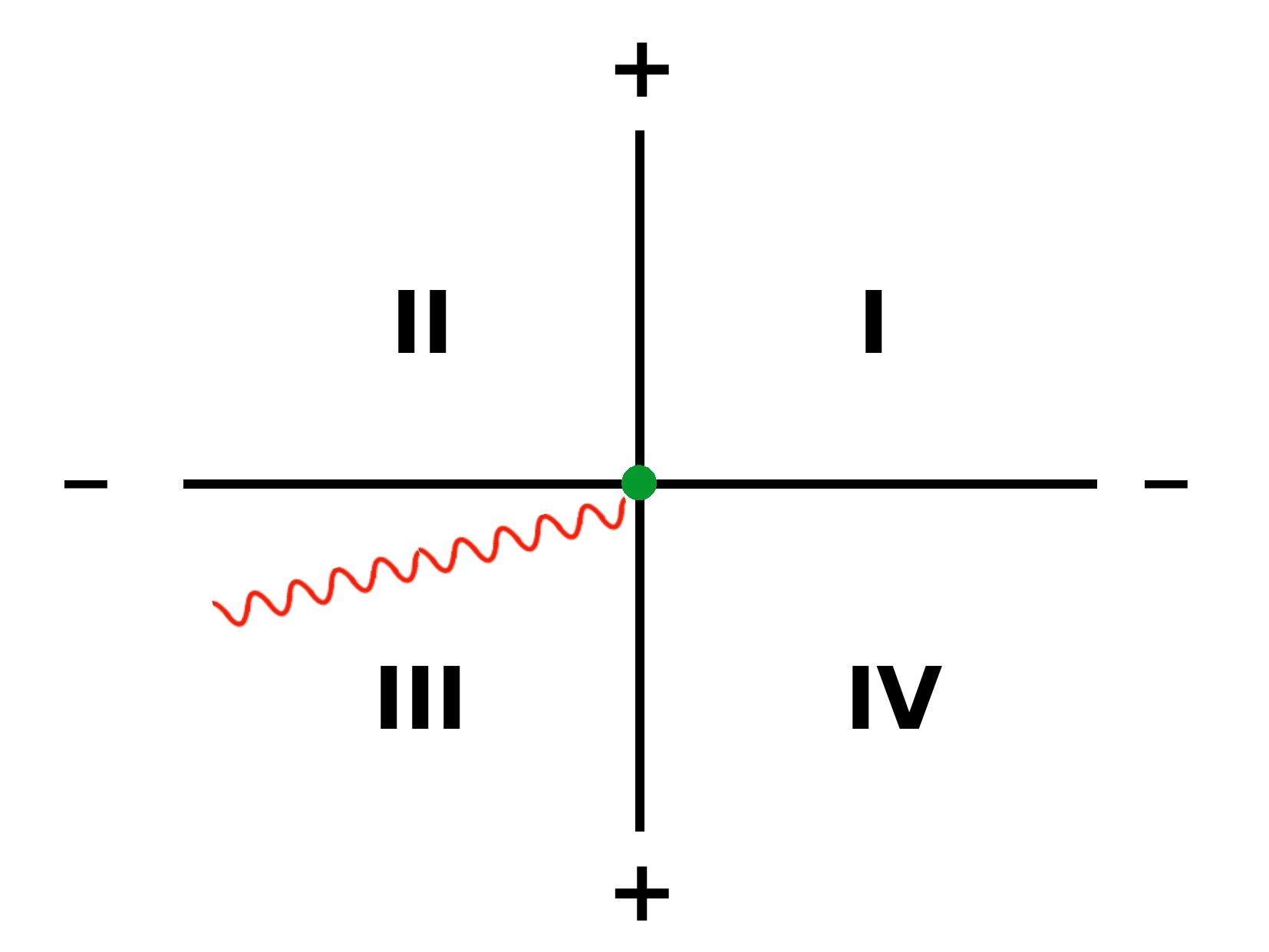}
 \caption{Stokes graph for the degenerate Weber equation.
   The green dot is a double turning point.
   The black solid and red wave lines denote Stokes lines and a branch-cut, respectively.
}
\label{fig:DW_connection}
\end{figure}

We show the derivation of $G$ in Eq.(\ref{eq:rep_mass}) using a coordinate transform from the degenerate Weber (DW)-type Stokes graph to our potential form~\cite{Takei3,DDP2,Sueishi:2021xti,Bucciotti:2023trp}.
There is another method using Mellin transform, see Refs.~\cite{DP1,Zinn-Justin:2004qzw,Kamata:2021jrs}.

The DW-type Stokes graph offers a connection formula around a double turning point.
In general, any connection formulas are firstly defined in a local coordinate and then mapped to a coordinate used in a given problem.
The local DW-type connection formula is defined by the following Schr\"{o}dinger equation:
\be
\widehat{\cal L} := - \hbar^{2} \pd_y^2 + \frac{y^2}{4} -\kappa \hbar, \qquad \widehat{\cal L} \widehat{\psi}(y,\hbar) = 0, \label{eq:deg_web_L}
\ee
where $\kappa \in {\mathbb R}$.
The Stokes graph is shown in Fig.~\ref{fig:DW_connection}, and the connection matrices, $\widehat{\cal M}^{\bullet \rightarrow \bullet}$, are expressed by~\cite{DDP2,Kawai1,Takei3,Bucciotti:2023trp}
\be
 \widehat{\cal M}^{{\rm IV} \rightarrow {\rm I}} &=&
\begin{pmatrix}
  1 & 0 \\
  i  \frac{\sqrt{2 \pi} e^{+ \pi i \kappa} \hbar^{+\kappa}}{\Gamma(1/2-\kappa)} & 1
\end{pmatrix}, \qquad
\widehat{\cal M}^{{\rm I} \rightarrow {\rm II}} =
\begin{pmatrix}
  1 & i  \frac{\sqrt{2 \pi} \hbar^{-\kappa}}{\Gamma(1/2+\kappa)} \\
  0 & 1 
\end{pmatrix}, \nl \nl
 \widehat{\cal M}^{{\rm II} \rightarrow {\rm III}} &=&
\begin{pmatrix}
  1 & 0 \\
  i \frac{\sqrt{2 \pi} e^{- \pi i \kappa} \hbar^{+\kappa}}{\Gamma(1/2-\kappa)} & 1
\end{pmatrix}, \qquad
\widehat{\cal M}^{{\rm III} \rightarrow {\rm IV}} =
\begin{pmatrix}
  1 & i \frac{\sqrt{2 \pi}e^{-2 \pi i \kappa} \hbar^{-\kappa}}{\Gamma(1/2+\kappa)} \\
  0 & 1 
\end{pmatrix}. \label{eq:DW_conn_form_k}
\ee
In addition, the branch-cut matrix $\widehat{T}$ is given by
\be
\widehat{T} =
\begin{pmatrix}
 - e^{+2 \pi i \kappa} & 0 \\
  0 &- e^{-2 \pi i \kappa} 
\end{pmatrix}. \label{eq:DW_conn_br_k}
\ee
By a coordinate transform $y = y(x)$, the connection matrices in the $x$-coordinate, ${\cal M}^{\bullet \rightarrow \bullet}$, take the form that
\be
 {\cal M}^{{\rm IV} \rightarrow {\rm I}} &=&
\begin{pmatrix}
  1 & 0 \\
  i \frac{C_-}{C_+} \frac{\sqrt{2 \pi} e^{+ \pi i F} \hbar^{+F}}{\Gamma(1/2-F)} & 1
\end{pmatrix}, \qquad
{\cal M}^{{\rm I} \rightarrow {\rm II}} =
\begin{pmatrix}
  1 & i \frac{C_+}{C_-} \frac{\sqrt{2 \pi} \hbar^{-F}}{\Gamma(1/2+F)} \\
  0 & 1 
\end{pmatrix}, \nl \nl
 {\cal M}^{{\rm II} \rightarrow {\rm III}} &=&
\begin{pmatrix}
  1 & 0 \\
  i \frac{C_-}{C_+} \frac{\sqrt{2 \pi} e^{- \pi i F} \hbar^{+F}}{\Gamma(1/2-F)} & 1
\end{pmatrix}, \qquad
{\cal M}^{{\rm III} \rightarrow {\rm IV}} =
\begin{pmatrix}
  1 & i \frac{C_+}{C_-} \frac{\sqrt{2 \pi}e^{-2 \pi i F} \hbar^{-F}}{\Gamma(1/2+F)} \\
  0 & 1 
\end{pmatrix}. \label{eq:DW_conn_form}
\ee
and the branch-cut matrix $T$ is given by
\be
T =
\begin{pmatrix}
 - e^{+2 \pi i F} & 0 \\
  0 &- e^{-2 \pi i F} 
\end{pmatrix}. \label{eq:DW_conn_br}
\ee
The non-perturbative ${\frak B}$ in Eq.(\ref{eq:rep_mass}) is given by the above connection formula and takes the form as
\be
   {\frak B} = \frac{C_-}{C_+} \frac{\sqrt{2 \pi} {\frak B}_0 e^{+ \pi i F} \hbar^{+F}}{\Gamma(1/2-F)}, \qquad {\frak B}_0 := e^{-\frac{S_{\frak B}}{\hbar}}, \label{eq:B_CC}
\ee
where $S_{\frak B} \in {\mathbb R}_{>0} $ is determined by distance of two turning points defining the ${\frak B}$-cycle and $S_{\frak B} = 2/3$ in our case.
Hence, in order to know the specific form of $G$ in Eq.(\ref{eq:rep_mass}), our task is to compute $C_{\pm}$ by the coordinate transform.

Those are available by constructing the wavefunctions of Eqs.(\ref{eq:sc_mass_g_resc})(\ref{eq:deg_web_L}) specifically.
Here, $Q(x)$ and $\widehat{Q}(y)$ are defined as $Q(x) = x^2 - x^4 - \widetilde{E} \hbar$ and $\widehat{Q}(y) = \frac{y^2}{4}  - \kappa \hbar$, respectively.
One should be careful that $S_{\rm od}(\hbar)$ given by them contains not only odd powers of the $\hbar$-expansion but also even powers due to the energy terms proportional to $\hbar$.
The coordinate transform can be obtained from the relation of $Q(x)$ and $\widehat{Q}(y)$ given by
\be
Q(x,\hbar) = \left[ \pd_x y(x,\hbar) \right]^2 \widehat{Q}(y(x,\hbar),\hbar) - \frac{\hbar^2}{2} \{ y(x,\hbar);x\}, \label{eq:Qx_Qy}
\ee
with the Schwarzian derivative defined as
\be
\{ y(x,\hbar),x\} := \frac{\pd_x^3 y(x,\hbar)}{\pd_x y(x,\hbar)}  - \frac{3}{2} \left(  \frac{\pd_x^2 y(x,\hbar)}{\pd_x y(x,\hbar)} \right)^2.
\ee
By denoting $S_{\rm od}(x)$ and $\widehat{S}_{\rm od}(y)$ as solutions of Riccati equation defined by $Q(x)$ and $\widehat{Q}(y)$, respectively, the wavefunctions constructed from $S_{\rm od}(x)$ and $\widehat{S}_{\rm od}(y)$ are also related to each other as
\be
\psi_{\pm}(x,\hbar) = C_{\pm}(\hbar) [\pd_{x} y(x,\hbar)]^{-1/2} \widehat{\psi}_\pm(y(x,\hbar),\hbar), \label{eq:psi_coord_trans}
\ee
where the sign, ``$\pm$'', in the wavefunctions is a label to distinguish the upper and lower components.
From Eq.(\ref{eq:psi_coord_trans}), the coefficients, $C_\pm(\hbar)$, can be obtained as
\be
C_\pm(\hbar) = \lim_{x \rightarrow 0} [\pd_{x} y(x,\hbar)]^{1/2} \frac{\psi_{\pm}(x,\hbar)}{\widehat{\psi}_\pm(y(x,\hbar),\hbar)}. \label{eq:Cpm_def}
\ee
In our analysis, we define the upper and lower components of the wavefunctions such that $\psi_\pm(x,\hbar) = \frac{\exp [\mp \int dx S_{\rm od}(x,\hbar)]}{\sqrt{S_{\rm od}(x,\hbar)}}$ and $\widehat{\psi}_\pm(y,\hbar) = \frac{\exp [\pm \int dy \widehat{S}_{\rm od}(y,\hbar)]}{\sqrt{\widehat{S}_{\rm od}(y,\hbar)}}$ to adjust to the asymptotic behavior in Fig.~\ref{fig:DW_connection}.
One can easily check that  $\kappa = -{\rm Res}_{y=0} \widehat{S}_{\rm od}(y,\hbar) = -{\rm Res}_{x=0} S_{\rm od}(x,\hbar)$.
From Eq.(\ref{eq:Qx_Qy}), one can obtain $y(x,\hbar)$ and $\kappa(\hbar)$ as
\be
y(x,\hbar) &=&  \frac{i}{\sqrt{2}} \left[ 2 x - \frac{\widetilde{E}}{4} x \hbar - \frac{49 + 16 \widetilde{E}^2}{64} x \hbar^2 \right. \nl
  && \qquad \left. + \frac{\widetilde{E} ( 11 + \widetilde{E}^2)}{32x^3} \hbar^3 -\frac{\widetilde{E} ( 1379+ 226 \widetilde{E}^2) }{512} x \hbar^3 + O(\hbar^4)  \right] + O(x^2), \\
 \kappa(\hbar) &=& - \frac{\widetilde{E}}{2} - \frac{3 (1 + \widetilde{E}^2)}{16} \hbar - \frac{5 \widetilde{E} (17 + 7 \widetilde{E}^2)}{128} \hbar^2  - \frac{105(19 + 50 \widetilde{E}^2 + 11 \widetilde{E}^4)}{2048} \hbar^3 + O(\hbar^4). \label{eq:kappa_app} \nl
\ee
Thus, Eq.(\ref{eq:Cpm_def}) gives
\be
&& C_+(\hbar) \nl
&=& \exp \left[ \frac{\widetilde{E}^2}{16} \hbar + \frac{\widetilde{E}(55 + 23 \widetilde{E}^2)}{256} \hbar^2 + \frac{441 + 5388 \widetilde{E}^2 + 1091 \widetilde{E}^4}{6144} \hbar^3 + O(\hbar^4)
  \right] 2^{\frac{F(\hbar)}{2}} e^{\frac{\pi}{2} i F (\hbar)} \nl
&=& \frac{1}{C_-(\hbar)}, \label{eq:CC_sol}
\ee
where we used the fact that $F(\hbar) = \kappa(\hbar)$ by Eqs.(\ref{eq:F_G})(\ref{eq:kappa_app}).
Substituting Eq.(\ref{eq:CC_sol}) into Eq.(\ref{eq:B_CC}) and comparing with Eq.(\ref{eq:rep_mass}) give the form of $G$ in Eq.(\ref{eq:F_G}).

\section{Alien calculus for energy spectra} \label{sec:DDP_energy}
We formulate Stokes automorphism (or DDP formula) for the transseries solution of the energies in Sec.~\ref{sec:PT_sym_mass} by performing alien calculus.
We omit the subscript ``$0$'', e.g. ${\frak S}_0^{\nu} \rightarrow {\frak S}^{\nu}$ and only address the case that $\arg(g) = 0$. 
Since expanding the Stokes automorphism in Eq.(\ref{eq:Stokes_alien}) around $\nu=0$ gives
\be 
   {\frak S}^{\nu} 
   =  1 + \nu \bul{\Delta} + \frac{\nu^2}{2} (\bul{\Delta})^2 + O(\nu^3), \label{eq:Snu_app}
\ee
our main task is to calculate action of $(\bul{\Delta})^{n \in {\mathbb N}}$ to the transseries solutions order by order.

Before discussion about our case,  we suppose a function $f(E(\hbar),\hbar)$ with an indirect dependence of $\hbar$ in the variable $E=E(\hbar)$ and consider action of the alien derivative to $f(E(\hbar),\hbar)$.
In this case, the alien derivative can be decomposed into two parts, i.e. actions to $E(\hbar)$ and to $f(E,\hbar)$ with a fixed $E$.
Using the chain rule, it can be expressed by\footnote{
This procedure is the same methodology to split a total derivative into partial derivatives for a function with indirect independences~\cite{Bucciotti:2023trp}.  
}
\be
&& \bul{\Delta}[f](E(\hbar),\hbar)
= \left. \frac{\pd f(E,\hbar)}{\pd E} \right|_{E=E(\hbar)} \bul{\Delta}[E(\hbar)] + \bul{\Delta}_\hbar [f](E(\hbar),\hbar), \\
&& \bul{\Delta}_\hbar [f](E(\hbar),\hbar):= \left. \bul{\Delta} [f(E,\hbar)] \right|_{E=E(\hbar)}.
\ee
Below, we use a simplified notation as
\be
&& \bul{\Delta}[f](E(\hbar),\hbar)\, \rightarrow \, \bul{\Delta}[f], \qquad \bul{\Delta}_\hbar[f](E(\hbar),\hbar)\, \rightarrow \, \bul{\Delta}_\hbar[f], \nl
&& \left. \frac{\pd^n f}{\pd E^n} \right|_{E=E(\hbar)} \, \rightarrow \, \pd_E^n, \qquad \quad \, \bul{\Delta}[E(\hbar)] \, \rightarrow \, \bul{\Delta}[E].
\ee
Notice that
\be
[\bul{\Delta}_\hbar, \pd_E ] = 0, \qquad \pd_E \bul{\Delta}[E] = \bul{\Delta}_\hbar \circ \bul{\Delta}[E] = 0,
\ee
where $[A,B] := AB - BA$.
By repeating the same calculus, the higher order derivatives to $f(E(\hbar),\hbar)$ can be derived as
\be
(\bul{\Delta})^2 [f]  &=& \left[ (\bul{\Delta}[E])^2 \pd_E^2 +  (\bul{\Delta})^2[E] \pd_E + 2 \bul{\Delta}[E] \bul{\Delta}_\hbar \pd_E + (\bul{\Delta}_\hbar)^2 \right] [f], \\
(\bul{\Delta})^3 [f] 
&=&  \left[ (\bul{\Delta}[E])^3 \pd_E^3 + 3 (\bul{\Delta})^2[E] \bul{\Delta}[E] \pd_E^2 + 3 (\bul{\Delta}[E])^2 \bul{\Delta}_\hbar \pd_E^2 + 
  (\bul{\Delta})^3[E] \pd_E   \right. \nl
&& \left. +  3 (\bul{\Delta})^2[E] \bul{\Delta}_\hbar \pd_E  + 3 \bul{\Delta}[E] (\bul{\Delta}_\hbar)^2 \pd_E + (\bul{\Delta}_\hbar)^3 \right] [f], \\
 \vdots && \nl
 (\bul{\Delta})^n [f] &=& \sum_{\substack{k_1,\cdots,k_n = 0; \\ k_1+2 k_2 + \cdots +n k_n \le n}} \left[ \prod_{p=1}^{n} \frac{(n - \sum_{\ell=1}^{p-1} \ell k_\ell )!}{(n - \sum_{\ell=1}^{p} \ell k_\ell )!(p!)^{k_p} k_p!} ((\bul{\Delta})^p[E])^{k_p}  (\bul{\Delta}_\hbar)^{- p k_p} \pd_E^{k_p} \right] (\bul{\Delta}_\hbar)^{n}[f]. \label{eq:alienDnf} \nl
\ee

We apply the above alien calculus to our case in Sec.~\ref{sec:PT_sym_mass} by beginning with
\be
&&   {\frak D}_{\cal PT} \propto 1 + A, \label{eq:DPT_app} \\  
&&  {\frak S}^{\nu}[A] = A (1 + B)^{-2 \nu}, \qquad {\frak S}^{\nu}[B] = B. \qquad  (\nu \in {\mathbb R}) \label{eq:DDP_app}
\ee
Our strategy to obtain Stokes automorphism for the energy solution is determining its transformation law to make the QC (\ref{eq:DPT_app}) invariant, i.e. keeping zero, under the DDP formula of cycles.
Remind that, in the DDP formula for the cycles (\ref{eq:DDP_app}), the energy $E$ (or $\widetilde{E} = E/\hbar$) is a \textit{free parameter}.
In this sense, the Stokes automorphism in Eq.(\ref{eq:DDP_app}) should be described only by $\bul{\Delta}_\hbar$ without $\bul{\Delta}[E] \pd_E$, and thus Eq.(\ref{eq:DDP_app}) is expressed by
\be
(\bul{\Delta}_\hbar)^{n \in {\mathbb N}}[ \log A] =
\begin{cases}
  - 2  \log \left( 1 + B \right) & \mbox{for} \quad n=1 \\
  0 & \mbox{otherwise}
\end{cases}, \qquad 
(\bul{\Delta}_\hbar)^{n \in {\mathbb N}}[B] = 0. \label{eq:DDP_alien_cycles}
\ee
From Eqs.(\ref{eq:alienDnf})(\ref{eq:DDP_alien_cycles}), actions of the (total) alien derivative to the $A$-cycle is given by
\be
\bul{\Delta} [\log A] 
&=& \bul{\Delta} [E] \pd_E \log A  - 2 \log (1+ B), \label{eq:Del1logA} \\
(\bul{\Delta})^2 [\log A]  &=& \left[ (\bul{\Delta}[E])^2 \pd_E^2 +  (\bul{\Delta})^2[E] \pd_E \right] \log A - 4 \bul{\Delta}[E] \pd_E \log(1+B), \label{eq:Del2logA} \\
(\bul{\Delta})^3 [\log A] &=&  \left[ (\bul{\Delta}[E])^3 \pd_E^3 + 3 (\bul{\Delta})^2[E] \bul{\Delta}[E] \pd_E^2 
  +  (\bul{\Delta})^3[E] \pd_E \right] \log A \nl
&& -6 \left[ (\bul{\Delta})^2[E]  \pd_E + (\bul{\Delta}[E])^2 \pd_E^2 \right]  \log (1+B)
, \\
 \vdots && \nl
 (\bul{\Delta})^n [\log A] &=& \sum_{\substack{k_1,\cdots,k_n = 0; \\ k_1+2 k_2 + \cdots + n k_n \le n}} \left[ \prod_{p=1}^{n} \frac{(\sum_{\ell=p}^{n} \ell k_\ell )!}{(\sum_{\ell=p+1}^{n} \ell k_\ell )!(p!)^{k_p} k_p!} ((\bul{\Delta})^p[E])^{k_p}  \pd_E^{k_p} \right] \log A \nl
 && -2
\sum_{\substack{k_1,\cdots,k_{n-1} = 0; \\ k_1+2 k_2 + \cdots +(n-1) k_{n-1} \le n-1}} \left[ \prod_{p=1}^{n-1} \frac{(1+\sum_{\ell=p}^{n-1} \ell k_\ell )!}{(1+ \sum_{\ell=p+1}^{n-1} \ell k_\ell )!(p!)^{k_p} k_p!} ((\bul{\Delta})^p[E])^{k_p} \pd_E^{k_p} \right] \log (1+B). \label{eq:DelnlogA} \nl 
\ee
Requiring $(\bul{\Delta})^{n \in {\mathbb N}}[{\frak D}_{\cal PT}] = 0$ is equivalent to $(\bul{\Delta})^{n \in {\mathbb N}} [\log A] = 0$.
Hence, solving Eqs.(\ref{eq:Del1logA})-(\ref{eq:DelnlogA}) to be zero determines the alien derivative to the energy, $(\bul{\Delta})^{n \in {\mathbb N}}[E]$, recursively.
We replace the Airy-type cycles with the DW-type as $(A,B) \rightarrow ({\frak A},{\frak B})$, and thus Eqs.(\ref{eq:Del1logA})(\ref{eq:Del2logA}), for example, leads to
\be
\bul{\Delta} [\widetilde{E}_{\cal PT}] &=&  - \frac{\log (1+ {\frak B})}{\pi i \pd_{\widetilde{E}} F} = O({\frak B}), \label{eq:Del_EPT}\\
(\bul{\Delta})^2[\widetilde{E}_{\cal PT}] &=&  \frac{1}{\pi^2 (\pd_{\widetilde{E}} F)^3} \left( \pd_{\widetilde{E}}^2 F -  \pd_{\widetilde{E}} F  \pd_{\widetilde{E}} \right) \left[ \log (1+ {\frak B}) \right]^2 = O({\frak B}^2, {\frak B} \pd_{\widetilde{E}} {\frak B}), \label{eq:Del_EPT2}
\ee
where we used ${\frak A} = e^{-2 \pi i F}$.
If one wants to identify non-perturbative contributions from each the singular point like Eq.(\ref{eq:Stokes_alien}), it is easily obtained by expanding Eq.(\ref{eq:Del_EPT}) by ${\frak B}$.
It can be written as
\be
&& \bul{\Delta}_{w_\ell}  [\widetilde{E}_{\cal PT}]
= (-1)^{\ell} \frac{{\frak B}^\ell}{\ell \pi i \pd_{\widetilde{E}} F}, \qquad \Gamma(\theta = 0) = \left\{ w_\ell = \frac{2}{3}\ell, \ell \in {\mathbb N} \right\}. \label{eq:Del_EPT_exp}
\ee
The case of higher derivatives, $(\bul{\Delta})^n_{w_\ell}  [\widetilde{E}_{\cal PT}]$, is also obtained in the same way.
Finally, substituting specific forms of $F$ and ${\frak B}$ in Eqs.(\ref{eq:rep_mass})(\ref{eq:F_G}) and the ${\cal PT}$ energy solution in Eq.(\ref{eq:E0_mass}) into Eqs.(\ref{eq:Del_EPT})(\ref{eq:Del_EPT2}) gives
\be
\bul{\Delta} [\widetilde{E}_{\cal PT}] &=& - 2 i \sigma \left[ 1 -\frac{q (q+6)}{8} \hbar +  \frac{q^4+q^3-102 q^2-43 q-134}{128} \hbar^2 \right. \nl
  &&  \left. \qquad \ \ -\frac{q \left(q^5-15 q^4-184 q^3+4371 q^2+2400 q+20484\right)}{3072} \hbar^3 + O(\hbar^4) \right],  \nl
&& + \pi i \sigma^2 \left[ 1  - \frac{q(q + 3)}{4} \hbar + \frac{2 q^4+q^3-51 q^2-43 q-67}{64} \hbar^2  \right. \nl
  &&  \left. \qquad \ \  -  \frac{q \left(2 q^5-15 q^4-92 q^3+996 q^2+1200 q+5121\right)}{768} \hbar^3 + O(\hbar^4) \right] + O(\sigma^3), \label{eq:DEPT_res1} \nl \\ 
(\bul{\Delta})^2 [\widetilde{E}_{\cal PT}] &=& \sigma^2 \left[ 4 \zeta   + \left( 2q + 3 \right) \hbar  - q(q + 3)\zeta \hbar \right. \nl
  &&  \left. \quad \   - \frac{8 q^3 + 3 q^2 - 102 q - 43}{16} \hbar^2 + \frac{2 q^4+q^3-51 q^2-43 q-67}{16} \zeta \hbar^2 \right. \nl
  && \left. \quad \ + \frac{12 q^5-75 q^4-368 q^3+2988 q^2+2400 q+5121}{192} \hbar^3 \right. \nl
    && \left. \quad \  -  \frac{q \left(2 q^5-15 q^4-92 q^3+996 q^2+1200 q+5121\right)}{192} \zeta \hbar^3  + O(\hbar^4) \right] + O(\sigma^3),  \label{eq:DEPT_res2} \nl
\ee
where $q \in 2 {\mathbb N}_0 + 1$ is the energy level, and $\sigma$ and $\zeta:= {\rm Re}[\zeta_\pm]$ are defined from Eq.(\ref{eq:sig_zeta}).
Notice that $(\bul{\Delta}) [\widetilde{E}_{\cal PT}]$ and $(\bul{\Delta})^2 [\widetilde{E}_{\cal PT}]$ are imaginary and real values, respectively.
The higher order derivatives can be found in the similar way.

\bibliographystyle{utphys}
\bibliography{PT_symmetry.bib}

\end{document}